\documentclass[12pt,a4paper]{article}
\usepackage[dvips]{graphicx}

\newcommand{\be}{\begin{eqnarray} }
\newcommand{\ee}{\end{eqnarray} }
\newcommand{\beq}{\begin{equation} }
\newcommand{\eeq}{\end{equation} }

\newcommand{\simleq}{\; \raisebox{-0.4ex}{\tiny$\stackrel
{{\textstyle<}}{\sim}$}\;} 
\newcommand{\ab}{{(\alpha,\beta)}}
\newcommand{\palka}{{\Bigl |}}

\begin{document}
\begin{center}
        {\large
        NLO QCD method of the polarized SIDIS data analysis
        }
\end{center}
\begin{center}
\vskip 1.5cm
{ \large A.N.~Sissakian}
\footnote{E-mail address: sisakian@jinr.ru},  
{\large O.Yu.~Shevchenko}
\footnote{E-mail address: shev@mail.cern.ch},
{\large O.N.~Ivanov}
\footnote{E-mail address: ivon@jinr.ru}\\
\vspace{1cm}
{\it Joint Institute for Nuclear Research}
\end{center}
\begin{abstract}
Method of polarized semi-inclusive deep inelastic scattering (SIDIS) data analysis in
the next to leading order (NLO) QCD is developed. 
Within the method one first  directly extracts in NLO few first 
truncated (available to measurement) Mellin moments of the quark helicity distributions. 
Second, using these moments as an input to the proposed modification
of the Jacobi polynomial expansion method (MJEM), one eventually reconstructs the local quark
helicity distributions themselves.
All numerical tests demonstrate that  MJEM allows us 
to reproduce with the high precision the input local  distributions even inside
the narrow  Bjorken $x$ region  accessible for experiment. 
It is of importance that only four first input  moments
are sufficient to achieve a good quality of reconstruction.
The application of the method to the simulated SIDIS data on the pion production
is considered. The obtained results encourage one that the proposed NLO method can 
be successfully applied to the SIDIS data analysis. 
The  analysis of HERMES data on pion production is performed. 
To this end the pion difference asymmetries are constructed from the measured by HERMES
standard semi-inclusive spin asymmetries.
The LO results of the valence distribution reconstruction are in a good accordance with
the respective leading order SMC and HERMES results, while the NLO results are in agreement 
with the existing NLO parametrizations on these quantities.
\end{abstract}


\section{Introduction}
One of very important topics in the modern  high energy physics is the  investigation  of the partonic spin structure of nucleon.
In this connection, nowadays, there is a huge growth of interest to the SIDIS experiments with
longitudinally polarized beam and target such as SMC \cite{smc}, HERMES \cite{hermes}, COMPASS \cite{compass}.
It is of importance that the SIDIS experiments, where one identifies the hadron in the final state,
provide us with the additional information on the partonic spin structure in comparison with
the usual DIS experiments. Namely, on the contrary to the DIS data,  the SIDIS data allows us to extract the sea and valence quark helicity
distributions in separation.

At the same time it is argued (see, for example, Ref. \cite{our1}) that to obtain the reliable
distributions at relatively low average $Q^2$ available to the modern
SIDIS experiments\footnote{For example, HERMES  data \cite{hermes} on semi-inclusive asymmetries 
is obtained at $Q^2_{\rm mean}=2.5\,GeV^2$ }, 
the leading order (LO) analysis is not sufficient  and NLO
analysis is necessary. In Ref. \cite{our2} it  was proposed the procedure allowing the direct 
extraction from the SIDIS data  of the first moments (truncated to the accessible for measurement Bjorken $x$ 
region) of the quark helicity distributions in NLO QCD.
However, in spite of the special importance of the first moments\footnote{Let us recall that namely these
quantities, first moments, are of the most importance for solution of the 
proton spin puzzle because namely these quantities compose the nucleon spin}, it is certainly very
desirable to have the procedure of reconstruction in NLO QCD of the 
polarized densities themselves. At the same time, it is extremely difficult to extract the local
in Bjorken $x$ ($x_B$) distributions directly,  because of the
double convolution product entering the NLO QCD expressions for the semi-inclusive asymmetries\footnote{
So, on the contrary to LO, where direct extraction of PDFs is possible,  it seems at first sight
that dealing with SIDIS asymmetries in NLO one can not avoid some fitting procedure.
However, the modern world SIDIS data provide us by the rather small number
of points for the measured asymmetries (and, besides, they suffer from the large
statistical errors). Thus, purely semi-inclusive data very weakly constrains  
the large number  of fit parameters entering NLO analysis  
(for example, twenty free parameters are used in Ref. \cite{florian-new}). At the same time, the addition
of DIS data in analysis can not help us to solve the main task of SIDIS -- to extract the valence, 
sea and strange PDFs in separation.
} 
(see Ref. \cite{our2} and references therein). Fortunately, operating just  as in Ref. \cite{our2},
one can directly extract not only the first moments, but the Mellin moments of any required order.
Using the truncated moments of parton distribution functions (PDFs) and  applying the modified Jacobi polynomial expansion method (MJEM) proposed in Ref.  \cite{our3} 
one can reconstruct PDFs themselves in the entire accessible for measurement $x_B$ region.
In the brief letter \cite{our3}  MJEM was tested using only the simple numerical (idealized) examples,
where the exact values of the input moments   entering MJEM are known (see Section 2).
However, in the conditions of the experiment we have at our disposal only rather small number of the
measured  asymmetry values (one point for each bin with the rather wide bin widths at the middle and large $x_B$).
Thus, extracting the moments from the measured asymmetries one calculates the integrals over
$x_B$ using rather small\footnote{For example, HERMES used 9 bins for the region $0.023<x<0.6$
and COMPASS used 12 bins for the region $0.003<x<0.7$.} number of points. 
So, because of this problem, even for the data obtained with very high precision
(small errors) the extracted moments always suffer from the deviation  from their true values. 
In this extended paper we investigate this problem in detail (see Section 3). 
In Section 3 MJEM is tested with the simulations corresponding
to the kinematics of the HERMES experiment, where the  accessible $x_B$ region  is the most narrow 
in comparison with SMC and COMPASS regions.

After the testing the proposed method is applied to the NLO QCD analysis of the HERMES
data (Section 4).
Notice that although the method is quite general, 
within this paper we deal only with the SIDIS data on the  pion production.
The point is that here we would like first of all to see how well the method itself works.
That is why, for a moment, 
we do not like to deal with the such poorly known objects as $D_q^{K^\pm}$ and $D_g^h$ 
fragmentation functions which additionally introduce the big uncertainties in the analysis
results. For example, the analysis performed in Ref. \cite{florian-new} shows that 
the different choices of parametrizations for these fragmentation functions lead to
the strong disagreement in the obtained results (about 30\% for valence quarks and about 100\%
for sea quarks).
From this point of view 
the most attractive objects  are the difference asymmetries \cite{frankfurt} (for details on
difference asymmetry in NLO see \cite{christova, our2} and references therein),
where the fragmentation functions are cancel out in LO, while in NLO the difference asymmetry  
has only weak dependence of the difference of the favored and unfavored pion fragmentation functions 
(known with a good precision).
In Section 4 the pion difference asymmetries are constructed from the measured by HERMES
standard semi-inclusive spin asymmetries. Using as a starting point the constructed in such
a way difference asymmetries, the (preliminary) reconstruction in NLO QCD of the valence
PDFs from the HERMES data is performed.

\section{MJEM  and the usual JEM in comparison. Numerical tests.}
In this section we, for the sake of  selfconsistence and clarity, represent in more detail the results of Ref. \cite{our3}.

There exist several methods   allowing to reconstruct the local in $x_B$ quantities
(like structure functions, polarized and unpolarized quark distributions, etc)
knowing only  finite  
number of numerical values of their  Mellin moments.
All of them use the expansion of the local
quantity in the series over the orthogonal polynomials (Bernstein, Laguerre, Legendre, Jacobi)  
-- see Ref. \cite{sidorov} and references therein.
The most successful in applications (reconstruction of the local distributions
from the evolved with GLAP moments and investigation of $\Lambda_{\rm QCD}$) occurred the
Jacobi polynomial expansion method (JEM)
proposed in the pioneer work by Parisi and Sourlas \cite{parisi}
and elaborated\footnote{JEM with respect to polarized quark densities
was first applied in Ref. \cite{sidorov2}} in Refs. \cite{barker} and \cite{sidorov}.

The local in $x_B$ functions (structure functions or quark distributions) are expanded in the
double series over the Jacobi polynomials and Mellin moments (see the Appendix):
\be
\label{sj}
 &  & F(x)\simeq F_{N_{max}}(x)=  \omega^\ab(x)\sum\nolimits_{k=0}^{N_{max}}\Theta_k^\ab(x)\nonumber\\
 &  & \times\sum\nolimits_{j=0}^kc_{kj}^{\ab}M(j+1),
\ee
where $\omega^\ab(x)=x^\beta(1-x)^\alpha$ and $N_{max}$ is the number 
of moments left\footnote{Expansion (\ref{sj}) becomes exact
when $N_{max}\rightarrow\infty$. However, the advantage of JEM is that even truncated series
with the small number of used  moments $N_{max}$ and properly fixed parameters $\alpha,\beta$ 
gives the good results (see, for example, \cite{sidorov})} in the expansion.
For what follows it is of importance that the moments entering Eq. (\ref{sj}) are the {\it full} moments,
i.e., the integrals over the entire Bjorken $x$ region $0<x<1$:
\be
\label{fullmoment}
M[j]=\int_0^1dx x^{j-1} F(x).
\ee
Till now nobody investigated the question of applicability of JEM 
to the rather narrow $x_B$ region available
to the modern polarized SIDIS experiments.
So, let us try to apply JEM to the reconstruction of $\Delta u_V(x)$ and  $\Delta d_V(x)$ in
the rather narrow $x_B$ region\footnote{We choose here the most narrow HERMES 
$x_B$ region where the difference between JEM and its modification MJEM (see below) application  becomes 
especially impressive. However, even with the more wide accessible $x_B$ region 
(for example, COMPASS region \cite{compass}) it is of importance to avoid the additional
systematical errors caused by the replacement of the full (unaccessible) moments 
in JEM (\ref{sj}) by the accessible  truncated moments.
} 
$ 
\label{region}
a=0.023<x<b=0.6
$
available to HERMES,
and to investigate
is it possible to safely replace the full moments (\ref{fullmoment}) by the truncated moments. 
To this end we perform the simple
test. We choose\footnote{Certainly, one can choose for testing any other parametrization.} GRSV2000{\bf NLO} (symmetric sea) parametrization \cite{grsv2000} at $Q^2=2.5\,GeV^2$.
Integrating the parametrizations on $\Delta u_V$ and $\Delta d_V$ over the HERMES $x_B$ region we calculate twelve truncated moments 
given by (c.f. Eq. (\ref{qtrunc}) below)
\be
\label{trm}
M'[j]\equiv M'_{[a,b]}[j]\equiv\int_a^bdx x^{j-1} F(x),
\ee
where we put  $F(x)=\Delta u_V(x)$ or  $F(x)=\Delta d_V(x)$ and choose $a=0.023,b=0.6$ .
Substituting these  moments in the  expansion
Eq. (\ref{sj}) with $N_{max}=12$, we look for optimal values of parameters $\alpha$ and $\beta$
corresponding to the minimal deviation of reconstructed curves for $\Delta u_V(x)$ and $\Delta d_V(x)$
from the input (reference) curves corresponding to input parametrization. To find these
optimal values $\alpha_{opt}$ and $\beta_{opt}$ we use the program MINUIT \cite{minuit}.
The results are presented in Fig. \ref{f1}. 
Looking at Fig. \ref{f1}, one can see that the curves strongly differ from each other      
even for the high number of used moments $N_{max}=12$.

Thus, the substitution of truncated moments instead of exact ones in the expansion
(\ref{sj}) is a rather crude approximation at least for HERMES $x_B$ region.
Fortunately it is possible to modify the standard JEM in a such way that new
series contains the truncated moments instead of the full ones. The new expansion 
looks as (see the Appendix)
\be
\label{mj}
&  & F(x)\simeq F_{N_{max}}(x)= \left(\frac{x-a}{b-a}\right)^\beta\left(1-\frac{x-a}{b-a}\right)^\alpha\nonumber\\
 & &  \times  \sum_{n=0}^{N_{max}} \Theta_n^\ab\left(\frac{x-a}{b-a}\right)
\sum_{k=0}^n c_{nk}^\ab
\frac{1}{(b-a)^{k+1}}\sum_{l=0}^k\frac{k!}{l!(k-l)!}M'[l+1](-a)^{k-l},
\ee
where we use the notation Eq. (\ref{trm})  
for the moments truncated to accessible for measurement $x_B$ region.
It is of great importance that now in the expansion enter not the full (unavailable) but the truncated (accessible) moments.
Thus, having at our disposal few first truncated moments extracted in NLO QCD (Eqs. (\ref{fmain}) below),
and applying MJEM, Eq. (\ref{mj}), one can reconstruct  the local 
distributions in the accessible for measurement
$x_B$ region.

To proceed let us clarify the important question about the boundary distortions.
The deviations of reconstructed with MJEM, Eq. (\ref{mj}), $F_{N_{max}}$ from F
near the boundary points are unavoidable since MJEM  is correctly defined in the entire
region $(a,b)$ except for the small vicinities of boundary points (see the Appendix). 
Fortunately,  $F_{N_{max}}$ and F are in very good agreement in the practically
entire accessible $x_B$ region,
while the boundary distortions are easily identified  and controlled since they are very sharp
and hold in very small vicinities of the boundary points (see Figs. \ref{f2}-\ref{f4} below). 
In this section we, for clarity, explicitly show these distortions in all figures.
In the next sections the all such distortions will be just cutted off.

Let us check how well MJEM works. To this end let us repeat the simple exercises
with reconstruction of the known GRSV2000{\bf NLO} (symmetric sea)
parametrization and compare the results of $\Delta u_V(x)$
and $\Delta d_V(x)$ reconstruction with the usual JEM and with the proposed MJEM. 
To control the quality of reconstruction we introduce the parameter\footnote{
Calculating  $\nu$ we just cut off the boundary
distortions which hold for MJEM in the small vicinities of the boundary points (see the Appendix),
and decrease the integration region, respectively.
To be more precise, one can apply after cutting  some extrapolation to the boundary points.
However, the practice shows that  the results on $\nu$ calculation are practically insensitive
to the way of extrapolation since the widths of the boundary distortion regions are very small (about $10^{-3}$).
} 
\be
\label{nu}
\nu=\frac{\int_a^b dx|F_{reconstructed}(x)-F_{reference}(x)|}{\int_a^b dx|F_{reference}|}\cdot 100\%,
\ee
where $F_{reference}(x)$ corresponds to the input parametrization and $F_{reconstructed}(x)\equiv F_{N_{max}}(x)$ in Eq. (\ref{mj}).
We first perform the reconstruction with very high number of moments 
$N_{max}=12$ 
and then with the small number $N_{max}=4$. Notice that the last
choice $N_{max}=4$
is especially important because of peculiarities of the data on asymmetries
provided by the SIDIS experiments.
Indeed, the number of used moments should be as small as possible
because 
first, the relative error $|\delta (M'[j])/M'[j]|$ on $M'[j]$ becomes higher with increase of $j$ and 
second, the high moments become very sensitive to the replacement of integration by the sum
over the bins. The results of $\Delta u_V(x)$ and $\Delta d_V(x)$ reconstruction
with  MJEM at $N_{max}=12$ and with application of both JEM and MJEM (in comparison) at $N_{max}=4$,
are presented in Figs. \ref{f2} and \ref{f3}.
It is seen (see Fig. \ref{f2}) that  for $N_{max}=12$ MJEM,  
on the contrary to the usual JEM (see Fig. \ref{f1}), 
gives the excellent agreement between the  reference and reconstructed curves. In the case
$N_{max}=4$ the difference in quality of reconstruction between JEM and MJEM (see Fig. \ref{f3}) becomes 
especially impressive. 
While for standard JEM the reconstructed and reference curves strongly 
differ from each other, the respective curves for MJEM are in a good 
agreement. Thus, one can conclude that dealing with the truncated, available to measurement,
$x_B$ region one should apply the proposed modified JEM to obtain
the reliable results on the local distributions.

Until now we looked for  the optimal values of parameters $\alpha$ and $\beta$ entering 
MJEM using explicit form of the reference curve (input parametrization). Certainly, in reality
we have no any reference curve to be used for optimization. However, one can extract from the data
in NLO QCD the first few moments (Eqs. (\ref{fmain}) below). Thus, we need some
criterion  of MJEM optimization which would use for optimization of $\alpha$ and $\beta$
only the known (extracted) moments entering MJEM.

On the first sight it seems to be natural to find the optimal values of $\alpha$ and $\beta$ minimizing
the difference of {\it reconstructed} with MJEM and input\footnote{In practice one should
reconstruct these input moments from the data using Eqs. (\ref{fmain}) (below). 
The reference ``twice-truncated''
moments (\ref{trm2}) should be reconstructed from the data in the same way.} ({\it entering}
MJEM (\ref{mj}))  moments.
However, it is easy to prove (see the Appendix) 
that this difference is equal to zero: 
\be
\label{meq}
M'_{[a,b]}[n]{\Biggl |}_{\rm reconstructed}= M'_{[a,b]}[n]{\Biggl |}_{input},\quad n\le N_{max},
\ee
i.e. all reconstructed moments with $n\le N_{max}$ are identically equal to the respective
input moments {\it for any} $\alpha$ and $\beta$. Fortunately, we can use for comparison
the reference  ``twice-truncated'' moments
\be
\label{trm2}
M''[n]\equiv M''_{[a+a',b-b']}[n]\equiv\int_{a+a'}^{b-b'} dx\, x^{n-1} F(x)\quad
(a<a+a'<b-b'<b),
\ee
i.e. the integrals over the region less than the integration region $(a,b)$ for the  
``once-truncated''  moments $M'_{[a,b]}$ entering MJEM (\ref{mj}). The respective optimization criterion 
can be written in the form
\be
\label{criterion}
\sum_{j=0}^{N_{max}}{\Bigl |}M''_{\rm (reconstructed)}[j]-M''_{\rm (reference)}[j]{\Bigl |}=min.
\ee
The ``twice truncated'' reference moments should be extracted in NLO QCD from the data in the same way
as the input (entering MJEM (\ref{mj})) ``once truncated'' moments. In reality
one can obtain ``twice-truncated'' moments using Eqs. (\ref{fmain}) (below) and removing, for example, first and/or last bin from the sum in Eq. (\ref{summ}) (below).

Let us now check how well the optimization criterion (\ref{criterion})
works. 
To this end we again perform the simple numerical
test. We choose GRSV2000{\bf NLO} parametrization at $Q^2=2.5\,GeV^2$  with both broken and  symmetric sea
scenarios. We then calculate four first ``once-truncated'' and four first ``twice-truncated'' moments 
defined by Eqs. (\ref{trm}) and (\ref{trm2}), and then substitute them in Eq. (\ref{mj}) and the optimization criterion 
(\ref{criterion}), respectively. To find the optimal values of $\alpha$ and $\beta$ we use the MINUIT \cite{minuit}
program. The results are presented in Fig. \ref{f4}.
It is seen that the optimization criterion works well for both symmetric and broken
sea scenarios.

Thus, the performed numerical tests show that the proposed modification of the 
Jacobi polynomial expansion method allows to reconstruct with a high precision
the quark helicity distributions in the accessible for measurement $x_B$ region.

\section{Reconstruction of
the valence quark helicity distributions from the simulated data.}
In this section the proposed NLO QCD method will be applied to the simulated data. 
The simulations give us a good tool for testing of the method since  here one knows {\it in advance}
the answer to be found -- the reference parametrization entering the generator as an input. 
At the same time 
the properly performed  simulations  (i.e., corresponding to the experimental statistics, 
binning and kinematical cuts) allow us finally adapt the method for application to the real
experimental data.

Let us first investigate the peculiarities of the n{\it th} moments extraction 
in the conditions of the real experimental binning (rather small number of bins covering the accessible 
for measurement $x_B$ region).

The simple extension of the procedure proposed in Ref. \cite{our2} gives for the n-th moments
$
\Delta_n q\equiv \int_0^1dx\, x^{n-1} q(x)
$ of the valence distributions
the equations
\be
\label{fmain}
\Delta_n u_V=\frac{1}{5}\frac{{\cal A}_p^{(n)}+{\cal A}_d^{(n)}}{L_{(n)1}-L_{(n)2}};
\quad  
\Delta_n d_V=\frac{1}{5}\frac{4
{\cal A}_d^{(n)}-{\cal A}_p^{(n)}}{L_{(n)1}-L_{(n)2}}.
\ee
Here the notation is absolutely analogous to one used in Ref.  \cite{our2}:
\begin{eqnarray}
\label{fap}
{\cal A}_p^{(n)}\equiv \int_0^1 dx\ x^{n-1}A_p^{\pi^+-\pi^-}
{\Biggl |}_Z(4u_V-d_V)\nonumber\\
\times
\int_Z^1 dz_h[1+\otimes \frac{\alpha_s}{2\pi}
{C}_{qq}\otimes]
(D_1-D_2), \\
\label{fad}
 {\cal A}_d^{(n)}\equiv \int_0^1 dx\ x^{n-1}
 A_d^{\pi^+-\pi^-}{\Biggl |}_Z(u_V+d_V)\nonumber\\
\times\int_Z^1 dz_h[1+\otimes
\frac{\alpha_s}{2\pi}
{C}_{qq}\otimes](D_1-D_2),
\end{eqnarray}
where $D_1$ ($D_2$) is favored (unfavored) pion fragmentation function,
the quantities $L_{(n)1}$, $L_{(n)2}$  are defined as 
\begin{eqnarray}
&&L_{(n)1}\equiv L_{(n)u}^{\pi^+}=L_{(n)\bar{u}}^{\pi^-}=L_{(n)\bar{d}}^{\pi^+}
=L_{(n)d}^{\pi^-},
\nonumber\\
&&L_{(n)2}\equiv L_{(n)d}^{\pi^+}=L_{{(n)}\bar{d}}^{\pi^-}=L_{(n)u}^{\pi^-}
=L_{(n)\bar{u}}^{\pi^+},\\
%
\label{lcoef}
&&L_{(n)q}^h\equiv \int_Z^1 dz_h\left[D_q^h(z_h)+
\frac{\alpha_s}{2\pi}\int_{z_h}^1
\frac{dz'}{z'}\ \Delta_n C_{qq}(z')D_q^h(\frac{z_h}{z'})\right],\nonumber
\ee
where 
\be
\Delta_n C_{qq}(z)\equiv \int_0^1dx\ x^{n-1} \Delta C_{qq}(x,z)\nonumber
\ee
are the n{\it th} moments of the polarized Wilson coefficients $\Delta C_{qq}(x,z)$ 
entering the 
NLO expressions for the  difference asymmetries $A_{p,d}^{\pi^+-\pi^-}$
(the respective experimental expressions via counting rates are given by Eq. (\ref{difas}) -- see below): 
\be
        \label{eold11}
A_p^{\pi^+-\pi^-}(x,Q^2){\Bigl |}_Z=\frac{(4\Delta u_V-\Delta d_V)
\int_Z^1 dz_h[1+\otimes\frac{\alpha_s}{2\pi}\Delta C_{qq}\otimes]
(D_1-D_2)}
{(4u_V-d_V)\int_Z^1 dz_h[1+\otimes\frac{\alpha_s}{2\pi}{C}_{qq}
\otimes ](D_1-D_2)},\nonumber
\ee
\be
\label{eold12}
A_d^{\pi^+-\pi^-}(x,Q^2){\Bigl |}_Z=\frac{(\Delta u_V+\Delta d_V)
\int_Z^1 dz_h[1+\otimes\frac{\alpha_s}{2\pi}\Delta C_{qq}\otimes]
(D_1-D_2)}
{(u_V+d_V)\int_Z^1 dz_h[1+\otimes\frac{\alpha_s}{2\pi}{C}_{qq}
\otimes ](D_1-D_2)}.\nonumber
\ee

It should be noticed that in reality one can measure the asymmetries only in the restricted
Bjorken $x$ region $a<x<b$, so that the approximate equations for the truncated moments (c.f. Eq. (\ref{trm}))
\be
\label{qtrunc}
\Delta'_n q\equiv M'[n]\equiv \int_a^bdx\, x^{n-1} \Delta q(x)
\ee
of the valence distributions have the form Eq. (\ref{fmain}) 
with the replacement of the full integrals in Eq. (\ref{fap})
by the sums over bins covering the accessible region $a<x<b$:
\be
\label{summ}
{\cal A}_p^{(n)}\simeq \sum_{i=1}^{N_{bins}} x^{n-1}\Delta x_i\,  A_p^{\pi^+-\pi^-}(x_i)
{\Bigl |}_Z
(4u_V-d_V)(x_i)\int_Z^1 dz_h[1+\otimes \frac{\alpha_s}{2\pi}
{C}_{qq}\otimes]
(D_1-D_2),
\ee
and analogously for ${\cal A}_d^{(n)}$.

The  approximation  to Eq. (\ref{fap}) given by Eq. (\ref{summ}) is based on the assumption
that {\it all} integrated quantities are the constants\footnote{Operating in such a way
one puts the $x$-dependent measured quantity (difference asymmetries $A_{p(d)}^{\pi^+-\pi^-}(x_i)$ here) 
to be equal its mean value in the i{\it th} bin, 
while the $x$-dependent rest is calculated in the point $x_i\equiv \langle x_i\rangle$.} within each bin. 
This is well-known ``middle point'' numerical integration method.

However, it seems that there is a way to improve this approximation having in mind the real
experimental situation. The point is that the reality of an experiment compel us to
approximate by the constant within the bin
only the measured quantity (difference asymmetry here), that can be written as  
\be
\label{parametr}
A_p^{\pi^+-\pi^-}(x)\palka_Z=\sum_{i=1}^{N_{bins}}A_p^{\pi^+-\pi^-}(\langle x_i\rangle )\palka_Z\theta(x-x_{i-1})\theta(x_i-x),
\ee
where  $A_p^{\pi^+-\pi^-}(\langle x_i\rangle)\palka_Z$ is the mean value of asymmetry in i{\it th} bin, 
$x_0=a$, $x_{N_{bins}}=b$ and $\theta(x)$ is the usual step function. At the same time, there is 
no any need to approximate by the constant another $x$-dependent quantities (unpolarized valence PDFs
and Wilson coefficients here) entering the integrals over $x_B$ as the known input.
Thus, substituting Eq. (\ref{parametr}) in the initial integral equation Eq. (\ref{fap}) we get (c.f. Eq. (\ref{summ})) 
\be
\label{aint}
{\cal A}_{p}^{(n)}=\sum_{i=1}^{N_{bins}}A_p^{\pi^+-\pi^-}(\langle x_i\rangle){\Bigl |}_Z\int_{x_{i-1}}^{x_i}dx x^{n-1}(4u_V-d_V)(x)\int_Z^1 dz_h[1+\otimes \frac{\alpha_s}{2\pi}
{C}_{qq}\otimes]
(D_1-D_2).
\ee

Notice that the analogous way of the integral approximation was applied by the HERMES collaboration 
in Ref. \cite{hermes}, where
the moments $\Delta'_n q$ were reconstructed substituting 
extracted from the data (``measured'') quantities $(\Delta q/q)(x_i)$  
in the equation (see Eq. (46) in Ref. \cite{hermes})
\be
\Delta'_n q=\int_{0.023}^{0,6}dx\sum_{i=1}^{N_{bins}} \left[\frac{\Delta q}{q}(\langle x_i\rangle)\theta(x-x_{i-1})\theta(x_i-x)\right] x^{n-1} q(x).
\ee
To see the advantage of Eq. (\ref{aint}) application, let us 
compare it with the application of the integration procedure  given by  Eq. (\ref{summ}). 
To this end we perform 
absolutely idealized  LO test, where in each bin (we choose the HERMES binning) 
the value of asymmetry is directly calculated
from the given\footnote{We choose for illustration GRSV2000{\bf LO} (symmetric sea) parametrization. 
At the same time, it is easy to check that absolutely the same same picture holds for any other parametrization.} 
parametrization  on $\Delta u_V$ and $\Delta d_V$ using the theoretical LO expressions \cite{frankfurt} 
for
the difference asymmetries  
\be
\label{difaslo}
A_{p}^{\pi^+-\pi^-}=\frac{4\Delta u_V-\Delta d_V}{4 u_V- d_V};\quad A_{d}^{\pi^+-\pi^-}=\frac{\Delta u_V+\Delta d_V}{u_V+ d_V}.
\ee
For simplicity, within this test we put $\langle x_i\rangle=(x_{i}-x_{i-1})/2$ ($i=1, \ldots, 9; x_0=0.023,x_{9}=0.6$),
so that reconstructed with Eq. (\ref{difaslo}) values of  $\Delta u_V(\langle x_i\rangle)$ and 
$\Delta d_V(\langle x_i\rangle)$ exactly coincide with the respective input parametrization values in the points 
$\langle x_i \rangle$ -- see Fig \ref{fig-integr}.  Now we calculate four first moments using reduced 
to LO Eqs. (\ref{fmain}) and the 
integration procedures given by Eqs. (\ref{summ}) and  (\ref{aint}) and then we apply\footnote{
Here we find $\alpha_{opt}$ and $\beta_{opt}$ values requiring the minimal deviation of reconstructed 
with MJEM $\Delta u_V(\langle x_i\rangle)$ and $\Delta d_V(\langle x_i\rangle)$ from the 
reference parametrization values at the points $\langle x_i\rangle$. } MJEM to both sets 
of the obtained moments. Looking at Fig. \ref{fig-integr} one can see that reconstructed in this way curves 
strongly differ  from each other, and the curve obtained with application of the integration
procedure given by Eq. (\ref{aint}) is in much better agreement with the input (reference) parametrization.

Thus, following the results of just performed test, from now on
we will use namely Eq. (\ref{aint}) performing the 
moments calculations.

Let us now perform LO and NLO analysis of the simulated SIDIS data on $\pi^+$ and $\pi^-$ production with both
proton and deutron targets. To this end we use the PEPSI generator of polarized events \cite{pepsi}. 
The conditions of simulations
are presented in Table \ref{simtable} and correspond to the HERMES kinematics. 
Let us stress that all the cuts on $Q^2$, $x_F$, $W^2$ and $z_h$ in Table \ref{simtable} are the standard
physical\footnote{For example, the important cut on invariant mass $W^2>10\,GeV^2$ is applied 
by these collaborations to exclude the events coming from the resonance region.
} cuts applied by SMC, HERMES and COMPASS. 
The statistics $3\cdot10^6$ in Table \ref{simtable} is the total number of DIS events
for both proton and deutron targets and for both longitudinal polarizations. 

\begin{table}[h!]
        \caption{\footnotesize \footnotesize Simulation conditions. Here $x_{B}$ and  $x_F$
        are the Bjorken and Feynman variables, respectively, $z_h$ is the standard hadronic
        variable and $W$ is the invariant mass of the final hadronic state.
}         
\vskip0.3cm
 \begin{tabular}{cccc}
 \hline
 \hline
 $E_{lepton}$ & $x_{B}$ & $x_F$ & $z_h$ \\
\hline 
27.5 $GeV$ & $0.023<x_{B}<0.6$&$x_F>0.1$&$z_h>Z=0.2$\\
\hline\hline
 $W^2$ & $Q^2$ &       $Q^2_{mean}$ & Events\\
 \hline
$W^2>10\,GeV^2$  &$Q^2>1\,GeV^2$ &   2.4\,$GeV^2$  & $3\cdot10^6$\\
 \hline
 \end{tabular}
 \label{simtable}
\end{table}

Using the simulated data we construct the difference asymmetries (see Ref. \cite{our2} for details)
\be
\label{difas}
A^{\pi^+-\pi^-}_{p(d)}\palka_Z=
\frac{1}{P_BP_TfD}\frac{(N^{\pi^+}_
{\uparrow\downarrow}-N^{\pi^-}_{\uparrow\downarrow})L_{\uparrow\uparrow}-(N^{\pi^+}_
{\uparrow\uparrow}-N^{\pi^-}_{\uparrow\uparrow})L_{\uparrow\downarrow}}
{(N^{\pi^+}_{\uparrow\downarrow}-N^{\pi^-}_{\uparrow\downarrow})L_{\uparrow\uparrow}+
(N^{\pi^+}_{\uparrow\uparrow}-N^{\pi^-}_{\uparrow\uparrow})L_{\uparrow\downarrow}},
\ee 
where $N^{\pi^\pm}_{\uparrow\downarrow(\uparrow\uparrow)}$ are the counting rates integrated over $z_h$ in 
the region $Z=0.2<z_h<1$, 
$L_{\uparrow\downarrow(\uparrow\uparrow)}=N_{\uparrow\downarrow(\uparrow\uparrow)}/
\sigma_{\uparrow\downarrow(\uparrow\uparrow)}$  are the luminosities, 
and the quantities $p_B$, $p_T$, $f$ are equal to unity in the conditions of simulations
with PEPSI.

We first perform the LO analysis of the simulated difference asymmetries. The important peculiarity of 
LO analysis is that in this case one can perform the extraction of $\Delta u_V$ and $\Delta d_V$ in two ways.
First is the direct extraction where one applies Eqs. (\ref{difaslo}) in each bin -- points with
error bars in Fig. \ref{lodirect}. The second method is the proposed one, where MJEM
is applied to the LO extracted moments -- dashed line in Fig. \ref{lodirect}. 
The moments used in MJEM are extracted  
from the simulated difference asymmetries  with application
of reduced to LO Eqs. (\ref{fmain}-\ref{lcoef}), (\ref{aint})  and are  presented in Table \ref{losim}. 
\begin{table}[h!]
        \caption{\footnotesize Results for LO extracted truncated moments for the simulations with the entering
        PEPSI two different parametrizations: 
        GRSV2000LO (symmetric sea) parametrization (top) and GRSV2000LO (broken sea) 
        parametrization (bottom). For comparison the respective reference (obtained by direct
        integration of entering PEPSI input parametrizations) moments are also presented.}
        \begin{tabular}{ccccc}
                \hline
                \hline
                 & \multicolumn{2}{c}{$\Delta'_n u_V$} & \multicolumn{2}{c}{$\Delta'_n d_V$}\\
n&Extracted&Reference &Extracted&Reference \\
\hline
 1 &  0.7042 $\pm$ 0.0124 &0.7176& -0.2568 $\pm$ 0.0271&-0.2618\\
 2 &  0.1489 $\pm$ 0.0037 &0.1477& -0.0439 $\pm$ 0.0079&-0.0482 \\
 3 &  0.0467 $\pm$ 0.0016 &0.0457& -0.0118 $\pm$ 0.0033&-0.0135 \\
 4 &  0.0179 $\pm$ 0.0007 &0.0173& -0.0041 $\pm$ 0.0015&-0.0048 \\
 \hline\hline

                 & \multicolumn{2}{c}{$\Delta'_n u_V$} & \multicolumn{2}{c}{$\Delta'_n d_V$}\\
n&Extracted&Reference &Extracted&Reference \\
\hline
 1 &  0.5346 $\pm$ 0.0123 &0.5255&-0.0952 $\pm$ 0.0274&-0.1103\\
 2 &  0.1318 $\pm$ 0.0036 &0.1282&-0.0297 $\pm$ 0.0081&-0.0331\\
 3 &  0.0434 $\pm$ 0.0015 &0.0425&-0.0098 $\pm$ 0.0034&-0.0107\\
 4 &  0.0167 $\pm$ 0.0007 &0.0166&-0.0037 $\pm$ 0.0015&-0.0039\\
 \hline
\end{tabular}
\label{losim}
\end{table}

Looking at Fig. \ref{lodirect}, one can see that the input (reference) parametrization
slightly deviates from both  the directly extracted values of $\Delta u_V$ and $\Delta d_V$
and the reconstructed with MJEM curve. These deviations are unavoidable and are caused 
by the specific character of the events generation with PEPSI. Our experience shows that the
asymmetries reconstructed from the generated events always slightly differ from the respective 
asymmetries calculated from the input parametrizations (with application of Eq. (\ref{difaslo}) in LO).
One the other hand, comparing the directly extracted  and the reconstructed
with MJEM $\Delta u_V$ and $\Delta d_V$, one can see  that they are in a good agreement 
with each other. Thus, the performed LO testing encourage us that the proposed method of 
PDFs extraction  could be successfully applied.

Let us now clarify the important point concerning application of the
optimization criterion Eq. (\ref{criterion}) which we use  to find the optimal 
values  $\alpha_{opt}$ and $\beta_{opt}$ 
of the entering MJEM parameters $\alpha$ and $\beta$ (see Section 2 for details).
All over the paper, applying the optimization criterion,  
we simultaneously use for each $j$ ($j=1, \ldots 4$) in the sum two extracted twice-truncated 
moments. These moments 
correspond to  two (sufficiently large and overlapping) integration regions, covering respectively
the  bins from first to seven and from third to last ninth. 
We make such choice because  on the one hand, one should take into account in the criterion
the whole accessible integration region, and, on the other hand, the ``twice-truncated'' moments
should essentially differ from the ``once truncated'' moments  for the well-working of the 
minimization procedure (see the respective  discussion just after Eq. (\ref{meq})).

Let us now perform the NLO analysis of the simulated data. We again use as an input two different
parametrizations GRSV2000NLO (symmetric sea) and GRSV2000NLO (broken sea). The conditions of
simulation are presented in Table \ref{simtable}. 
We first extract the truncated moments using  Eqs. (\ref{fmain})-(\ref{lcoef}) and (\ref{aint}).
The results are presented in Table \ref{nlosim}. Using these moments and applying MJEM, 
we reconstruct in NLO $\Delta u_V(x)$ and $\Delta d_V(x)$ with the results presented in Fig. \ref{nlo-comparison}. 
Comparing Fig. \ref{nlo-comparison} with  Fig. \ref{lodirect}, 
one can see that quality of reconstruction in NLO
is not worse than the quality of LO reconstruction.
The slight deviations of reconstructed and input curves as before (c.f. Fig. \ref{lodirect})
are explained by the unavoidable deviations of the simulated with PEPSI asymmetries from their
reference (corresponding to the input parametrization) values.

\begin{table}[h!]
        \caption{\footnotesize Results for NLO extracted truncated moments for the simulations with the entering
        PEPSI two different parametrizations: 
        GRSV2000NLO (symmetric sea) parametrization (top) and GRSV2000NLO (broken sea) 
        parametrization (bottom). For comparison the respective reference (obtained by direct
        integration of entering PEPSI input parametrizations) moments are also presented.}
        \begin{tabular}{ccccc}
                \hline
                \hline
                 & \multicolumn{2}{c}{$\Delta'_n u_V$} & \multicolumn{2}{c}{$\Delta'_n d_V$}\\
\hline
n&Extracted&Reference & Extracted&Reference\\ 
 1 &  0.7369 $\pm$ 0.0133&0.7507 &   -0.2577 $\pm$ 0.0293& -0.2760 \\
 2 &  0.1507 $\pm$ 0.0039&0.1545 &   -0.0423 $\pm$ 0.0085& -0.0490 \\
 3 &  0.0449 $\pm$ 0.0016&0.0471 &   -0.0109 $\pm$ 0.0033& -0.0133\\
 4 &  0.0163 $\pm$ 0.0007&0.0176 &   -0.0037 $\pm$ 0.0015& -0.0045 \\
 \hline\hline
                 & \multicolumn{2}{c}{$\Delta'_n u_V$} & \multicolumn{2}{c}{$\Delta'_n d_V$}\\
\hline
n&Extracted&Reference &  Extracted&Reference\\ 
 1 &  0.5860 $\pm$0.0134 &0.5701 &-0.1045 $\pm$0.0300 &-0.1137\\
 2 &  0.1392 $\pm$0.0039 &0.1381 &-0.0314 $\pm$0.0088 &-0.0367\\
 3 &  0.0433 $\pm$0.0015 &0.0448 &-0.0101 $\pm$0.0034 &-0.0121\\
 4 &  0.0159 $\pm$0.0007 &0.0172 &-0.0037 $\pm$0.0016 &-0.0045\\
 \hline
\end{tabular}
\label{nlosim}
\end{table}

The remark concerning very important peculiarity of application in NLO of 
the optimization criterion Eq. (\ref{criterion}) should be made here.
The crucial point for the optimization criterion  
 is the proper choice of the initial\footnote{These are starting values 
for the MIGRAD algorithm implemented in MINUIT package \cite{minuit}.} values of $\alpha$ and
$\beta$. Indeed, the experience shows that if these initial values are too far away from 
the real $\alpha_{opt}$ and $\beta_{opt}$ to be found, then the MINUIT program can ``fall'' into
some wrong local minimum and produce the false values of $\alpha_{opt}$ and $\beta_{opt}$.
Fortunately in LO we can compare the reconstructed with MJEM
curve with the reference (directly extracted) values of PDFs and unambiguously find 
the optimal values of $\alpha_{opt}$ and $\beta_{opt}$. On the other hand, 
it is natural to use $\alpha_{opt}$ and $\beta_{opt}$ obtained within LO analysis as the initial
(starting) values for the application of optimization criterion Eq. (\ref{criterion}) in NLO.
The simulations demonstrate (see Fig. \ref{nlo-comparison}) that with the such choice 
of initial $\alpha$ and $\beta$ the minimization procedure performed in NLO analysis 
unambiguously finds the proper  values of $\alpha_{opt}$ and $\beta_{opt}$. As a result 
the obtained with MJEM curves are in a good agreement with the input (reference) parametrizations.
 At the same time, looking at
Figs. \ref{comb-sym} and \ref{comb-br}, one can see that the behavior of both
LO and NLO extracted curves is in a good agreement with the respective behavior 
of the input (reference) parametrizations.

Thus, all performed in this section studies  show that the proposed  method can be successfully
applied to the polarized PDFs extraction in NLO QCD.

\section{NLO QCD analysis of the HERMES data on pion production.}
\subsection{Construction of the difference asymmetries from the HERMES data on pion production}
Let us now  apply the proposed method to the HERMES SIDIS data on the pion production.
Within this paper we would like first of all
to test the applicability of the method to the experimental data analysis. That is why, for a moment, 
we do not like to deal with the such poorly known objects as $D_q^{K^\pm}$ and $D_g^h$ fragmentation functions. 
As it was discussed before (see the Introduction), from this point of view 
the most attractive objects  are the difference asymmetries. At the same time the difference asymmetries are still not constructed\footnote{At  present the such analysis is performed by HERMES collaboration}.
So, let us apply a trick and express the difference asymmetry given by Eq. (\ref{difas})
via the standard virtual photon SIDIS asymmetries
\be
A_{p(d)}^{\pi^\pm}\palka_Z=\frac{1}{P_BP_TfD}\frac{N^{\pi^\pm}_{\uparrow\downarrow}L_{\uparrow\uparrow}-N^{\pi^\pm}_{\uparrow\uparrow}L_{\uparrow\downarrow}}
{N^{\pi^\pm}_{\uparrow\downarrow}L_{\uparrow\uparrow}+N^{\pi^\pm}_{\uparrow\uparrow}L_{\uparrow\downarrow}},
\nonumber
\ee
which were measured by HERMES \cite{hermes}.
Namely, in each {\it i}th bin the difference asymmetries given by Eq. (\ref{difas}) can be rewritten as 
\be
\label{expansatz}
A^{\pi^+-\pi^-}(x_i)=\frac{R_i^{+/-}}{R_i^{+/-}-1}A^{\pi^+}(x_i)-
\frac{1}{R_i^{+/-}-1}A^{\pi^-}(x_i),
\ee
where the quantity $R^{+/-}_i$ is defined as 
\be
R^{+/-}_i\equiv
\frac{N^{\pi^+}_{i\uparrow\downarrow}L_{\uparrow\uparrow}+N^{\pi^+}_{i\uparrow\uparrow}L_{\uparrow\downarrow}}
{N^{\pi^-}_{i\uparrow\downarrow}L_{\uparrow\uparrow}+N^{\pi^-}_{i\uparrow\uparrow}L_{\uparrow\downarrow}}.
\nonumber
\ee
It is easy to see that the ratio  $R^{+/-}_i$ can be rewritten as 
\be
\label{unpolri}
R^{+/-}_i=\frac{\sigma^{\pi^+}_{\uparrow\downarrow}(x_i)+\sigma^{\pi^+}_{\uparrow\uparrow}(x_i)}
{\sigma^{\pi^-}_{\uparrow\downarrow}(x_i)+\sigma^{\pi^-}_{\uparrow\uparrow}(x_i)}=\frac{\sigma^{\pi^+}_{unpol}(x_i)}{\sigma^{\pi^-}_{unpol}(x_i)}=\frac{N^{\pi^+}_i}{N^{\pi^-}_i},
\ee
and, thus, can be
 taken from the unpolarized SIDIS data.
This relative quantity is well defined and extracted with the high precision object.
We take its value from the LEPTO generator of unpolarized events \cite{LEPTO}, which 
gives a good\footnote{
Dealing with LEPTO generator one should properly  tune \cite{hermes} 
the internal parameters of generator in order to achieve a proper description of 
the fragmentation process in the different experiments (HERMES here).
} description of the fragmentation processes. 
The calculations
show that the relative quantities $R_i^{+/-}$ remarkably weakly depend on statistics of simulated events.
Indeed, if one changes $N_{total}^{\pi^+}\palka_{LEPTO}$ from $10^5$ to $10^6$  then only 1-3\% deviation of $R^{+/-}_i$ (in dependence of bin number) occurs.
 Nevertheless, to be more precise, 
extracting the quantities $R_i^{+/-}$ for the HERMES statistics ($N^{\pi^+}_{total}=117.000$, 
$N^{\pi^-}_{total}=82.000$ for proton and $N^{\pi^+}_{total}=491.000$, 
$N^{\pi^-}_{total}=385.000$ for deutron targets, respectively \cite{hermes}) we 
preserve the condition
\be
N^{\pi^\pm}_{total}\palka_{LEPTO}\simeq N^{\pi^\pm}_{total}\palka_{HERMES}, \quad N_{total}^{\pi^\pm}\equiv \sum\nolimits_iN_{i}^{\pi^\pm},
\ee
performing the simulations with LEPTO.
The results for $R_i^{+/-}$ are presented in Table \ref{rvalues}. 

\begin{table}[h!]
        \caption{\footnotesize The obtained from the LEPTO generator results for the relative unpolarized quantity $R_i^{+/-}$ given by Eq. (\ref{unpolri}). }
\begin{tabular}{cc||c}
        \hline\hline
        & \multicolumn{1}{c||}{proton target} &  \multicolumn{1}{c}{deutron target}\\
        \hline
        $i$& $R^{+/-}_i$ &$R^{+/-}_i$\\ 
        \hline
1   &1.220   &1.150 \\ 
2   &1.270   &1.201 \\
3   &1.346  &1.229 \\
4   &1.436  &1.274 \\
5   &1.494  &1.315 \\
6   &1.569  &1.350 \\
7   &1.629  &1.407 \\
8   &1.669  &1.444 \\
9   &1.803  &1.556 \\
\hline
\end{tabular}
\label{rvalues}
\end{table}

Thus, using in Eq. (\ref{expansatz}) the results from Table \ref{rvalues} and HERMES results \cite{hermes}
on $A_{p,d}^{\pi^\pm}$ (see Tables XII and XIII in Ref. \cite{hermes}), one easily constructs
the difference asymmetries $A_{p,d}^{\pi^+-\pi^-}$. The results are presented in Fig. \ref{figasym}.

First, for the sake of testing (to check how well Eq. (\ref{expansatz}) works), we reconstruct the  valence PDFs in the leading order.
In LO  the  equations for the difference asymmetries take the  simple form given by Eq. (\ref{difaslo}).
With these equations we reconstruct $\Delta u_V$ and $\Delta d_V$ using the results on difference 
asymmetries presented in Fig. \ref{figasym}. The results  are shown in  Fig. \ref{her-nomjem}  , where we also 
plotted the respective results from Ref. \cite{hermes} (obtained with the purity method).
One can see that the results  obtained with both procedures are in a good agreement.

Let us now compare the LO extracted\footnote{As before (see Section 3), reconstructing the moments we apply the procedure of
integration given by Eq. (\ref{aint}) since it gives more precise reconstruction of the local PDFs
than the usually applied procedure  given by Eq. (\ref{summ}). } moments (truncated to the HERMES $x_B$ region) obtained with 
application of Eq. (\ref{expansatz}) (first line in the Table \ref{tablohermes} ) 
with the existing  LO results of SMC and HERMES
taken from the Table XI in Ref. \cite{hermes}.
One can see that the results  obtained with application of Eq. (\ref{expansatz})
(i.e., from the difference asymmetries plotted in Fig. \ref{figasym} )
are in a good accordance with both HERMES and SMC results.

\begin{table}[h!]
        \caption{\footnotesize LO extracted truncated moments obtained from the difference asymmetries
        constructed with application of Eq. (\ref{expansatz})
        in comparison with the existing LO results of SMC  and HERMES collaborations.The SMC moments are truncated to the HERMES $x_B$ region and are evolved to the HERMES $Q^2_{mean}=2.5\,GeV^2$ -- see Table XI  in Ref. \cite{hermes}. }
\begin{tabular}{ccccc}
        \hline\hline
\multicolumn{5}{c}{$\Delta'_n u_V$}\\
\hline
\hline
n  & 1 & 2 & 3 & 4\\
\hline
This paper & 0.510 $\pm$ 0.110   & 0.134$\pm$0.043 &0.048 $\pm$ 0.020   &0.020 $\pm$ 0.010 \\ 
HERMES     & 0.603 $\pm$ 0.071   & 0.144$\pm$0.014 & -/- & -/-  \\
SMC        & 0.614 $\pm$ 0.082   & 0.152$\pm$0.016 & -/- &-/- \\
        \hline\hline
\multicolumn{5}{c}{$\Delta'_n d_V$}\\
\hline\hline
n  & 1 & 2 & 3 & 4\\
\hline
This paper & -0.280$\pm$0.146 & -0.074$\pm$ 0.058&-0.026 $\pm$ 0.026&-0.011 $\pm$ 0.013\\
HERMES     & -0.172$\pm$0.068 & -0.047$\pm$ 0.012& -/- & -/-\\
SMC        & -0.334$\pm$0.112 & -0.056$\pm$ 0.026& -/- & -/-\\
\hline
\end{tabular}
\label{tablohermes}
\end{table}

Thus, the performed LO tests show that representation Eq. (\ref{expansatz}) for the difference asymmetry
can be successfully applied.

Notice also that even the LO  extraction
of $\Delta u_V$ and $\Delta d_V$ from the difference asymmetries is interesting in itself
as an alternative (complementary) possibility. Indeed, Eqs. (\ref{difaslo}) are free from the rather 
badly known fragmentation functions and purities.

\subsection{Reconstruction of the valence PDFs in NLO QCD}
Here, using the constructed difference asymmetries as a starting point and operating just 
as in Section 3, we will reconstruct  in NLO QCD  both the truncated Mellin moments
of the valence PDFs and the local PDFs themselves.

Let us first extract four first  
moments truncated to the HERMES $x_B$ region. 
The results are presented in Table \ref{tabnlohermes}. 
It is of importance that the proposed procedure  allows us to extract the moments in NLO directly, 
without  the commonly used assumptions like
$\Delta \bar u=\Delta\bar d=\Delta s =\Delta\bar s$ 
(see, for example, Refs. \cite{grsv2000,parametrizations}).
Notice that the first moments $\Delta_1 u_V$ and
$\Delta_1 d_V$ are very important in themselves because namely the first moments compose
the nucleon spin.
At the same time all four moments presented in Table \ref{tabnlohermes} are necessary
for the reconstruction of the local PDFs with MJEM application.

\begin{table}[h!]
        \caption{\footnotesize NLO extracted truncated moments obtained from the difference asymmetries
        constructed with application of Eq. (\ref{expansatz}).}
\begin{tabular}{ccccc}
        \hline\hline
n& 1 & 2 & 3 & 4\\
 $\Delta'_n u_V$  &0.555 $\pm$ 0.126  & 0.134 $\pm$ 0.047 &0.047 $\pm$ 0.020  &0.019 $\pm$ 0.10  \\
\hline\hline
$\Delta'_n d_V$&-0.302 $\pm$ 0.173&-0.076 $\pm$ 0.064&-0.025 $\pm$ 0.027&-0.010 $\pm$ 0.012 \\
\hline
\end{tabular}
\label{tabnlohermes}
\end{table}

Before application of MJEM in NLO, let us, for the sake of testing, reconstruct 
the local valence distributions in the leading order using LO moments from  Table \ref{tablohermes}. 
The results  are presented in Fig. \ref{her-lo-mjem}.  
One can see that 
reconstructed with MJEM Eq. (\ref{mj}) and optimization criterion  Eq. (\ref{criterion}) curve is in a good agreement with both HERMES results and with the results of direct LO extraction
from the difference asymmetries constructed with application of Eq. (\ref{expansatz}).

After the successful LO testing we apply MJEM in NLO QCD using the results for $\Delta'_n u_V$ 
and $\Delta'_n d_V$ from Table \ref{tabnlohermes}.  The results are presented in Fig. \ref{her-lonlo},
where also, for comparison, the respective LO results are plotted.
It is seen that the behavior of  NLO and LO curves with respect to each other  
is in agreement with the predictions of  existing parametrizations (see, for example, \cite{grsv2000}).

\subsection{Corrections caused by $Q^2$ dependence of asymmetries}
Until now we applied the approximation 
\be
A(x_i,Q_i^2)\simeq A(x_i,Q_{\rm mean}^2)\nonumber
\ee
commonly used (see Refs. \cite{smc,hermes}) for analysis of the DIS and SIDIS asymmetries. 
This approximation is in a good agreement 
even with the COMPASS data (see Fig. 5 in Ref. \cite{compass-dis}) and is  
especially suitable for the HERMES kinematics, where the ``shoulder'' in $Q^2$ is rather 
small ($1\,GeV^2\simleq Q^2\simleq 10\,GeV^2$; $Q^2_{mean}=2.5\,GeV^2$ -- see Tables XII and XIII in Ref. \cite{hermes}) 
in comparison with the SMC and COMPASS kinematics. 
Nevertheless, even for the HERMES kinematics we deal with, for more comprehensive analysis,
it is useful to estimate the corrections caused by 
the weak $Q^2$ dependence of the difference asymmetries.
So, let us estimate the shifts in all four NLO moments caused by the respective shifts
\be
\label{corr}
\delta_i A_{p,d}^{\pi^+-\pi^-}=A_{p,d}^{\pi^+-\pi^-}(x_i,Q^2_{mean})-A_{p,d}^{\pi^+-\pi^-}(x_i,Q_i^2)
\ee
in the difference asymmetries. The most simple way to estimate $\delta_i A_{p,d}^{\pi^+-\pi^-}$ 
is to use the maximal number of the latest available NLO parametrizations. Namely, we approximate 
r.h.s of Eq. (\ref{corr}) by the respective difference of ``theoretical'' asymmetries calculated 
with substitution of the different parametrizations to NLO equations for the difference asymmetries --
Eqs. (14), (15) in Ref. \cite{our2}.

Adding the  calculated in this way $\delta_i A_{p,d}^{\pi^+-\pi^-}$ to the initial experimental
asymmetries $A_{p,d}^{\pi^+-\pi^-}(x_i,Q_i^2)$, one estimates the  
evolved from $Q_i^2$ to $Q^2_{mean}$ asymmetries $A_{p,d}^{\pi^+-\pi^-}(x_i,Q_{mean}^2)|_{Evolved}$. 
Using the obtained in such a way
evolved asymmetries we extract the respective corrected moments of the valence PDFs 
$\Delta'_n q_V|_{Corrected}$  repeating  the procedure from Section 3. Then we compare the 
corrected moments $\Delta'_n q_V|_{Corrected}$  with the respective moments $\Delta'_n q_V$  
from the previous  section (obtained without corrections due to evolution)
and calculate the respective shifts $\delta(\Delta'_n q_V)=\Delta'_n q_V|_{Corrected}-\Delta'_n q_V$.
The results are presented in the Tables \ref{tabdeviuv} and \ref{tabdevidv}, 
where also the relative quantities
$\delta(\Delta'_n q_V)/\Delta'_n q_V$ are presented.

\begin{table}[h!]
        \caption{\footnotesize NLO results for $\Delta'_n u_V$ 
        corrected due to evolution (top) together with the respective absolute and relative deviations
        from the uncorrected moments (bottom). The corrections are 
        estimated using seven different NLO parametrizations. 
        The roman numbers I and II correspond to GRSV2000NLO parametrization for broken and symmetric 
        sea scenarios, respectively. The roman numbers III and IV correspond to Ref. \cite{florian97} for
        sets ii+ and ii- (symmetric and weakly broken sea scenarios).
        The rest of numbers V-VII corresponds to the NLO parametrizations from
        Ref. \cite{parametrizations} (in the order of citation). }
\begin{tabular}{cccccccc}
\hline\hline
\multicolumn{8}{c}{$\Delta'_n u_V|_{Corrected}$}\\
\hline
        n &I & II&III &IV &V  &VI  &VII  \\
        \hline
        1 &0.5495  &0.5464  &0.5555  &0.5551  &0.5473  &0.5588  &0.5457 \\
        2 &0.1364  &0.1367  &0.1378  &0.1377  &0.1368  &0.1387  &0.1363 \\ 
        3 &0.0459  &0.0460  &0.0463  &0.0463  &0.0460  &0.0467  &0.0459 \\
        4 &0.0182  &0.0182  &0.0183  &0.0183  &0.0182  &0.0185  &0.0182 \\

\hline
\hline
\multicolumn{8}{c}{Average $\Delta'_n u_V|_{Corrected}$}\\
\hline
n &  &  &  &  &  &  & \\

1  &\multicolumn{7}{c}{0.5437 $\pm$ 0.1266}\\
2  &\multicolumn{7}{c}{0.1348 $\pm$ 0.0475}\\
3  &\multicolumn{7}{c}{0.0453 $\pm$ 0.0199}\\
4  &\multicolumn{7}{c}{0.0179 $\pm$ 0.0089}\\

        \hline\hline
        \multicolumn{8}{c}{$\delta(\Delta'_n u_V)$}\\
        \hline
        n &I & II&III &IV &V  &VI  &VII  \\
\hline
1  &-0.0054  &-0.0085  & 0.0006  & 0.0002  &-0.0077  & 0.0039  &-0.0092 \\
2  &-0.0034  &-0.0031  &-0.0019  &-0.0021  &-0.0030  &-0.0010  &-0.0035 \\ 
3  &-0.0013  &-0.0012  &-0.0009  &-0.0009  &-0.0012  &-0.0005  &-0.0013 \\
4  &-0.0005  &-0.0005  &-0.0004  &-0.0004  &-0.0005  &-0.0002  &-0.0006 \\
\hline\hline
         \multicolumn{8}{c}{$\delta(\Delta'_n u_V)/\Delta'_n u_V$ (\%)}\\
         \hline
        n &I & II&III &IV &V  &VI  &VII  \\
        \hline
1 &-0.97&-1.53&0.11& 0.04& -1.38&0.71& -1.65\\
2 &-2.42&-2.21&-1.37&-1.47&-2.12&-0.73&-2.47\\
3 &-2.67&-2.54&-1.86&-1.95&-2.56&-1.14&-2.84\\
4 &-2.62&-2.62&-2.08&-2.14&-2.72&-1.28&-2.94\\
\hline\hline
\multicolumn{8}{c}{Average $\delta(\Delta'_n u_V)$}\\
\hline
n &  &  &  &  &  &  & \\

1  &\multicolumn{7}{c}{-0.0037}\\
2  &\multicolumn{7}{c}{-0.0026}\\
3  &\multicolumn{7}{c}{-0.0011}\\
4  &\multicolumn{7}{c}{-0.0004}\\
\hline

\end{tabular}
\label{tabdeviuv}
\end{table}

Notice that the considered procedure of the asymmetry evolution is quite similar to the procedure
used by SMC for the $\Gamma_{1p(d)}$ reconstruction (see Section V  in Ref. \cite{smc-evolution}).

\begin{table}[h!]
        \caption{\footnotesize NLO results for $\Delta'_n d_V$ 
        corrected due to evolution (top) together with the respective absolute and relative deviations
        from the uncorrected moments (bottom). The corrections are 
        estimated using seven different NLO parametrizations. 
        The roman numbers I and II correspond to GRSV2000NLO parametrization for broken and symmetric 
        sea scenarios, respectively. The roman numbers III and IV correspond to Ref. \cite{florian97} for
        sets ii+ and ii- (symmetric and weakly broken sea scenarios).
        The rest of numbers V-VII corresponds to the NLO parametrizations from
        Ref. \cite{parametrizations} (in the order of citation). }
\begin{tabular}{cccccccc}
\hline\hline
\multicolumn{8}{c}{$\Delta'_n d_V|_{Corrected}$}\\
\hline
        n &I & II&III &IV &V  &VI  &VII  \\
        \hline
        1 &-0.3130  &-0.3091  &-0.3197  &-0.3251  &-0.3062  &-0.3096  &-0.3064 \\
        2 &-0.0778  &-0.0779  &-0.0811  &-0.0822  &-0.0771  &-0.0780  &-0.0774 \\ 
        3 &-0.0256  &-0.0256  &-0.0267  &-0.0269  &-0.0254  &-0.0256  &-0.0255 \\
        4 &-0.0096  &-0.0098  &-0.0102  &-0.0102  &-0.0097  &-0.0097  &-0.0097 \\

\hline\hline
\multicolumn{8}{c}{Average $\Delta'_n d_V|_{Corrected}$}\\
\hline
n &  &  &  &  &  &  & \\

1  &\multicolumn{7}{c}{-0.3127 $\pm$0.1731 }\\
2  &\multicolumn{7}{c}{-0.0788 $\pm$0.0643 }\\
3  &\multicolumn{7}{c}{-0.0259 $\pm$0.0269 }\\
4  &\multicolumn{7}{c}{-0.0099 $\pm$0.0119 }\\

        \hline\hline
        \multicolumn{8}{c}{$\delta(\Delta'_n d_V)$}\\
        \hline
        n &I & II&III &IV &V  &VI  &VII  \\
\hline
1  &-0.0114  &-0.0075  &-0.0181  &-0.0235  &-0.0046  &-0.0080  &-0.0048 \\
2  &-0.0015  &-0.0017  &-0.0048  &-0.0059  &-0.0008  &-0.0018  &-0.0012 \\ 
3  &-0.0003  &-0.0004  &-0.0014  &-0.0017  &-0.0001  &-0.0003  &-0.0002 \\
4  &-0.0001  &-0.0001  &-0.0005  &-0.0006  &-0.0000  &-0.0000  &-0.0000 \\
\hline\hline
         \multicolumn{8}{c}{$\delta(\Delta'_n d_V)/\Delta'_n d_V$ (\%)}\\
         \hline
        n &I & II&III &IV &V  &VI  &VII  \\
        \hline
1 &3.77&2.48&5.99&7.78&1.53&2.65&1.58\\
2 &1.97&2.20&6.33&7.72&1.10&2.33&1.53\\
3 &1.31&1.47&5.71&6.74&0.52&1.35&0.95\\
4 &0.83&0.83&5.07&5.89&0.10&0.52&0.52\\
\hline\hline
\multicolumn{8}{c}{Average $\delta(\Delta'_n d_V)$}\\
\hline
n &  &  &  &  &  &  & \\

1  &\multicolumn{7}{c}{-0.0111}\\
2  &\multicolumn{7}{c}{-0.0025}\\
3  &\multicolumn{7}{c}{-0.0007}\\
4  &\multicolumn{7}{c}{-0.0002}\\
\hline

\end{tabular}
\label{tabdevidv}
\end{table}

It is of importance that we use for estimations the set of essentially  different\footnote{
They correspond to the different sea scenarios (symmetric sea, weakly and strongly broken light quark sea), 
different details  of calculations and different choice of ansatz for PDFs.}
NLO parametrizations  and some of them, for example GRSV2000 (broken sea) and GRSV2000 (symmetric sea),
differ from each other very strongly. However, one can see from the Tables \ref{tabdeviuv}, \ref{tabdevidv} that independently 
of the chosen parametrization the corrections for moments caused by evolution are very small
(negligible in comparison with the statistical errors). To be precise, one can just include 
$\delta(\Delta'_n q_V)|_{Average}$ (see  Tables \ref{tabdeviuv} and \ref{tabdevidv}) 
in the systematical error.

Let us now reconstruct the local valence PDFs applying MJEM to the corrected moments 
$\Delta'_n q_V|_{Corrected}$ from the Tables \ref{tabdeviuv} and \ref{tabdevidv}.
The results are presented in 
Fig. \ref{her-nlo-corrections}.  One can see that the curves corresponding to the different ways of correction
(different used parametrizations) are very close to each other. The averaged over the different 
corrections curve (dashed line) also very insignificantly differs from the initial (obtained without corrections) curve (solid line).

Thus, the performed in this Section analysis demonstrates that  at least for the HERMES kinematics
we deal with here,
the results are very insensitive to the corrections on the difference asymmetries caused by evolution.

\section{Conclusions and prospects}
Thus, in this paper the method of polarized SIDIS data analysis in NLO QCD is developed. 
The main peculiarity of the method is that its application is based on two  subsequently 
applied procedures.
First one directly extracts in NLO few first 
truncated (available to measurement) Mellin moments of the quark helicity distributions. 
The obtained at this stage  results are very important and interesting in themselves. 
Indeed, the first moments are the main objects  for understanding of the nucleon spin 
structure since they compose the nucleon spin. 
Second, using the obtained at first stage truncated moments as an input to the modification
of the Jacobi polynomial expansion method, one eventually reconstructs the local quark
helicity distributions in the accessible for measurement $x_B$ region.

After successful testing we apply the proposed NLO method
to the HERMES data on the pion production. To this end the pion difference 
asymmetries are constructed for both proton and deutron targets. To construct the difference 
asymmetries we use the HERMES data on the usual virtual photon SIDIS asymmetries, and, also, the 
well known quantity -- ratio of unpolarized cross-sections for $\pi^+$ and $\pi^-$ production
which we take from the LEPTO generator of unpolarized events.
With the  constructed in such a way difference 
asymmetries the LO results of the valence distribution reconstruction are in a good accordance with
the respective leading order HERMES and SMC results, while the NLO results are in agreement 
with the existing NLO parametrizations on these quantities.
Nevertheless, the obtained results 
should be considered as the rather preliminary since we construct the difference
asymmetries in indirect way. 
At present the difference asymmetries are constructed by HERMES
and are expected to be available in the nearest future. Certainly, it is very desirable
to perform the NLO analysis of these directly constructed asymmetries and 
compare the results with the respective results presented in this paper.

In this paper we apply the proposed method to the difference asymmetries only. Let us recall 
once again that essential advantage of these asymmetries in comparison with any other ones
is the absence of fragmentation functions in LO and the weak dependence of well known
difference of favored and unfavored pion fragmentation functions in NLO. So, since within this
paper we mainly would like to investigate how well the method itself works, 
we, for a moment, prefer to deal namely with these very clean (from theoretical point of view) objects.  
In this connection it is of importance that  the measurement of the difference asymmetries is
one of the main topics of the physical program of E04-113 experiment planned at Jefferson Lab \cite{jlab}.
It is of importance that in this experiment the expected average $Q^2$ is also rather small (about 
$2\,GeV^2$). So, the NLO analysis in this experiment is also strongly required.

Certainly, in the nearest future we will apply the method to NLO analysis of all measured in 
the  SIDIS experiments asymmetries. In particular, it can allow us to extract in NLO  so important
quantity as the polarized strangeness in nucleon. 
Regretfully, the only existing today data on kaon production (HERMES experiment) 
suffers of large statistical errors
and, besides, the  accessible for measurement HERMES $x_B$ region is rather narrow. So,  to obtain
the reliable results on the such 
tiny quantity as $\Delta_1 s$ (as well as on $\Delta_1\bar u $ and $\Delta_1\bar d$) it is necessary
to perform the combined analysis, i.e., to analyze in NLO  the combined data of SMC, HERMES, 
COMPASS and the planned E04-113 (Jefferson Lab) experiments. This is one of the main subjects
of our future investigations.

\section{Acknowledgments}
The authors are grateful to N.~Akopov, E.C.~Aschenauer, R.~Bertini, M.P.~Bussa,
 O.~Denisov, D.~Hasch, H.E.~Jackson, 
 A.~Korzenev, V.~Krivokhizhin, E.~Kuraev,
 A.~Maggiora, C.A.~Miller, A.~Nagaytsev, A.~Olshevsky, 
G.~Piragino, G.~Pontecorvo,  I.~Savin, A.~Sidorov, O.~Teryaev and R.~Windmolders,
 for fruitful discussions. The work of O.S and  O.I. was supported by the
 Russian Foundation for Basic Research (project no. 05-02-17748).

\begin{center}
        {\Large\bf Appendix}
\end{center}
\renewcommand{\theequation}{A.\arabic{equation}}
\setcounter{equation}{0}

JEM is the expansion of $x$-dependent function (structure function or quark density)
in the series over Jacobi polynomials 
$\Theta^\ab_n(x)$ orthogonal with weight $\omega^{(\alpha,\beta)}(x)=x^\beta(1-x)^{\alpha}$ (see \cite{parisi}
-\cite{sidorov} for details):
\be
\label{a1}
 &  & F(x)\simeq F_{N_{max}}(x)=  \omega^\ab\sum\nolimits_{k=0}^{N_{max}}\Theta_k^\ab(x)\nonumber\\
 &  & \times\sum\nolimits_{j=0}^kc_{kj}^{\ab}M(j+1),
\ee

where 
\be
M[j]=\int_0^1dx\, x^{j-1} F(x)
\ee
and 
\be
\label{ortho}
\int_0^1dx\omega^\ab(x)\Theta^\ab_n(x)\Theta^\ab_m(x)=\delta_{nm}.
\ee
The details on the Jacobi polynomials 
\be
\label{razlojenie}
\Theta_k^\ab(x)=\sum_{j=0}^kc_{kj}^\ab x^j
\ee
 can be found in Refs. \cite{parisi} and
\cite{barker}. In practice one truncates the series (\ref{a1})
living in the expansion only finite number of moments $N_{max}$ -- see Eq. (\ref{mj}).
The experience shows \cite{sidorov} that even small $N_{max}$ gives good results.

The idea of modified expansion is to reexpand $F(x)$ in the series over the truncated
moments $M'_{[ab]}[j]$ given by Eq. (\ref{trm}), 
performing the rescaling $x\rightarrow a+(b-a)x$ which compress the entire region
$[0,1]$ to the truncated region $[a,b]$. To this end let us apply the following ansatz
\be
\label{e1}
F(x)=\left(\frac{x-a}{b-a}\right)^\beta\left(1-\frac{x-a}{b-a}\right)^\alpha\sum_{n=0}^\infty\tilde f_n \Theta_n^\ab\left(\frac{x-a}{b-a}\right)
\ee
and try to find the coefficients $\tilde f_n$. Multiplying both parts of Eq. (\ref{e1})
by $\Theta_k^\ab((x-a)/(b-a))$, integrating over x in the  limits $[a,b]$ and  performing the
replacement $t=(x-a)/(b-a)$, one gets
\be
\int_a^bdx\, F(x)\Theta_k^\ab\left(\frac{x-a}{b-a}\right)=
(b-a)\sum_{n=0}^\infty{\tilde f_n}\int_0^1dt\, t^\beta(1-t)^\alpha \Theta_n^\ab(t)\Theta_k^\ab(t),
\ee
so that with the orthogonality condition Eq. (\ref{ortho})
one obtains 
\be
\label{e2}
\tilde f_n=(b-a)^{-1}\int_a^b dx\, F(x) \Theta_n^\ab\left(\frac{x-a}{b-a}\right).
\ee
Substituting Eq. (\ref{e2}) in the expansion (\ref{e1}), and using Eq. (\ref{razlojenie}) one eventually gets 
\be
\label{apfinal}
 &  & F(x)  =  \left(\frac{x-a}{b-a}\right)^\beta\left(1-\frac{x-a}{b-a}\right)^\alpha\nonumber\\
 & &  \times  \sum_{n=0}^{\infty} \Theta_n^\ab\left(\frac{x-a}{b-a}\right)
 \sum_{k=0}^n c_{nk}^\ab
 \frac{1}{(b-a)^{k+1}}\sum_{l=0}^k\frac{k!}{l!(k-l)!}M'_{[a,b]}[l+1](-a)^{k-l},
\ee
where $M'_{[a,b]}[j]$ is given by Eq. (\ref{trm}).
Truncating in the exact Eq. (\ref{apfinal}) the infinite sum over $n$ to the sum $\sum_{n=0}^{N_{max}}$
one gets the approximate equation (\ref{mj}).

Let us prove the important property\footnote{The proof of  analogous property for the usual JEM can be
found in \cite{sidorov}} of the truncated moments reconstructed with MJEM.

For any $n\le N_{max}$
\be
\label{statement}
M'_{[a,b]}[n+1]{\Biggl |}_{reconstructed}=M'_{[a,b]}[n+1]{\Biggl |}_{input},
\ee
where
\be
M'_{[a,b]}[n]{\Biggl |}_{input}=\int_a^bdx x^{n-1} F(x),\quad
M'_{[a,b]}[n]{\Biggl |}_{reconstructed}=\int_a^bdx x^{n-1} F_{N_{max}}(x),
\ee
$F_{N_{max}}(x)$ is the function reconstructed with application of MJEM (\ref{mj}),
and $N_{max}+1$ is the number of the highest of moments used in  Eq. (\ref{mj}). 

To prove this statement we will need the inverse to (\ref{razlojenie}) expansion
\be
\label{rexpansion}
x^n=\sum_{k=0}^nd_{nk}^\ab\Theta_k^\ab(x),
\ee
with the obvious property of $d_{nk}^\ab$ coefficients
\be
\label{dprop}
\sum_{k=j}^nd_{nk}^\ab c_{kj}^\ab=\delta_{nj}.
\ee

Let us integrate Eq. (\ref{mj}) over $x$ in the limits $[a,b]$ with weight  $((x-a)/(b-a))^n$.
\be
\int_a^b dx \left(\frac{x-a}{b-a}\right)^nF_{N_{max}}(x)=\nonumber\\
\int_a^b dx \left(\frac{x-a}{b-a}\right)^n\left\{
\omega^\ab\left(\frac{x-a}{b-a}\right)\sum_{m=0}^{N_{max}} \Theta_m^\ab\left(\frac{x-a}{b-a}\right)\nonumber\right.\\
\left.
\times
(b-a)^{-1}\sum_{j=0}^m c_{mj}^\ab \int_a^b dz\, F(z)\left(\frac{z-a}{b-a}\right)^j\right\}.
\ee
Using expansion (\ref{rexpansion}) and orthogonality condition (\ref{ortho}), 
one easily gets 

\be
\int_a^bdx \left(\frac{x-a}{b-a}\right)^nF_{N_{max}}(x)=\nonumber\\
\sum_{m=0}^{N_{max}}\sum_{k=0}^n\delta_{km}\left[d_{nk}^\ab\sum_{j=0}^mc_{mj}^\ab\int_a^bdx\,F(x)\left(\frac{x-a}{b-a}\right)^j\right].
\ee
It is obvious that at $n\le N_{max}$  the Kronecker symbol $\delta_{km}$ reduces the sum  $\sum_{m=0}^{N_{max}}$ to the sum $\sum_{m=0}^n$. 
Thus, summing over $m$ with $\delta_{km}$, using 
the identity
\be
\sum_{k=0}^n\sum_{j=0}^kf_{jk}\equiv\sum_{j=0}^n\sum_{k=j}^nf_{jk}
\ee
and applying Eq. (\ref{dprop}), one get eventually:
\be
\label{final}
\int_a^bdx \left(\frac{x-a}{b-a}\right)^nF_{N_{max}}(x)=\int_a^bdx \left(\frac{x-a}{b-a}\right)^nF(x),\quad
{n\le N_{max}}
\ee
Setting $n=0$ in Eq. (\ref{final}) one obtains
\be
\label{n0eq}
M'_{[a,b]}[1]{\Biggl |}_{reconstructed}\equiv\int_a^bdx\,F_{N_{max}}(x)=\int_a^bdx\,F(x)\equiv M'_{[a,b]}[1]{\Biggl |}_{input}.
\ee
Putting then $n=1$ in (\ref{final}) and using (\ref{n0eq}) one gets
\be
M'_{[a,b]}[2]{\Biggl |}_{reconstructed}=M'_{[a,b]}[2]{\Biggl |}_{input}.
\ee
Operating in this way for all $n\le N_{max}$, one arrives at the equality (\ref{statement}) to be proved.

In conclusion, very important remark should be made here.
Notice that ansatz (\ref{e1}) (as well as the expansion Eq. (\ref{mj}) itself) is correctly defined inside the entire region $(a,b)$ except for
the small vicinities of boundary points (absolutely the same situation holds for the usual JEM, Eq. (\ref{a1}), applied
to the quark distributions in the region $(0,1)$). 
Thus, near the boundary points the deviations of reconstructed with MJEM function from its true values 
are unavoidable.
Fortunately, all numerical examples 
(see section 2) show that input and  reconstructed with MJEM functions are in very good agreement in the practically
entire considered $x_B$ region,
while the boundary distortions are easily identified  and controlled since they are very sharp
and hold in very small vicinities of the boundary points (see Figs. \ref{f2}-\ref{f4}).
Thus, to reconstruct the curve near the boundary points one should just
cut off these distortions and then extrapolate the rest to the boundaries of the considered $x_B$ region.

To be precise, it should be also noticed that 
the all proofs given in the Appendix are rather formal because of the boundary distortions problem. 
All equations in the Appendix (like, for example, Eq. (\ref{statement})) become exact only when the distortions are cutted off and extrapolation
to the boundaries is made. Fortunately, the practice show that  the distortion regions are so small
that the numerical results on the  integrals over the entire  region $[a,b]$
are practically insensitive to the way of extrapolation, so that   all equations in the Appendix   are valid with a high
numerical precision.

\newpage

\begin{figure}[h]
        \caption{\footnotesize Results  of $\Delta u_V(x)$ ($\alpha_{opt}=8.189221$, $\beta_{opt}=-0.99000$)
        and $\Delta d_V(x)$ ($\alpha_{opt}=-0.99000$, $\beta_{opt}=-0.387196$) reconstruction
         with the usual JEM. Solid line corresponds to the input (reference) parametrization GRSV2000NLO (symmetric sea).
         Dotted line corresponds to the distribution reconstructed with JEM for $N_{max}=12$.}
\begin{center}
{\includegraphics[height=4cm,width=6cm]{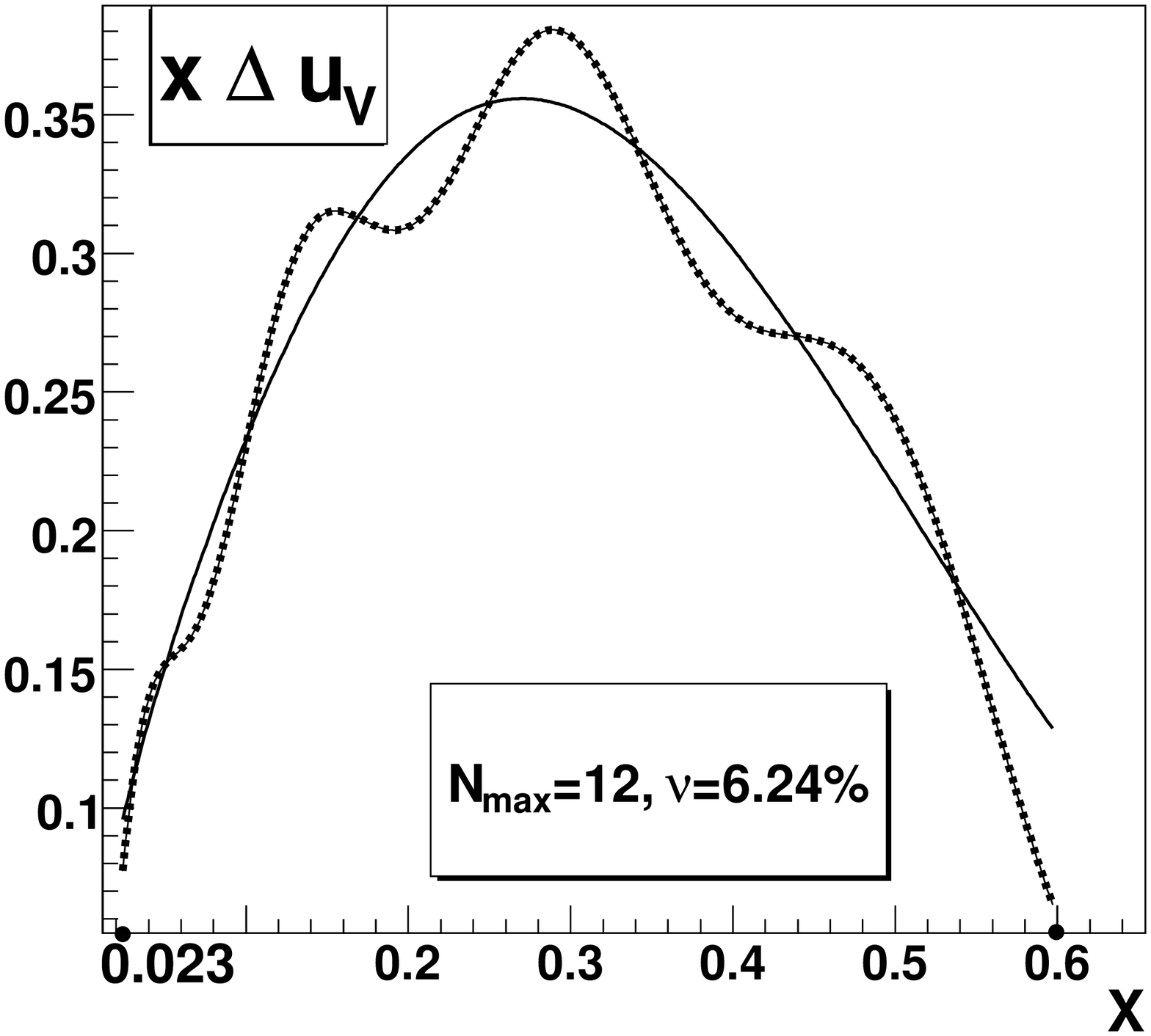}}
{\includegraphics[height=4cm,width=6cm]{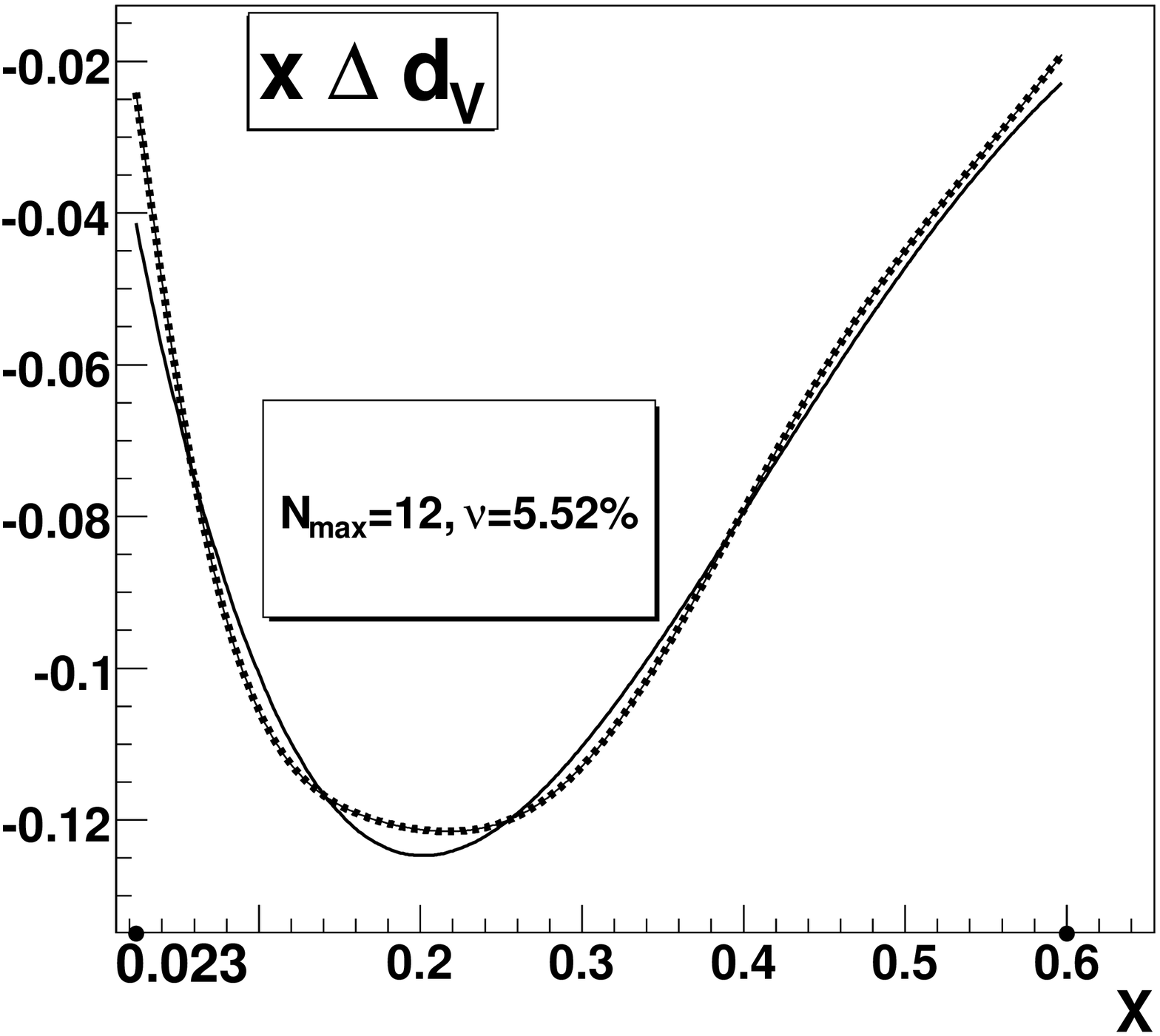}}
\end{center}
        \label{f1}
\end{figure}

\begin{figure}[h]
        \caption{\footnotesize Results of $\Delta u_V(x)$ ($\alpha_{opt}=-0.827885$, $\beta_{opt}=-0.011505$)
        and $\Delta d_V(x)$ ($\alpha_{opt}=-0.989752$, $\beta_{opt}=-0.012393$) reconstruction 
         with MJEM. Solid line corresponds to the input (reference) parametrization GRSV2000NLO (symmetric sea).
         Dotted line corresponds to the distribution reconstructed with MJEM for $N_{max}=12$.}
\begin{center}
{\includegraphics[height=4cm,width=6cm]{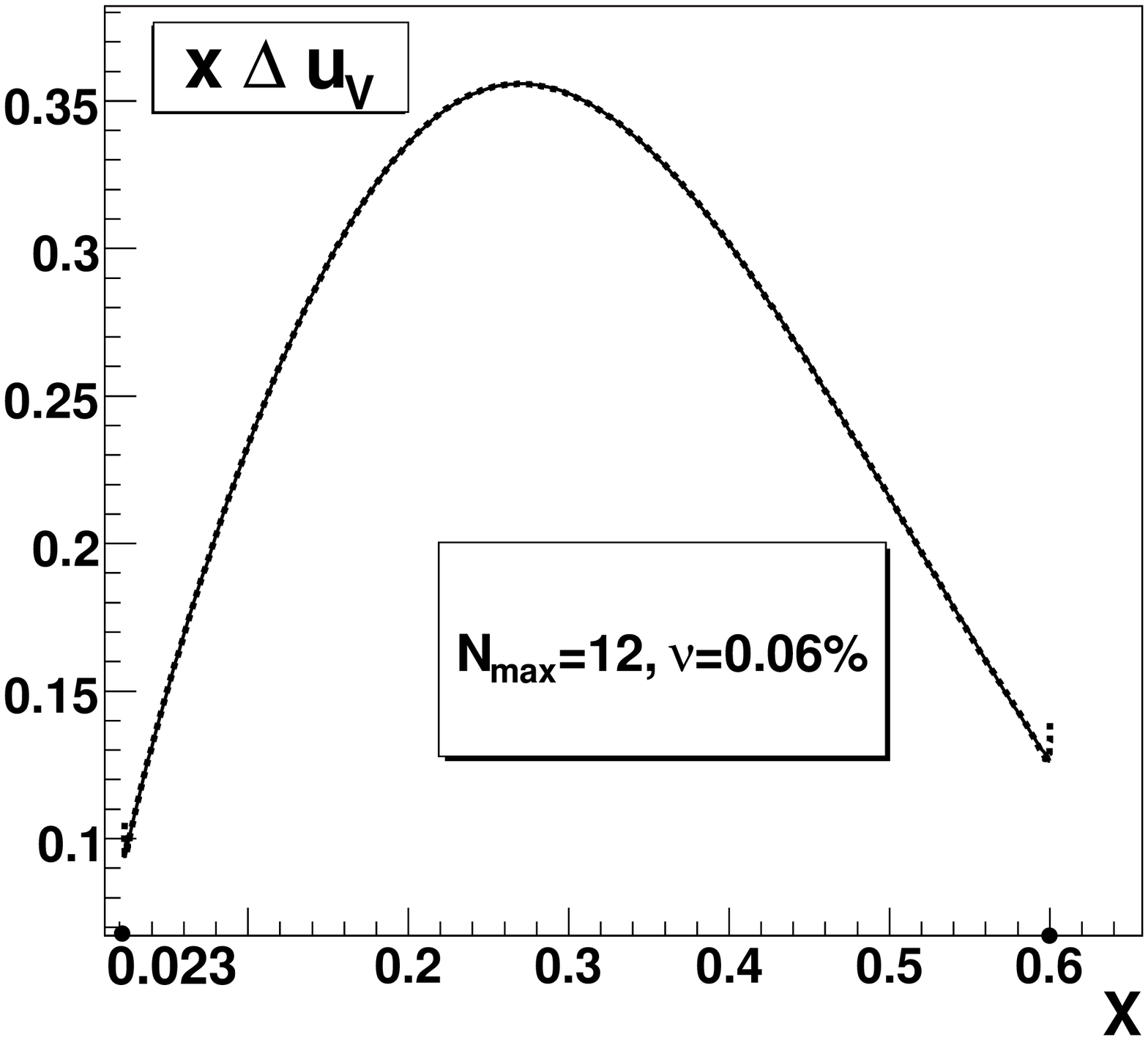}}
{\includegraphics[height=4cm,width=6cm]{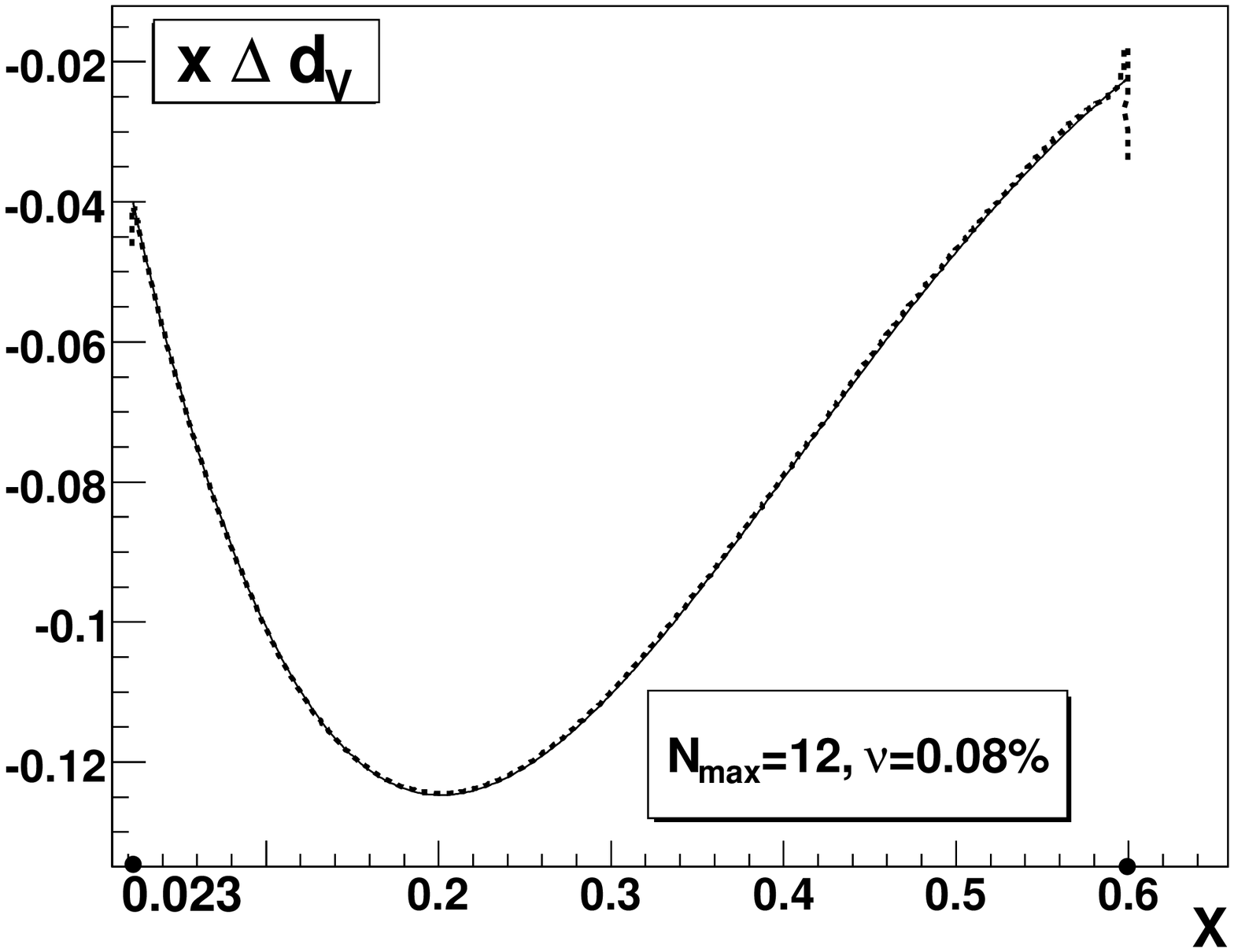}}
\end{center}
        \label{f2}
\end{figure}
\begin{figure}[h]
        \caption{\footnotesize  Results of the valence PDFs reconstruction with JEM and MJEM in comparison.
        Top part corresponds to  
        $\Delta u_V(x)$ ($\alpha_{opt}=-0.99$, $\beta_{opt}=0.054010$)
        and $\Delta d_V(x)$ ($\alpha_{opt}=0.174096$, $\beta_{opt}=0.162567$) 
        reconstructed with the usual JEM.
        Bottom part corresponds to 
        $\Delta u_V(x)$ ($\alpha_{opt}=-0.0025869$, $\beta_{opt}=-0.071591$)
        and $\Delta d_V(x)$ ($\alpha_{opt}=0.110331$, $\beta_{opt}=-0.049255$) reconstructed with MJEM.
        Solid lines correspond to input (reference) parametrization GRSV2000NLO (symmetric sea).
        Dotted lines correspond to the distributions reconstructed with JEM (top) and MJEM (bottom).}
\begin{center}
{\includegraphics[height=4cm,width=6cm]{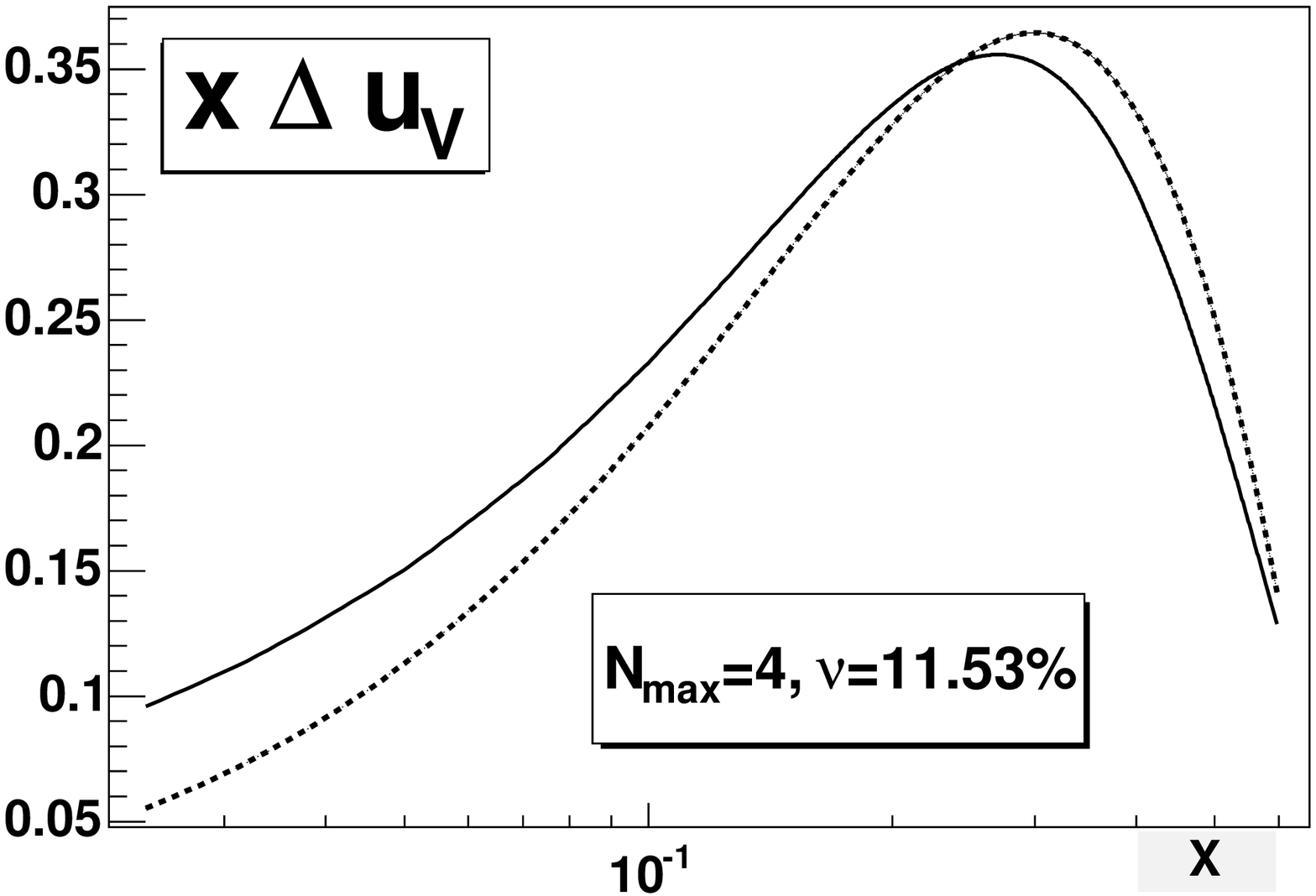}}
{\includegraphics[height=4cm,width=6cm]{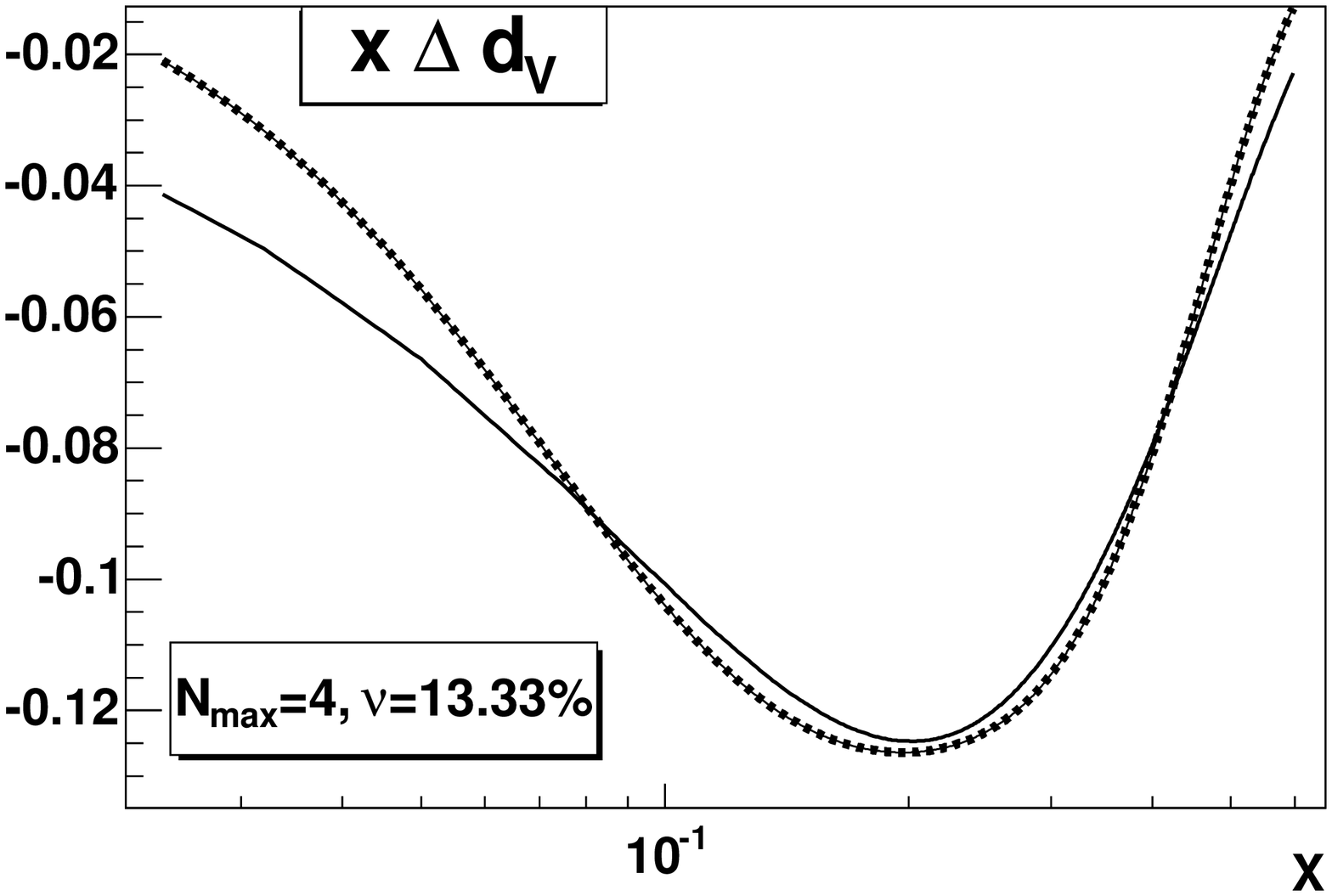}}
{\includegraphics[height=4cm,width=6cm]{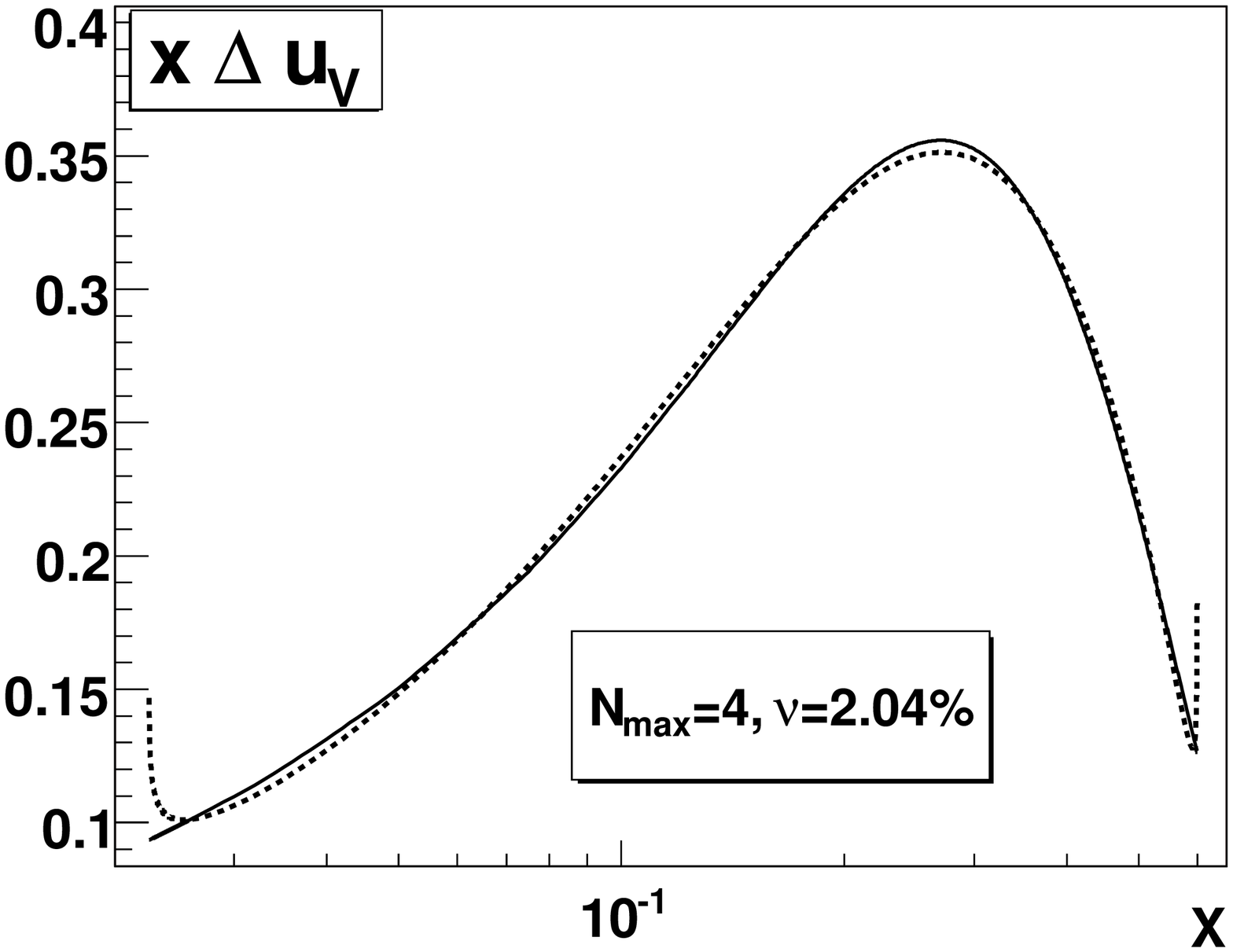}}
{\includegraphics[height=4cm,width=6cm]{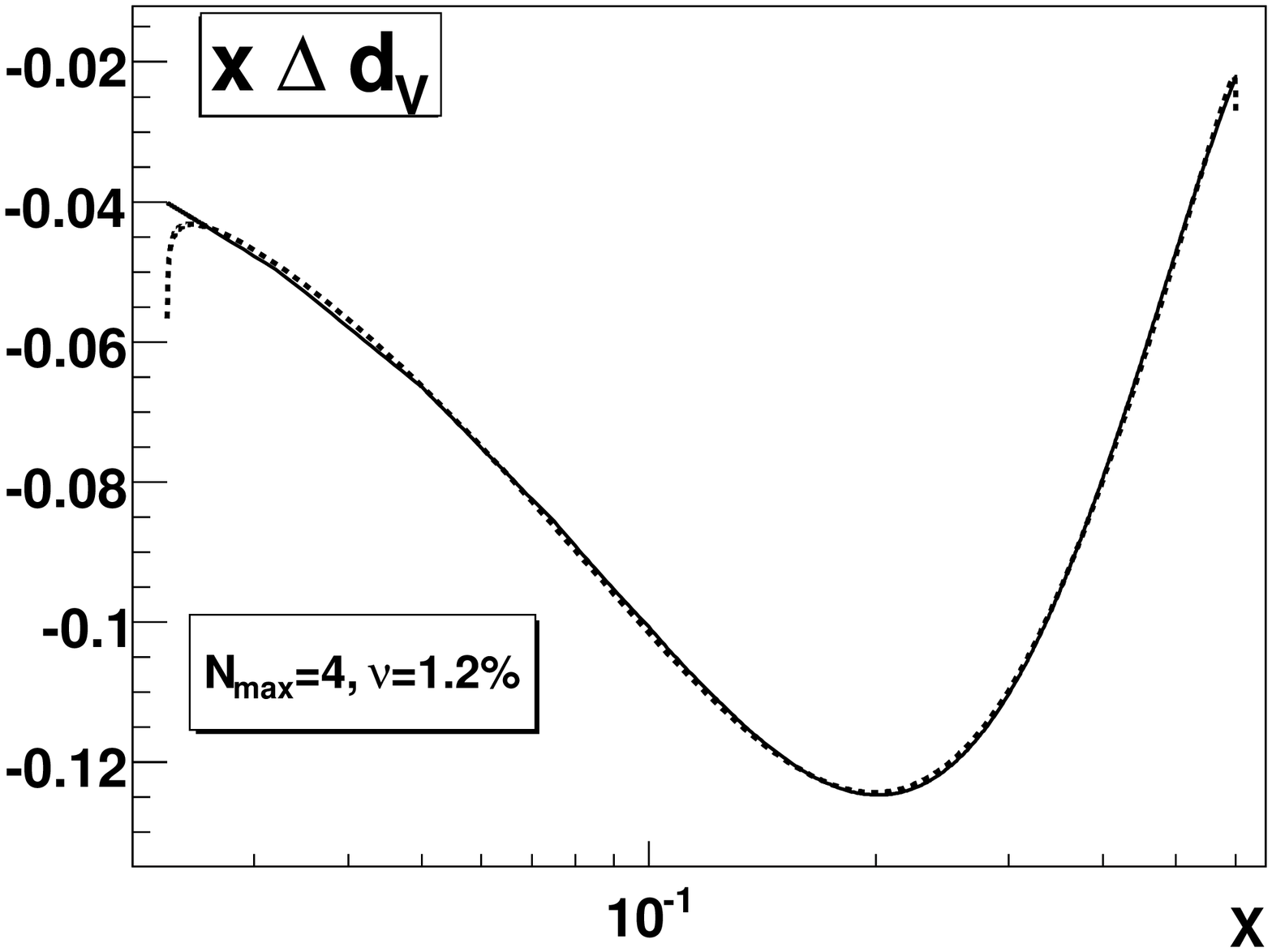}}
\end{center}
        \label{f3}
\end{figure}
\begin{figure}[h]
        \begin{center}
        \caption
        {\footnotesize 
        Results of $\Delta u_V$ and $\Delta d_V$ reconstruction   for  GRSV2000{\bf NLO} 
        parametrization for both symmetric (top) and
        broken sea (bottom) scenarios.
        Solid line corresponds to the reference curve (input parametrization).
        Dotted line is 
         reconstructed with MJEM and criterion (\ref{criterion}) 
        inside the accessible for measurement region ([0.023,0.6] here).
        Optimal values of parameters for symmetric sea scenario for $\Delta u_V$ are $\alpha_{opt}=-0.15555$,
        $\beta_{opt}=-0.097951$ and for $\Delta d_V$ are $\alpha_{opt}=-0.002750$, $\beta_{opt}=-0.07190$.
        Optimal values of parameters for broken sea scenario for $\Delta u_V$ are $\alpha_{opt}=-0.209346$,
        $\beta_{opt}=0.153417$ and for $\Delta d_V$ are $\alpha_{opt}=0.702699$, $\beta_{opt}=-0.293231$.
        }
{\includegraphics[height=4cm,width=6cm]{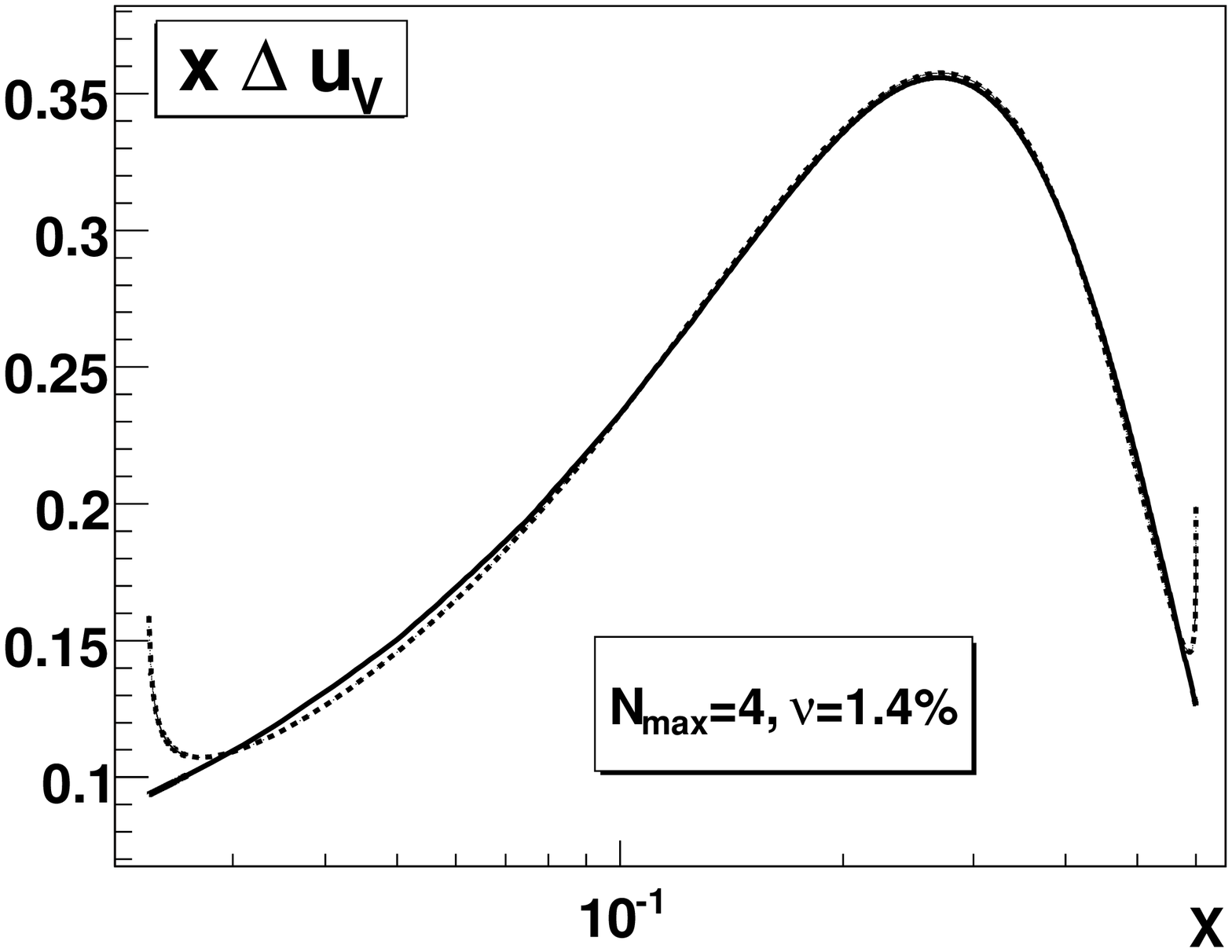}}
{\includegraphics[height=4cm,width=6cm]{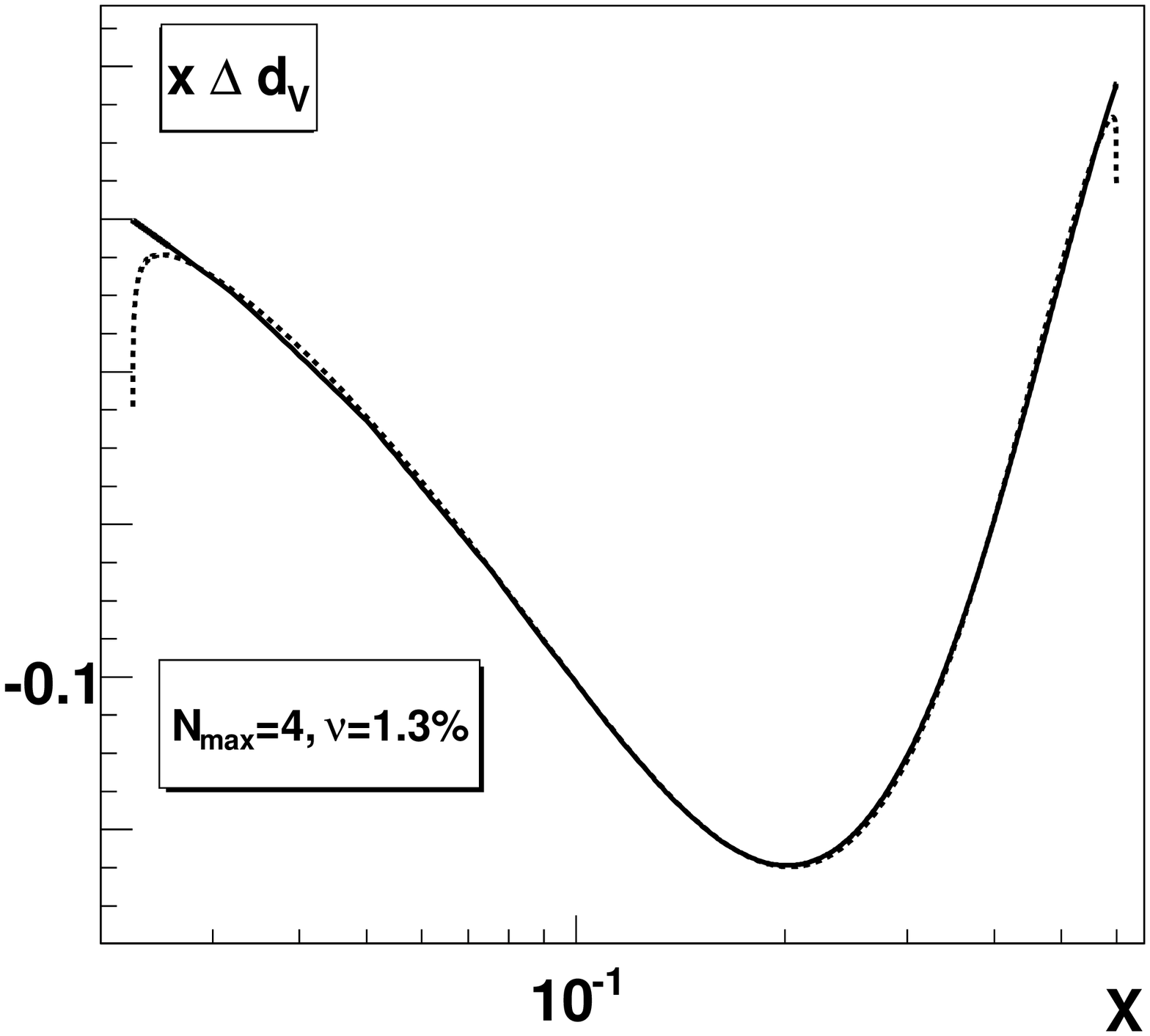}}
{\includegraphics[height=4cm,width=6cm]{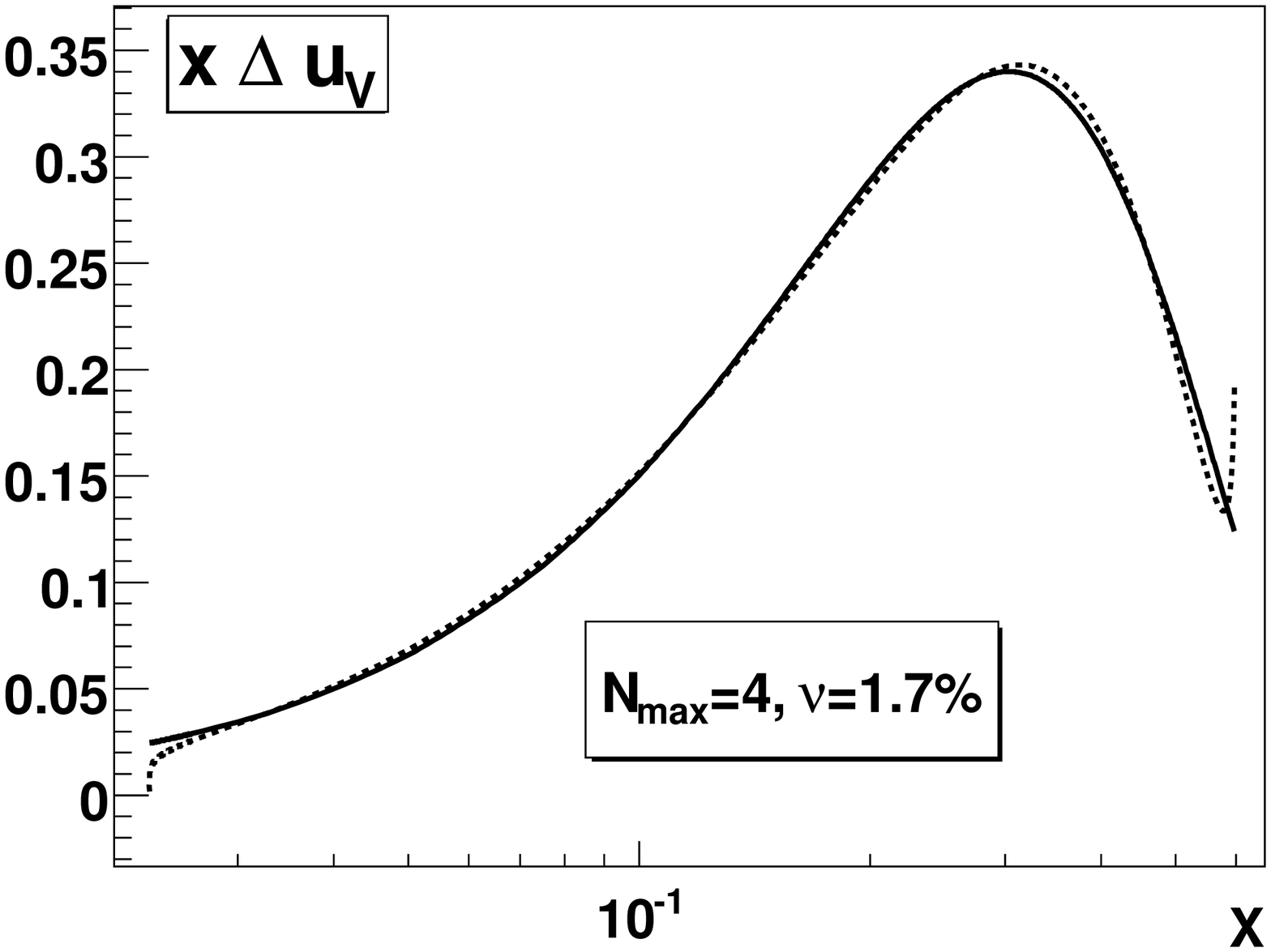}}
{\includegraphics[height=4cm,width=6cm]{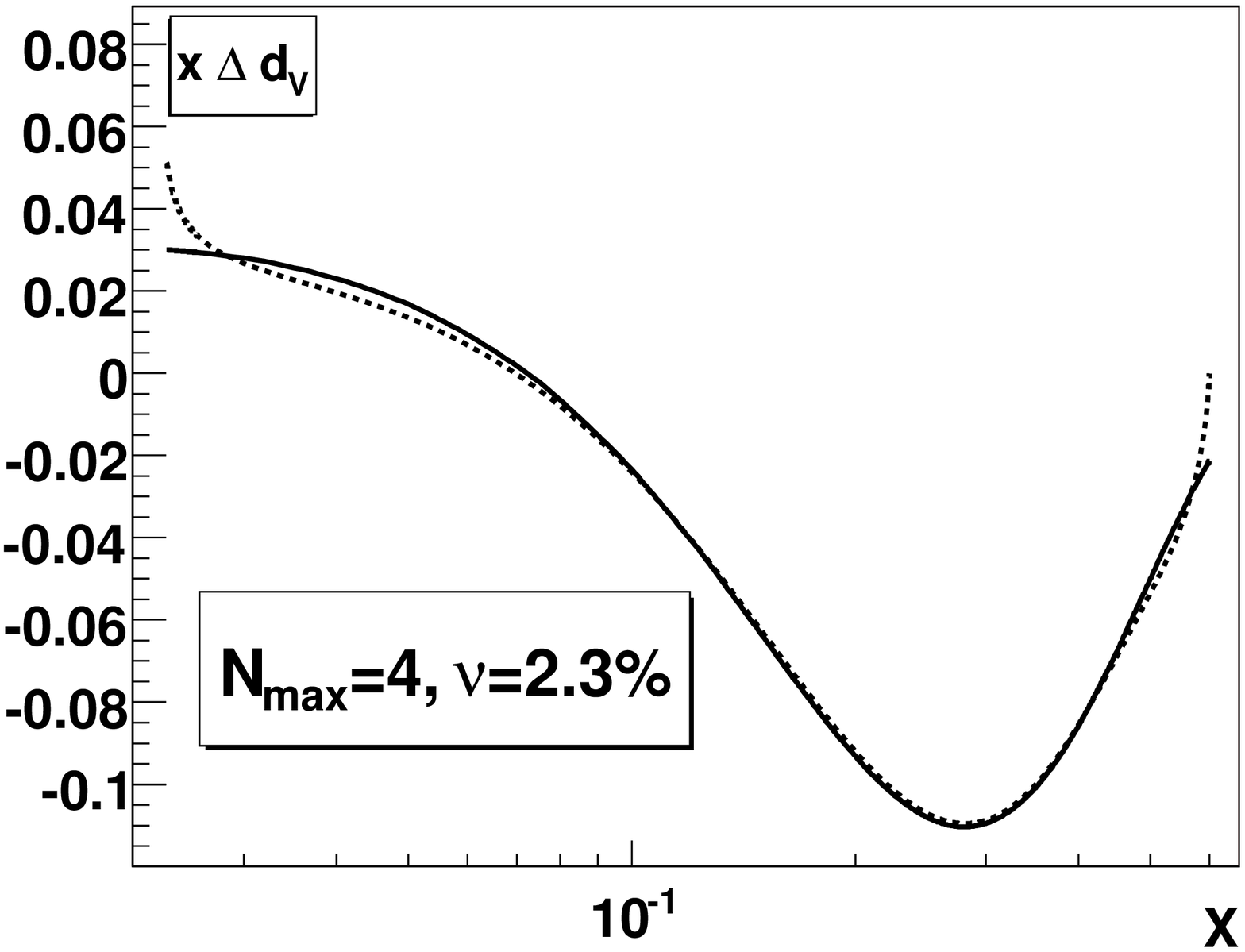}}
\label{f4}
\end{center}
\end{figure}

\begin{figure}
        \caption{\footnotesize (color online) Idealized LO testing of the integration procedures given by Eq. (\ref{summ}) 
        and Eq. (\ref{aint}). The input parametrizations GRSV2000LO with symmetric sea scenario (top) and
        broken sea scenario (bottom) are used for direct calculation of asymmetries with Eq. (\ref{difaslo}).
        Solid line corresponds to the input parametrization. Closed circles correspond to the values of input parametrization in the middle of each bin. 
        Broken line shows the way of integral approximation corresponding to application
        of Eq. (\ref{summ}). 
        Dashed line 
        is obtained with MJEM and application of  Eq. (\ref{aint}) for the moment calculation.
        Dot-dashed line is obtained with MJEM and application of Eq. (\ref{summ}) for the moment
        calculation. Parameters $\nu_1$ and $\nu_2$ are given by Eq. (\ref{nu}) and show
        the quality of reconstruction for the dashed and dot-dashed lines, respectively.
        }
{\includegraphics[height=4cm,width=6cm]{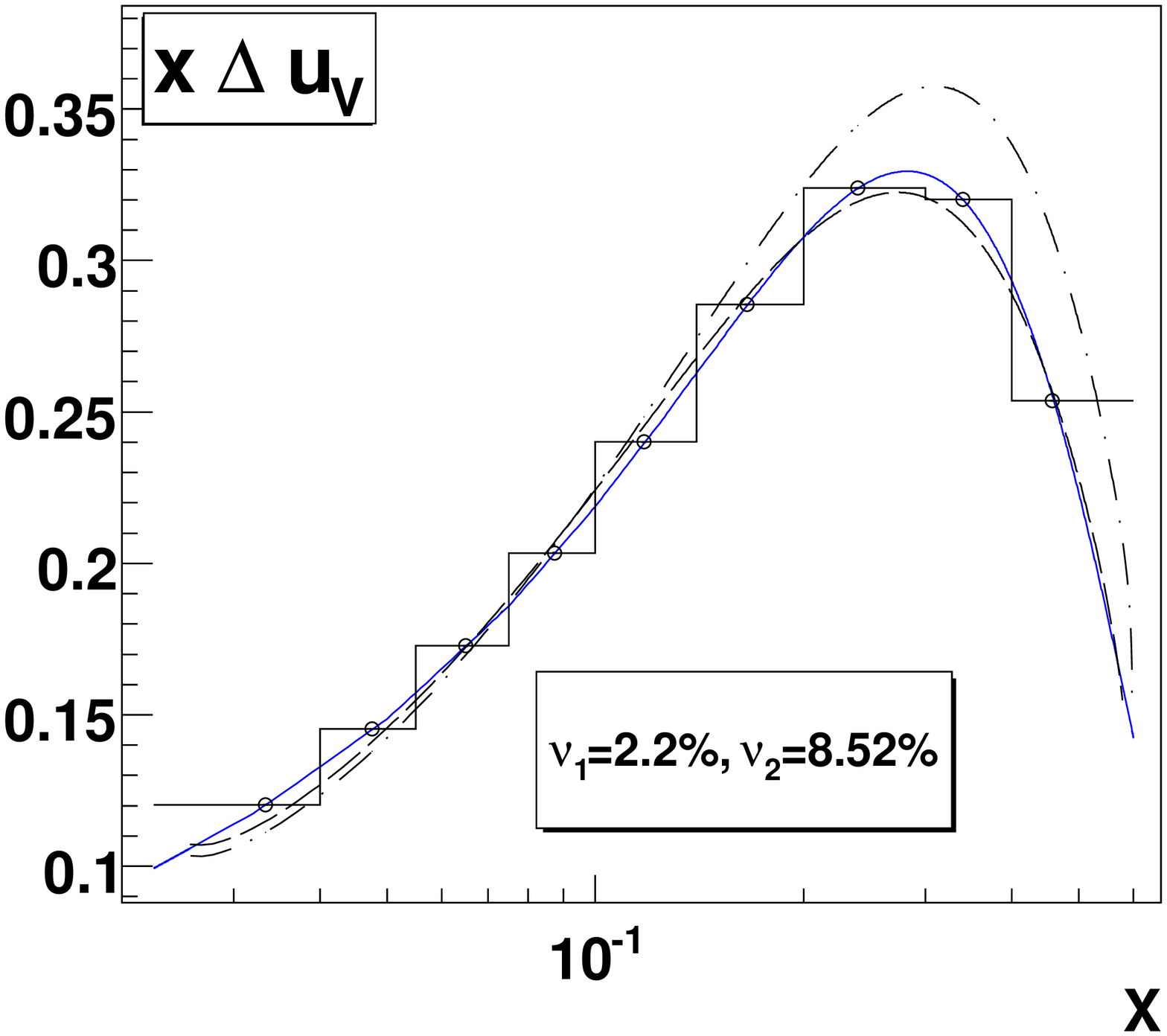}}
{\includegraphics[height=4cm,width=6cm]{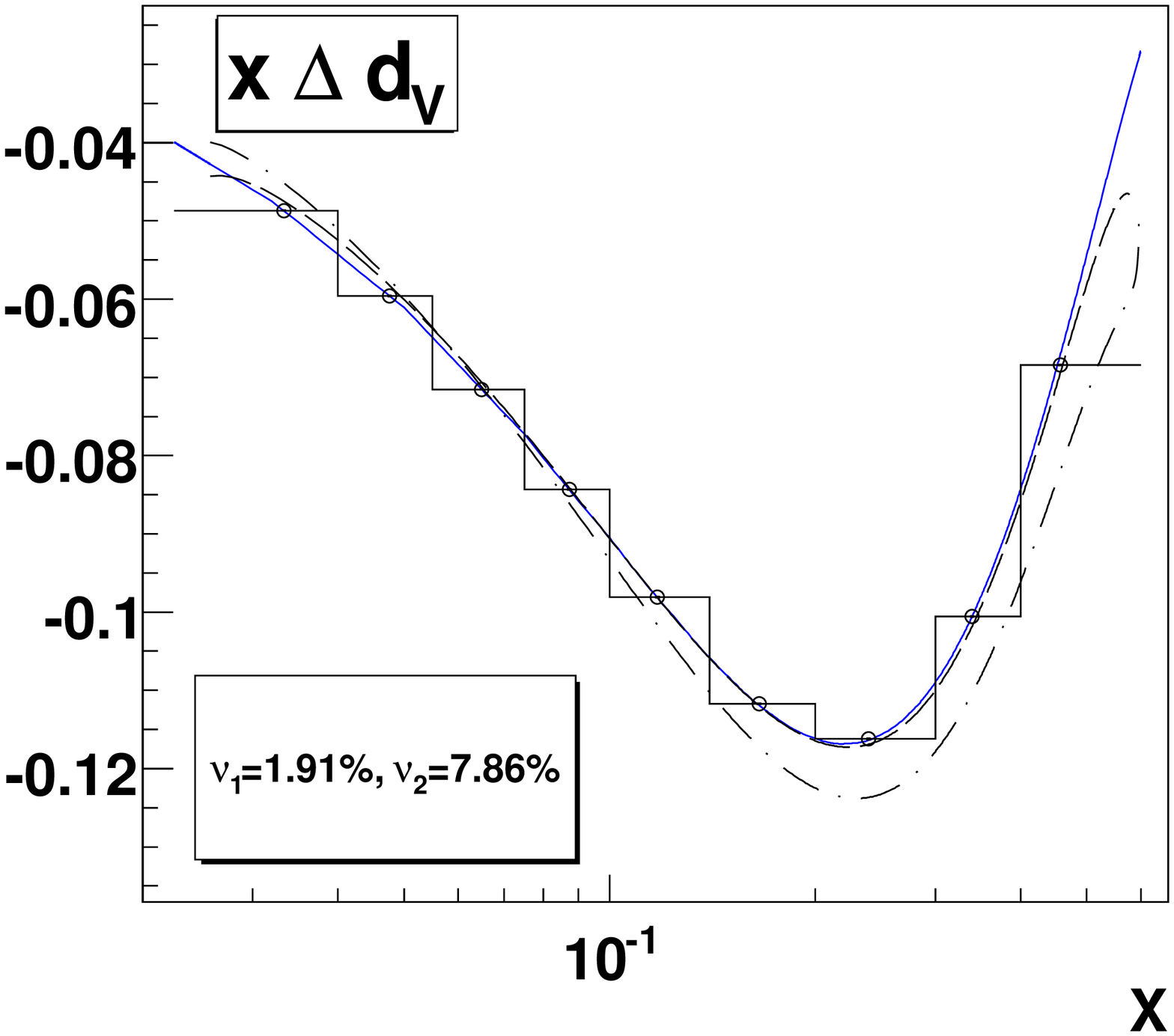}}\\
{\includegraphics[height=4cm,width=6cm]{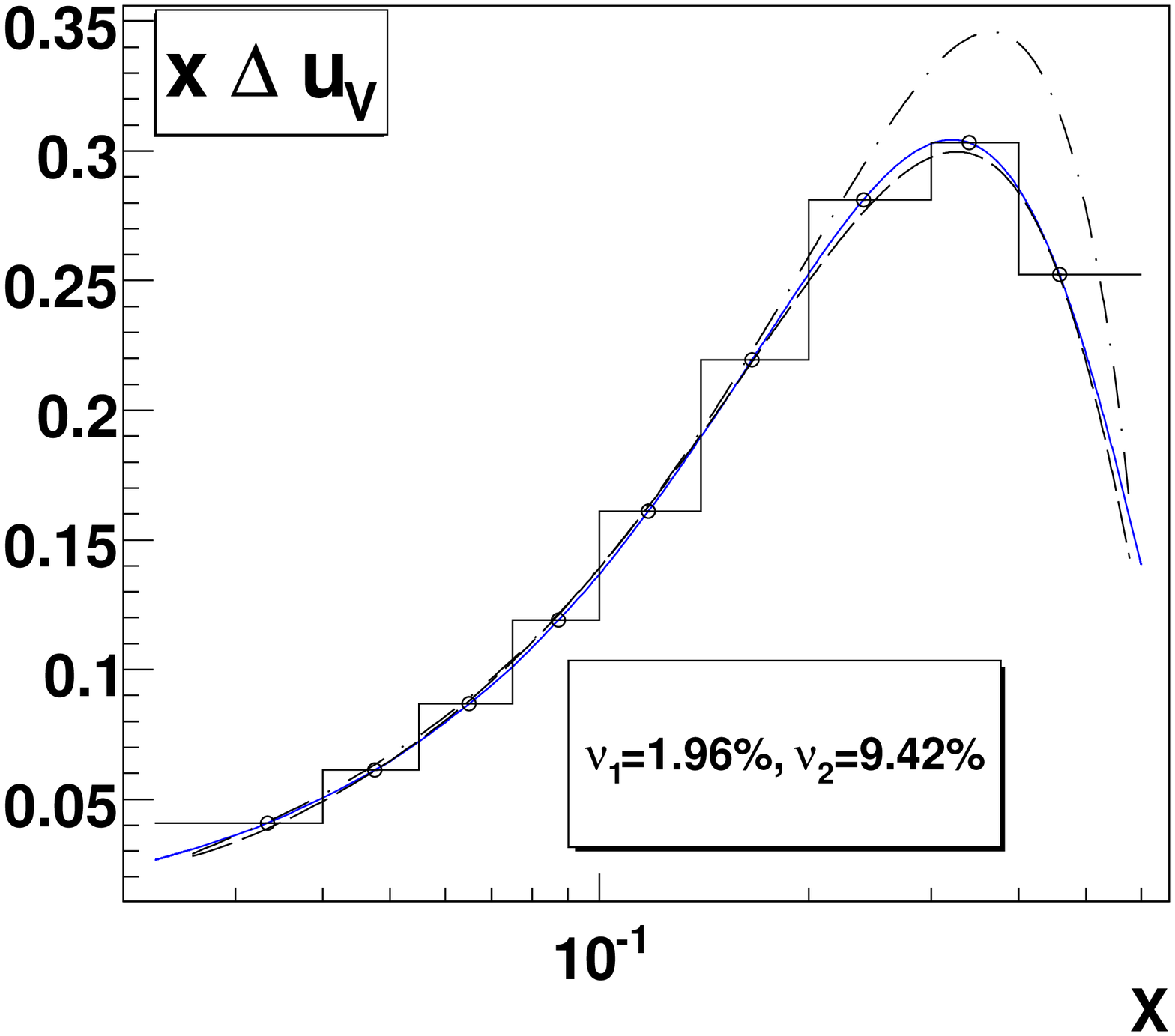}}
{\includegraphics[height=4cm,width=6cm]{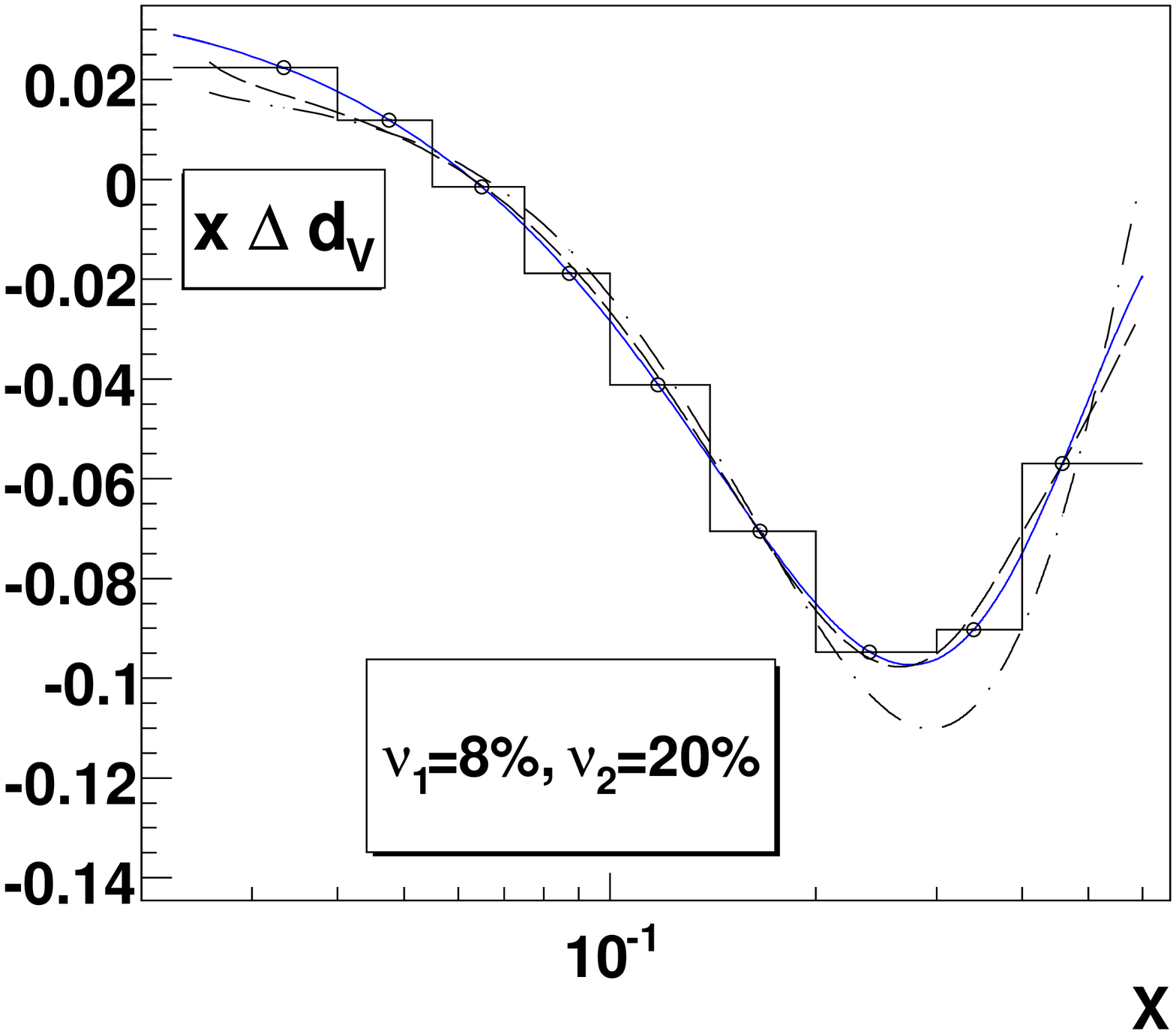}}
        \label{fig-integr}
\end{figure}

\begin{figure}
\caption{\footnotesize Results of  the simulated difference asymmetry analysis in LO.
GRSV2000LO parametrizations for  symmetric sea (top) and broken sea (bottom) scenarios 
are used for simulations.
Solid line corresponds to the input parametrization. Dashed line corresponds to the reconstructed with MJEM curve.
Points with error bars correspond to direct extraction of the valence distributions with Eq. (\ref{difaslo}). 
        }
\includegraphics[height=4.7cm,width=8.7cm]{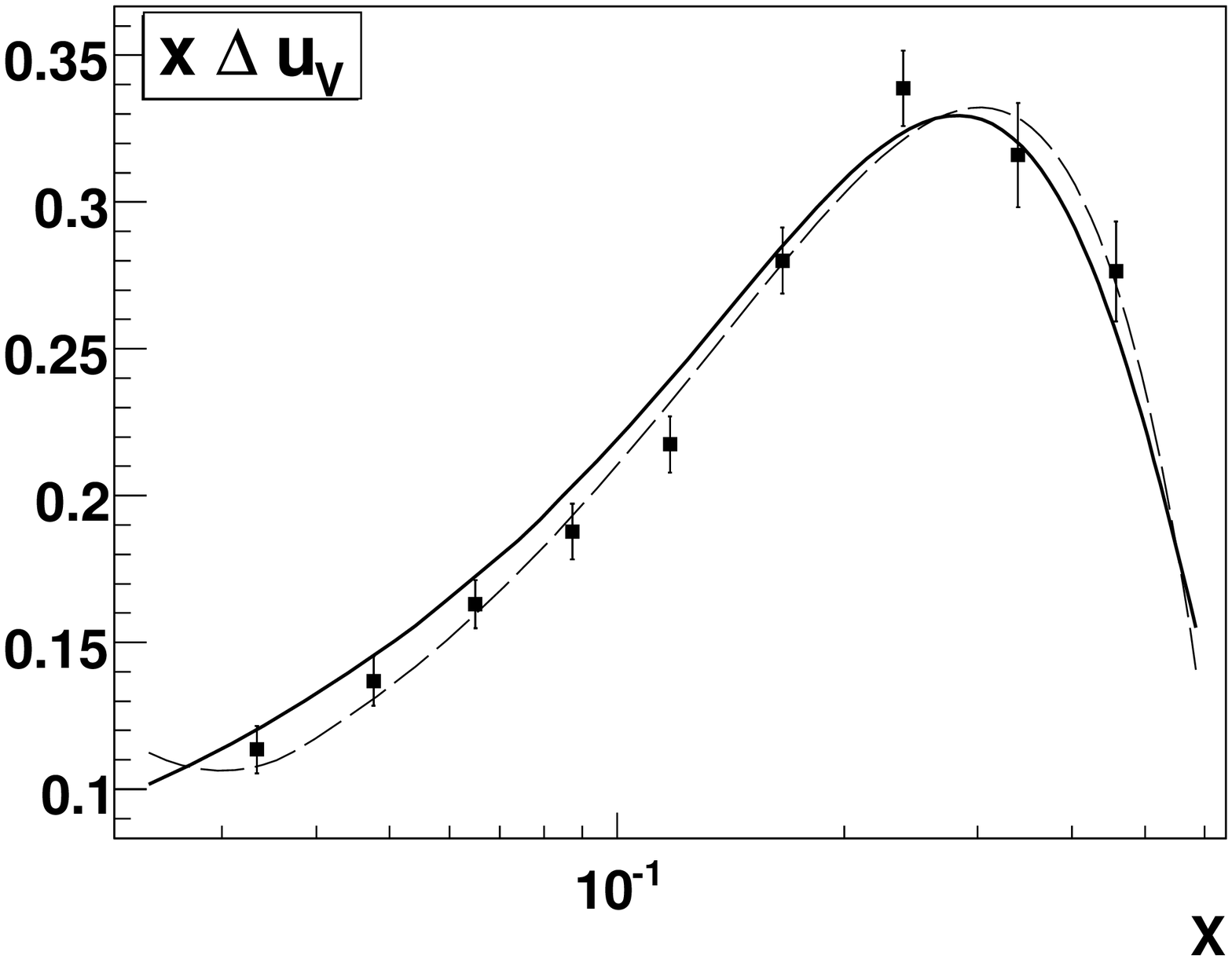}
\includegraphics[height=4.7cm,width=8.7cm]{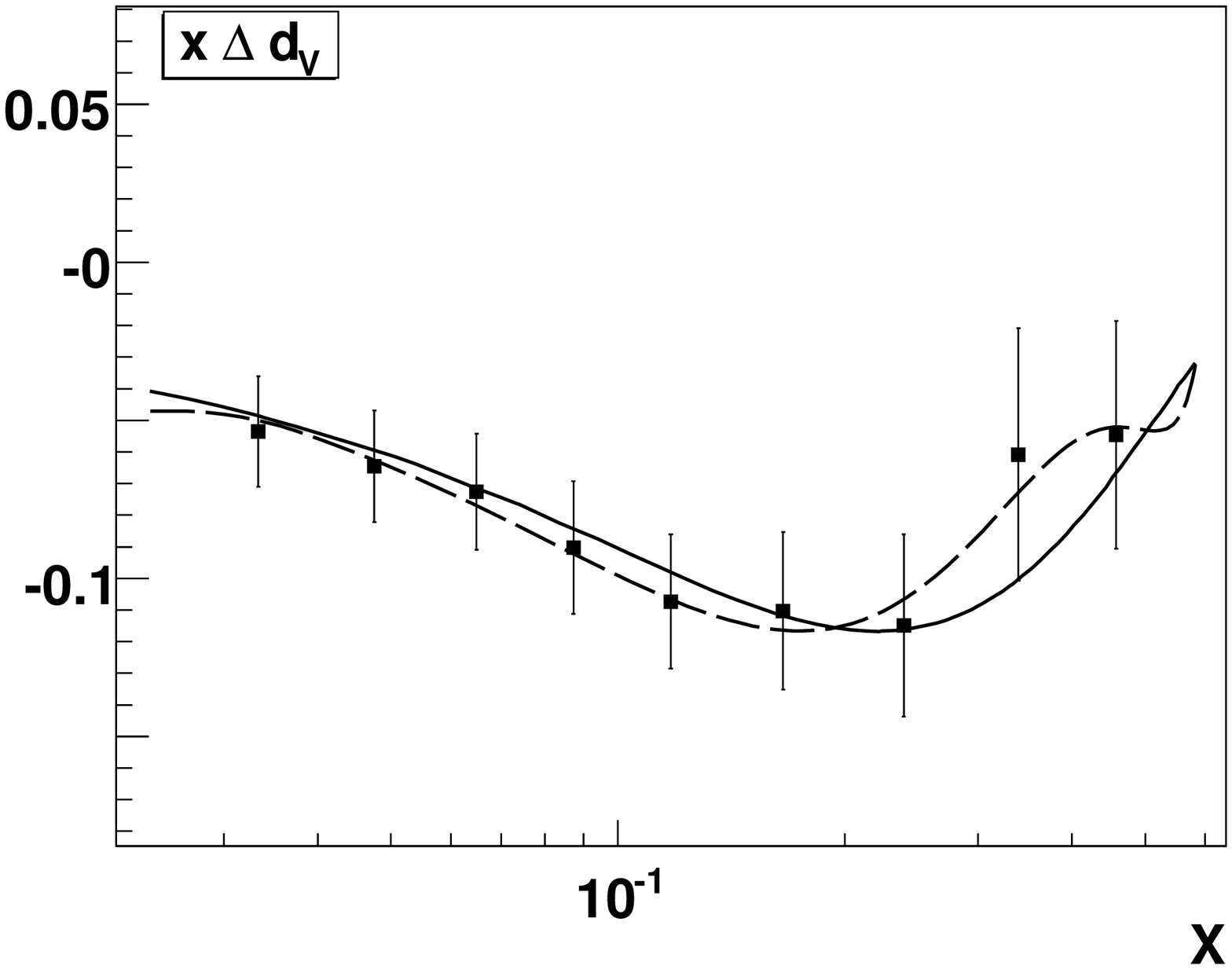}
\includegraphics[height=4.7cm,width=8.7cm]{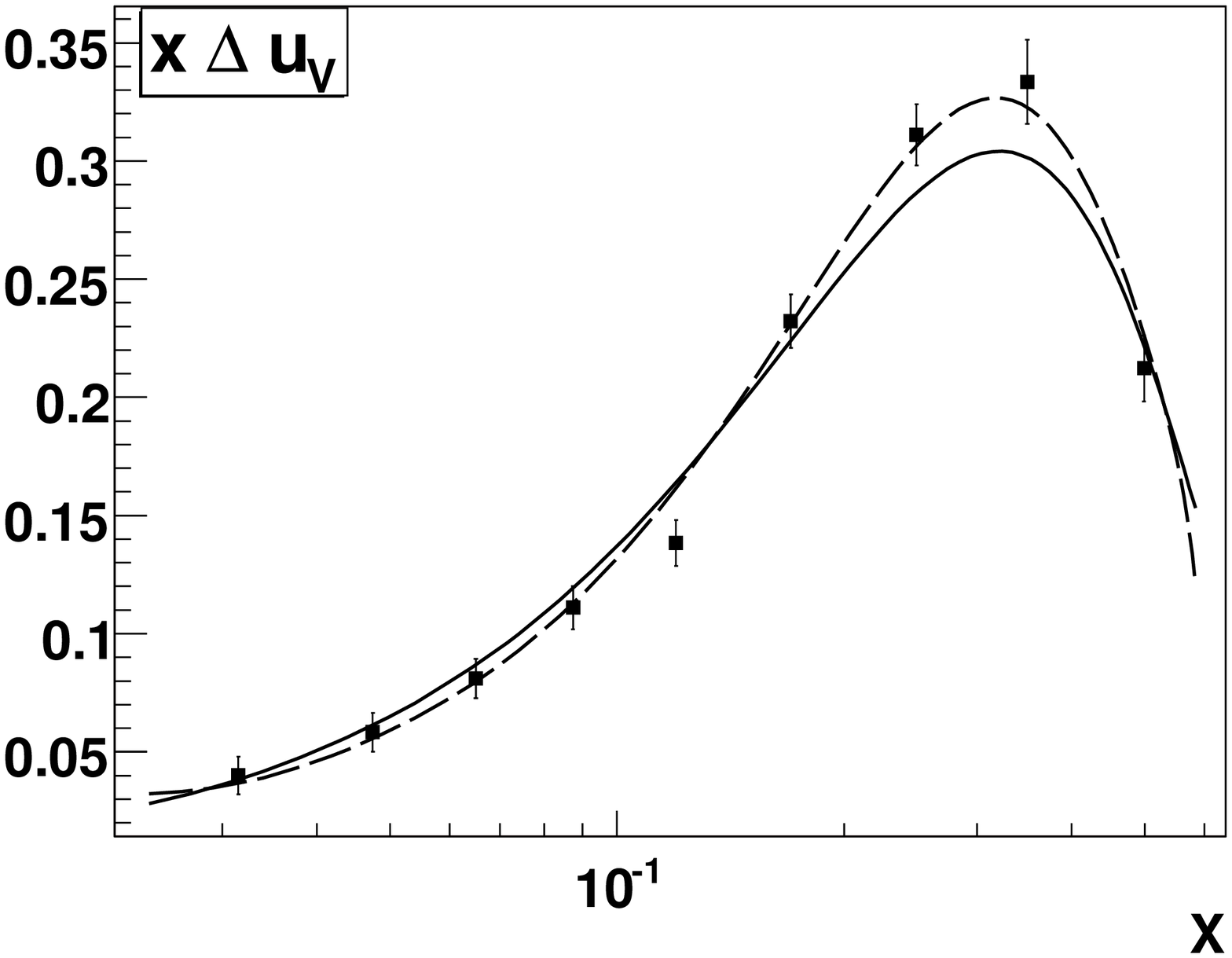}
\includegraphics[height=4.7cm,width=8.7cm]{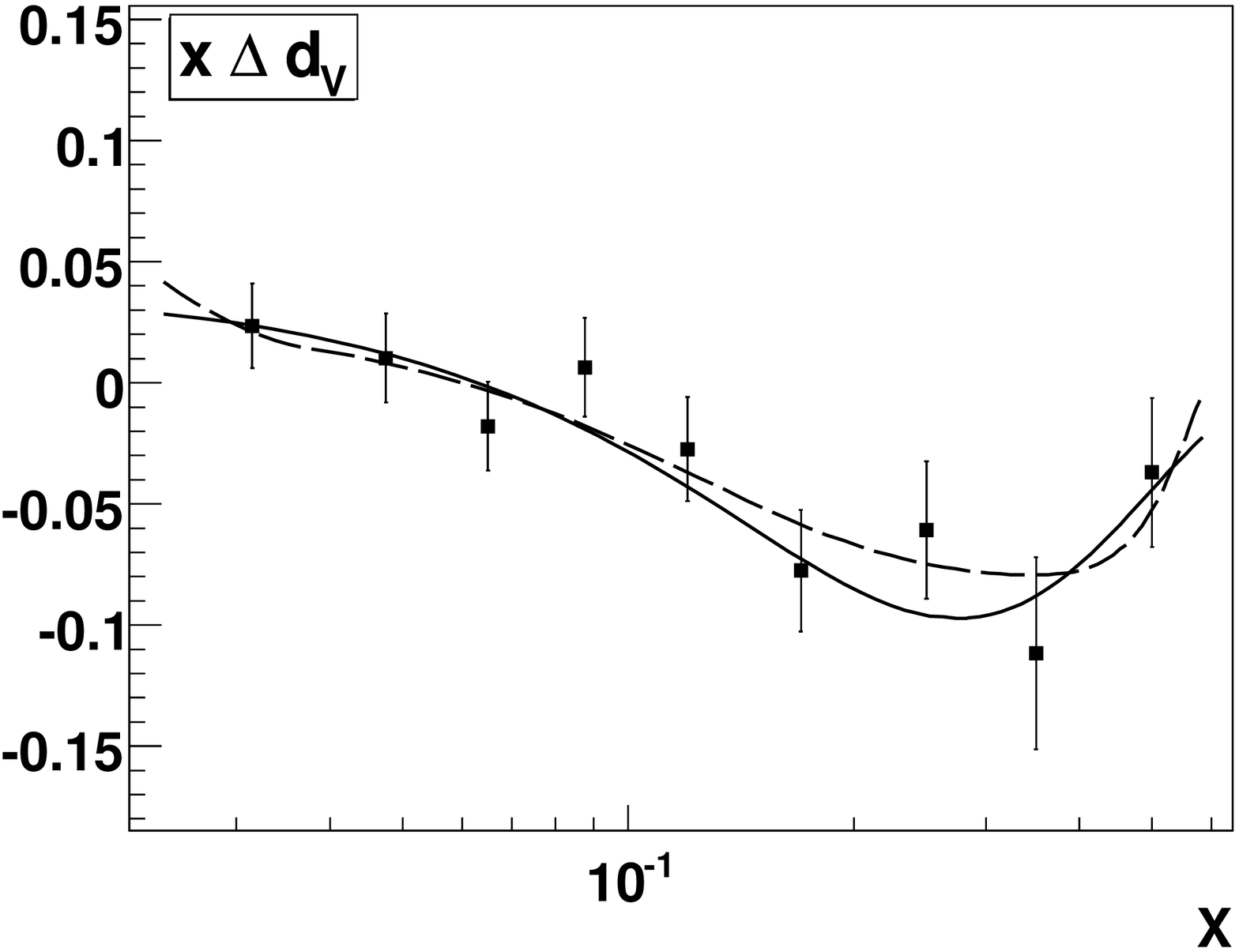}
\label{lodirect}
\end{figure}
\begin{figure}
\caption{\footnotesize 
Results of  the simulated difference asymmetry analysis in NLO.
GRSV2000NLO parametrizations for  symmetric sea (top) and broken sea (bottom) scenarios 
are used for simulations.
Solid line corresponds to the input parametrization. Dashed line corresponds to the reconstructed with MJEM curve.
}
\includegraphics[height=4.7cm,width=8.7cm]{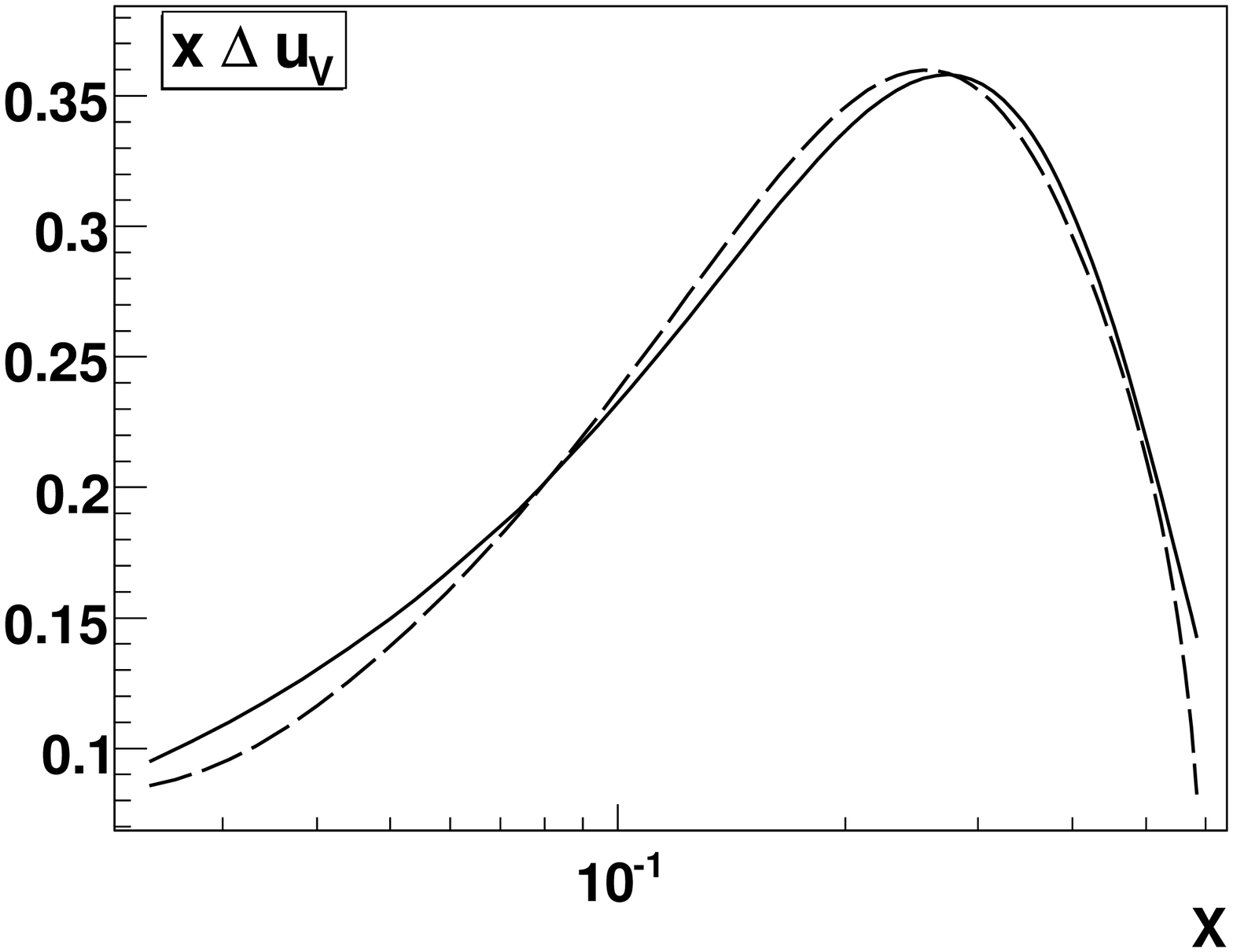}
\includegraphics[height=4.7cm,width=8.7cm]{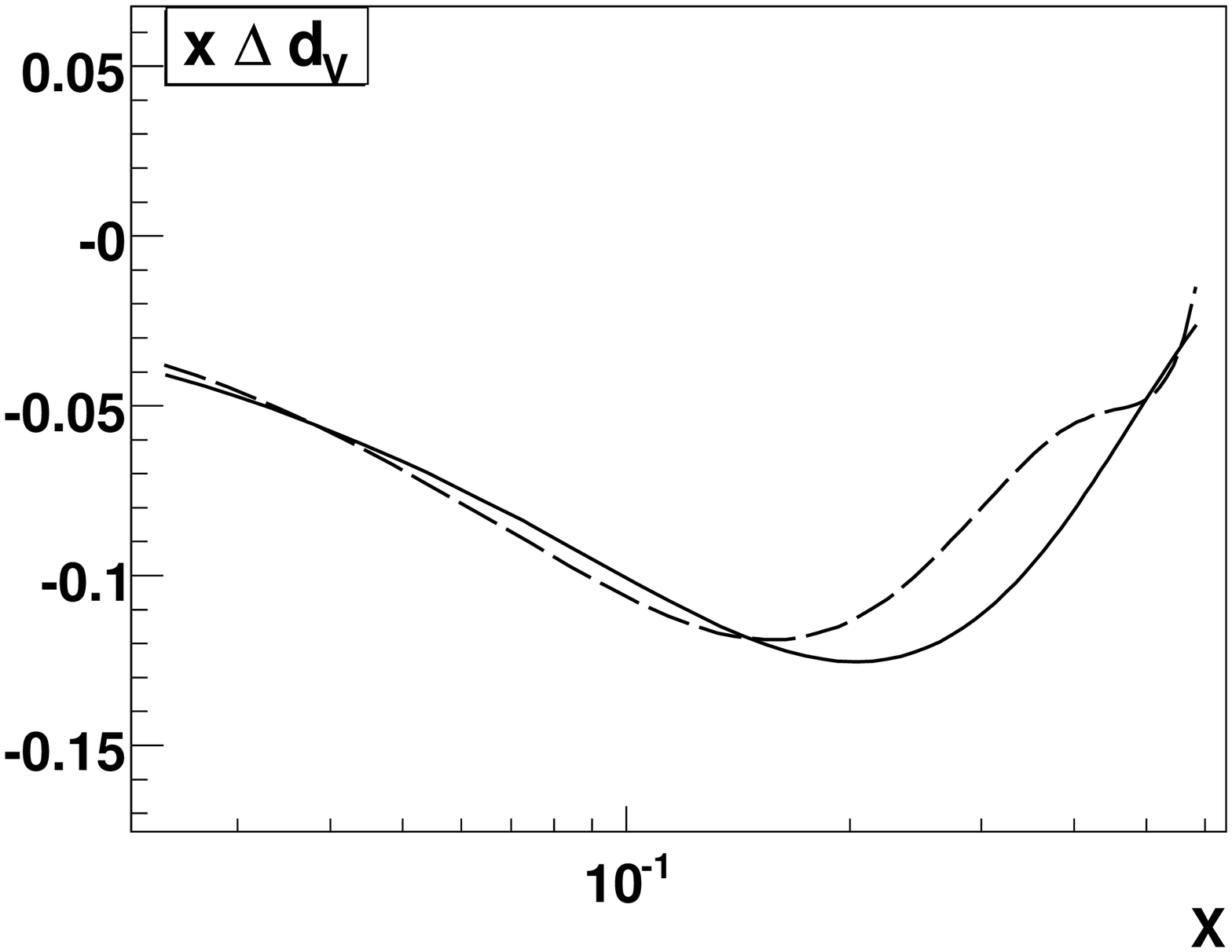}
\includegraphics[height=4.7cm,width=8.7cm]{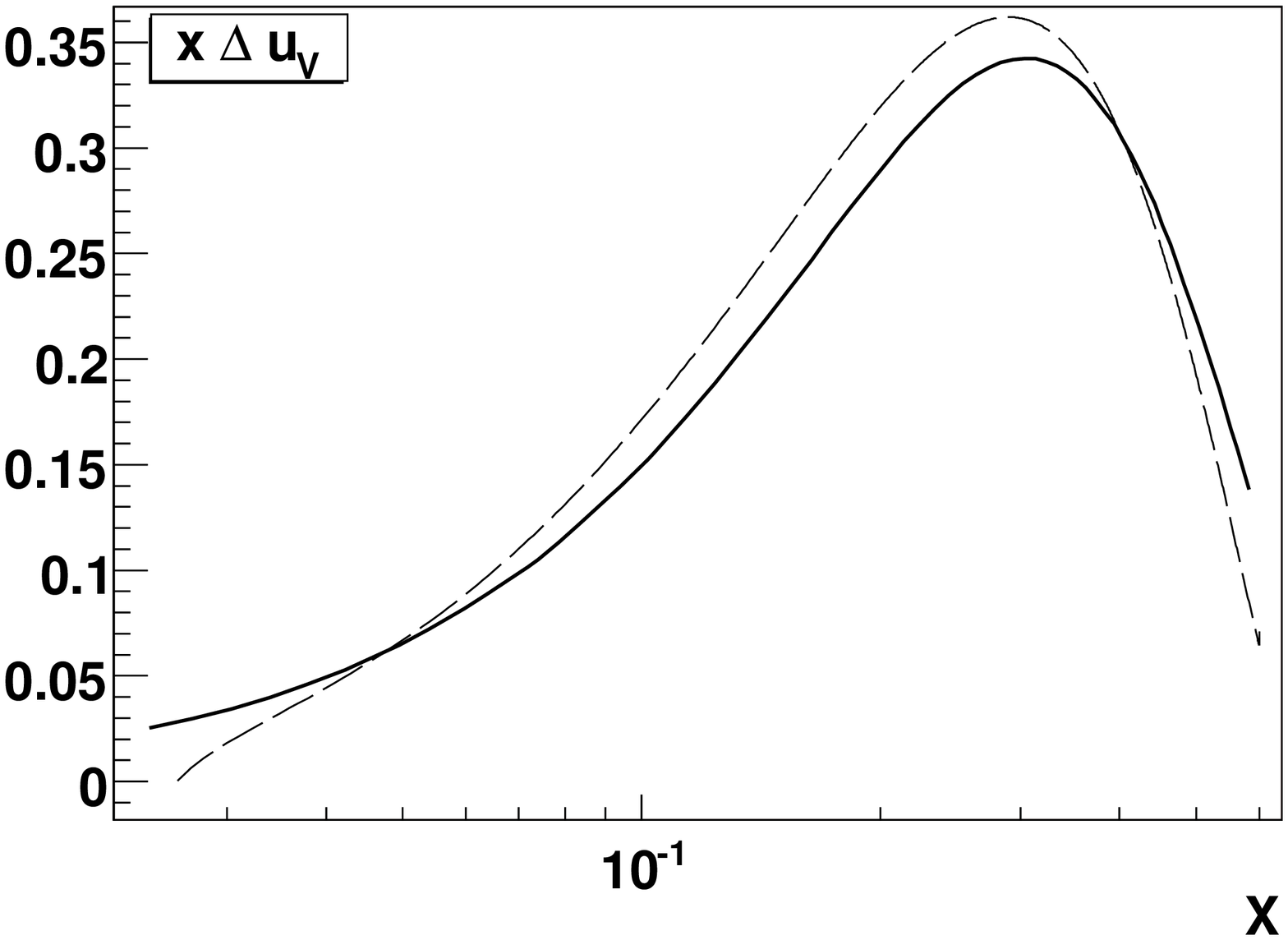}
\includegraphics[height=4.7cm,width=8.7cm]{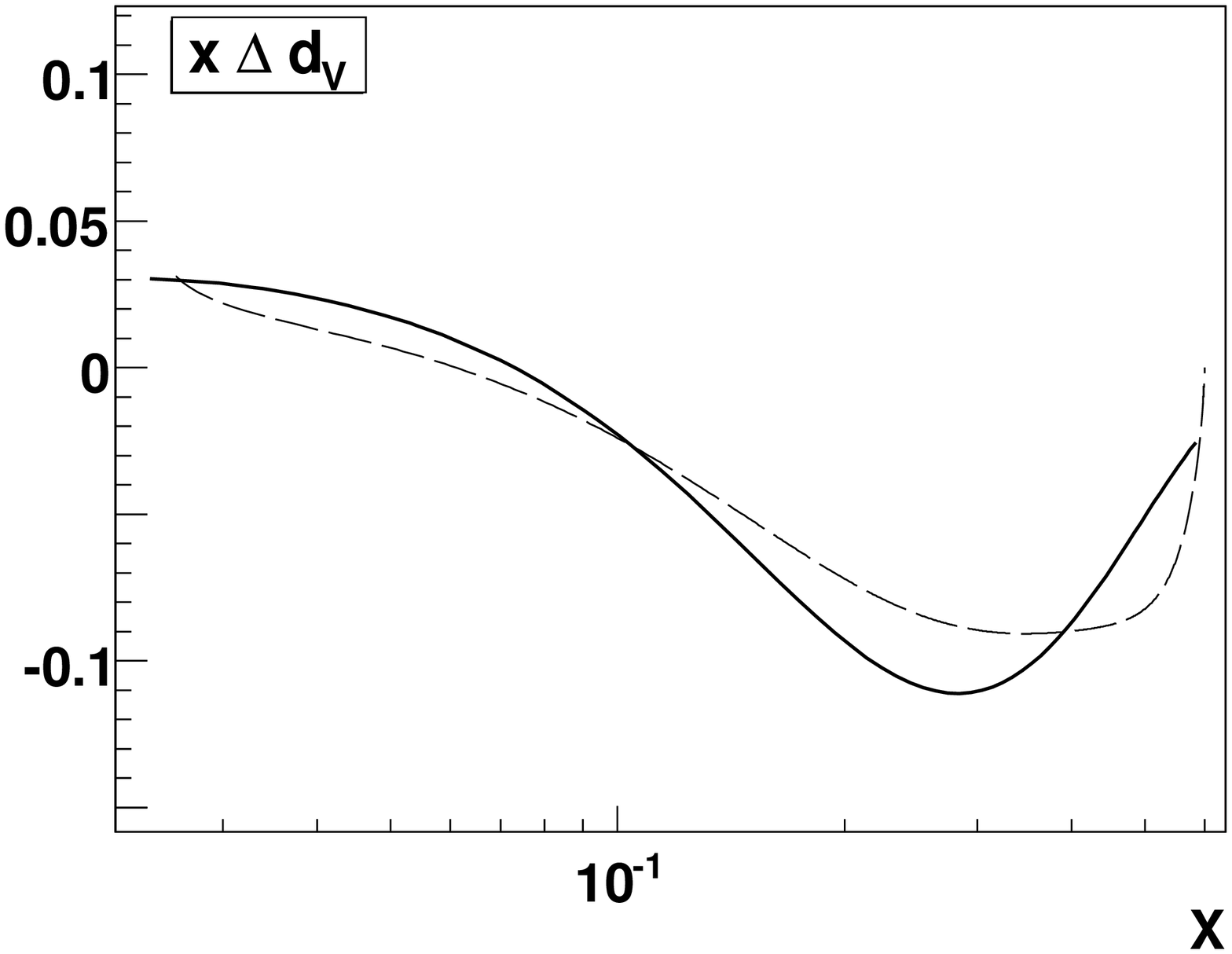}
\label{nlo-comparison}
\end{figure}

\begin{figure}
        \caption{\footnotesize Combined results of LO and NLO analysis (top) of the simulated difference asymmetries
        in comparison with the respective LO and NLO versions of GRSV2000 (symmetric sea) parametrization (bottom).
        Dashed and solid lines correspond to LO and NLO curves, respectively.
        }
\includegraphics[height=4.7cm,width=8.7cm]{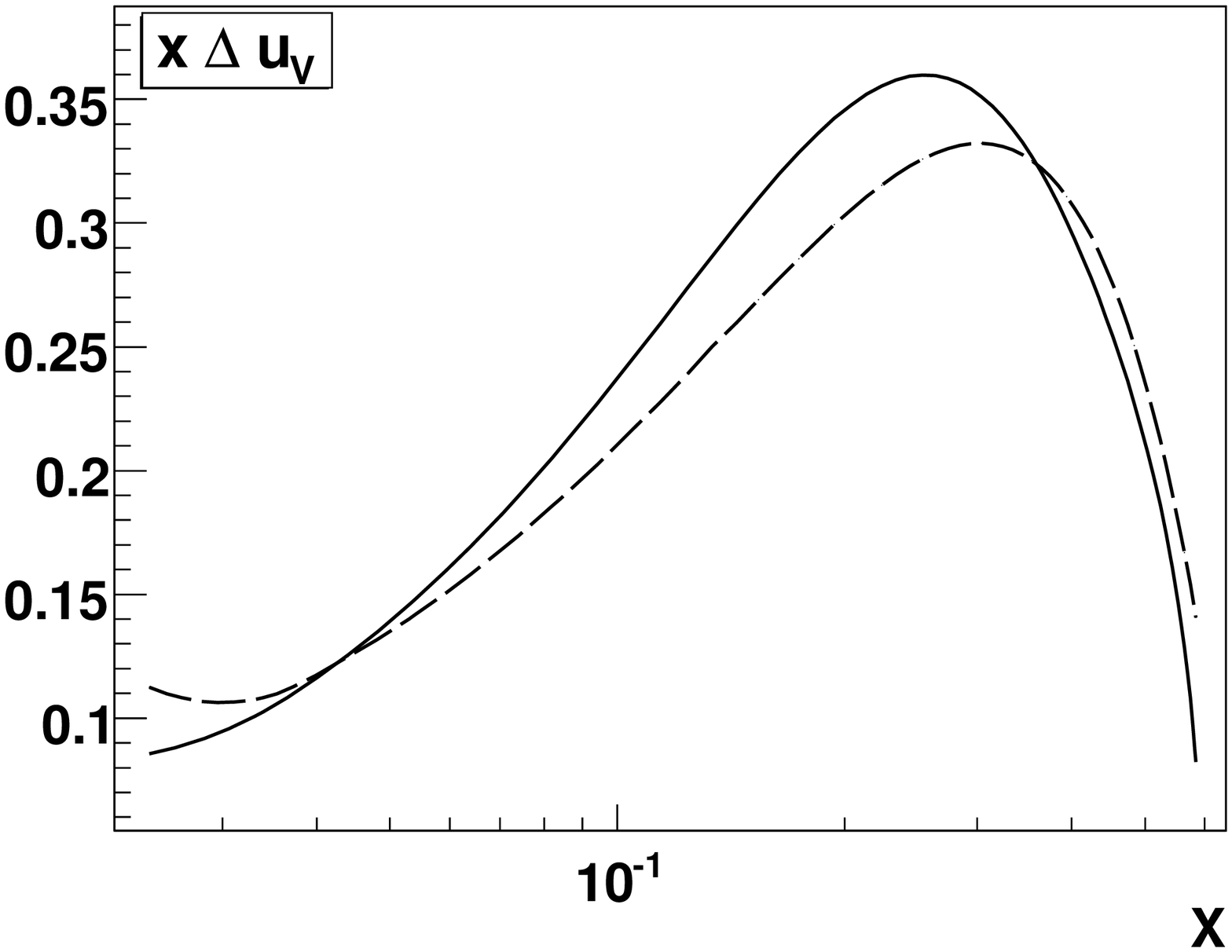}
\includegraphics[height=4.7cm,width=8.7cm]{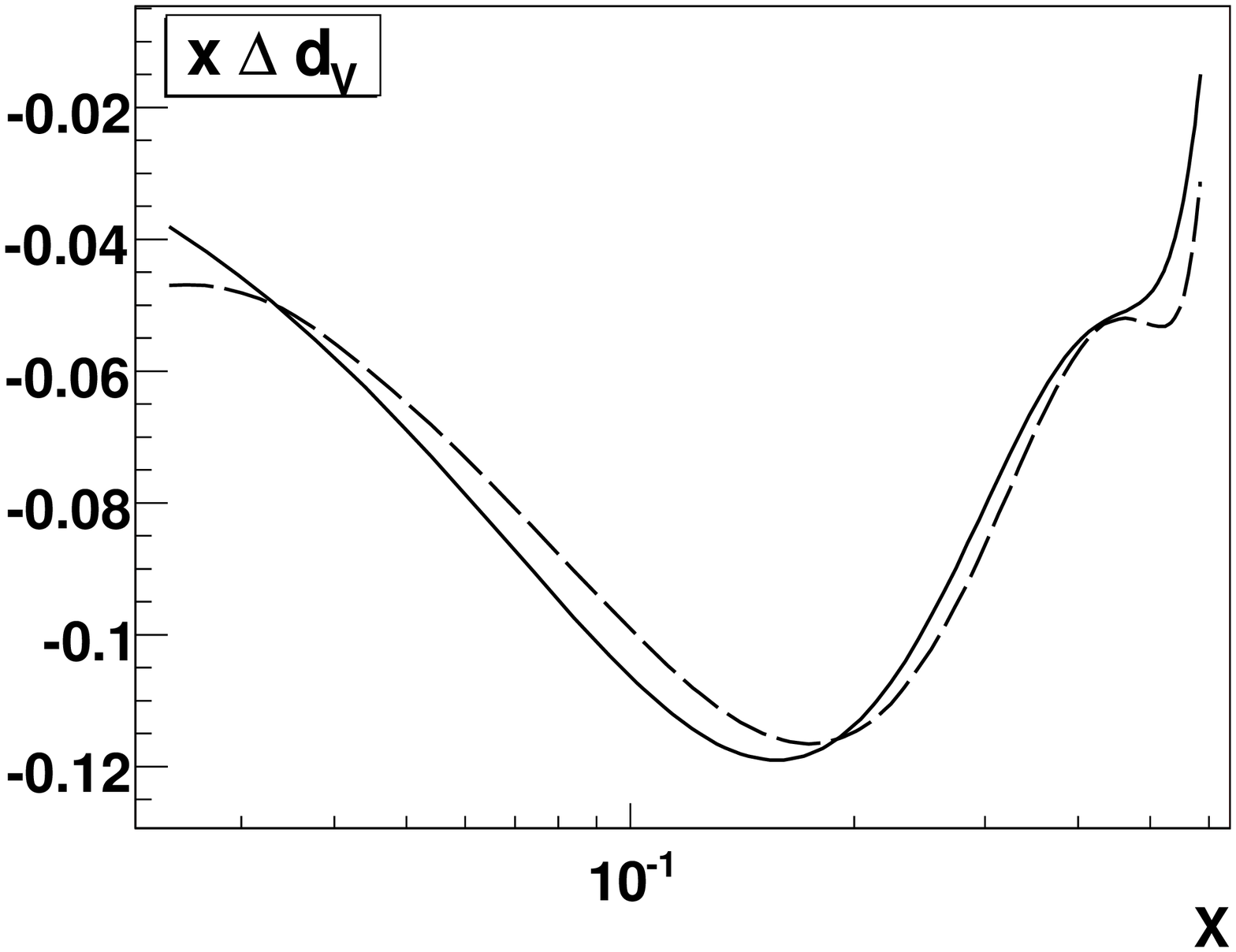}
\includegraphics[height=4.7cm,width=8.7cm]{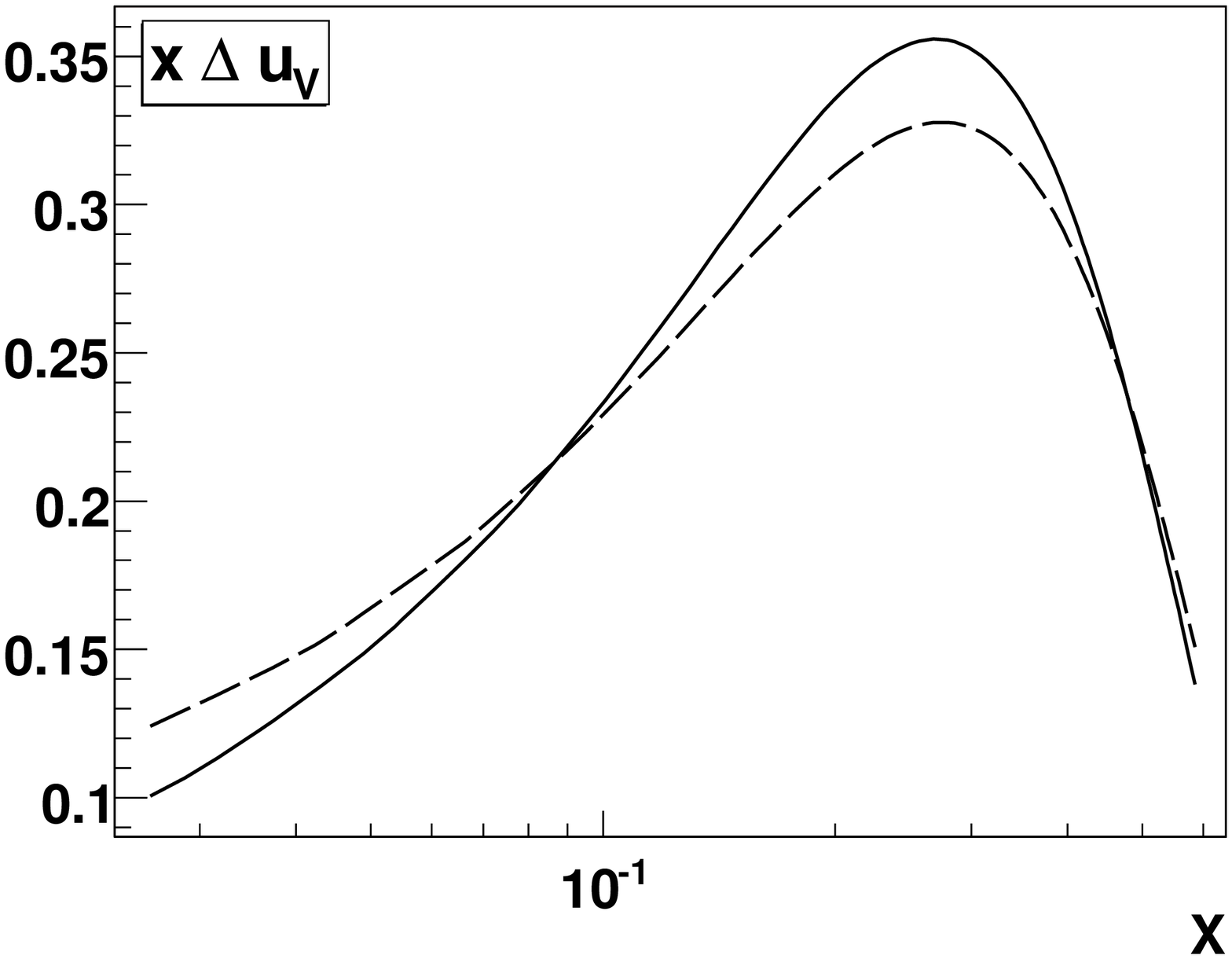}
\includegraphics[height=4.7cm,width=8.7cm]{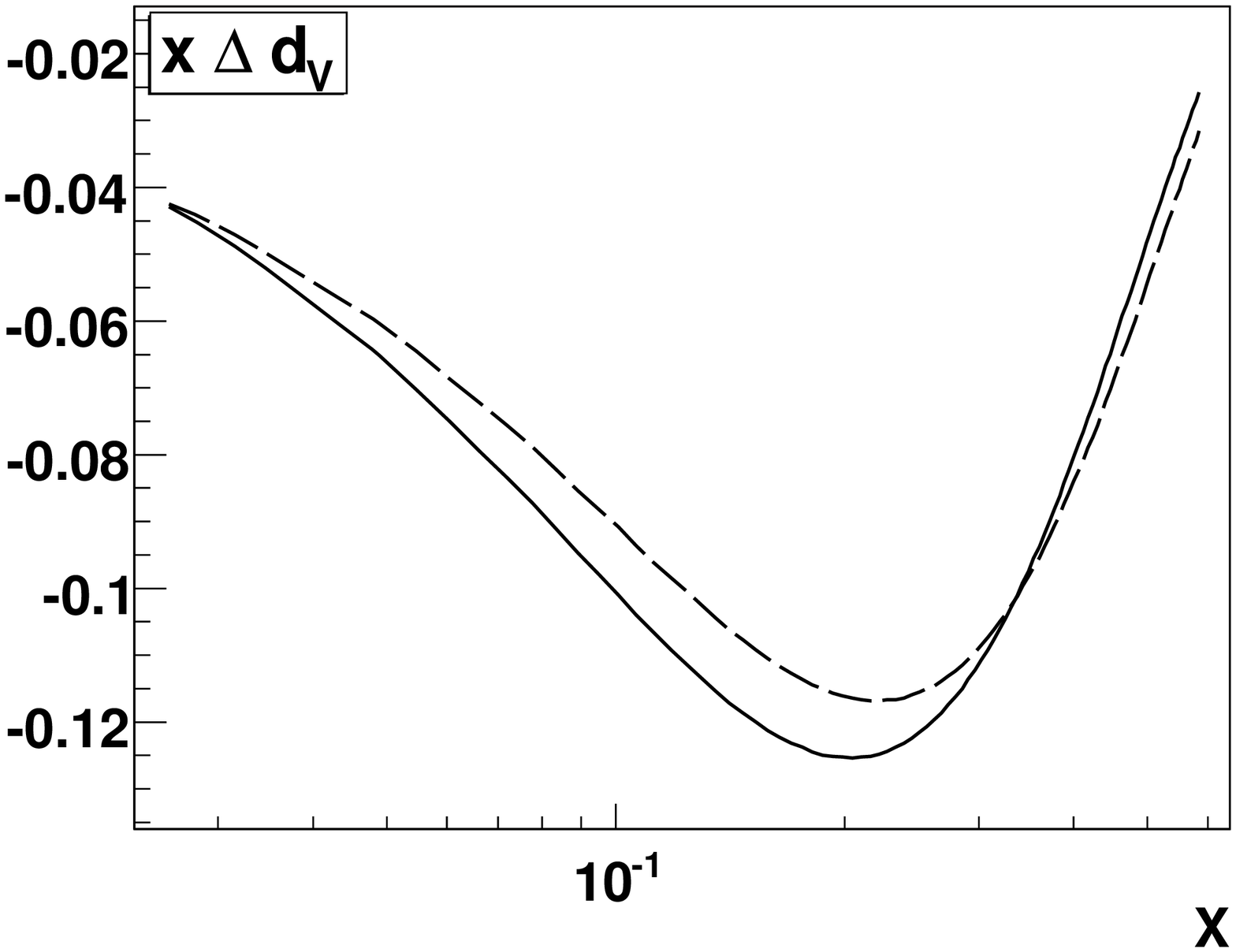}
\label{comb-sym}
\end{figure}
\begin{figure}
        \caption{\footnotesize Combined results of LO and NLO analysis (top) of the simulated difference asymmetries
                in comparison with the respective LO and NLO versions of GRSV2000 (broken sea) parametrization (bottom).
        Dashed and solid lines correspond to LO and NLO curves, respectively.}
\includegraphics[height=4.7cm,width=8.7cm]{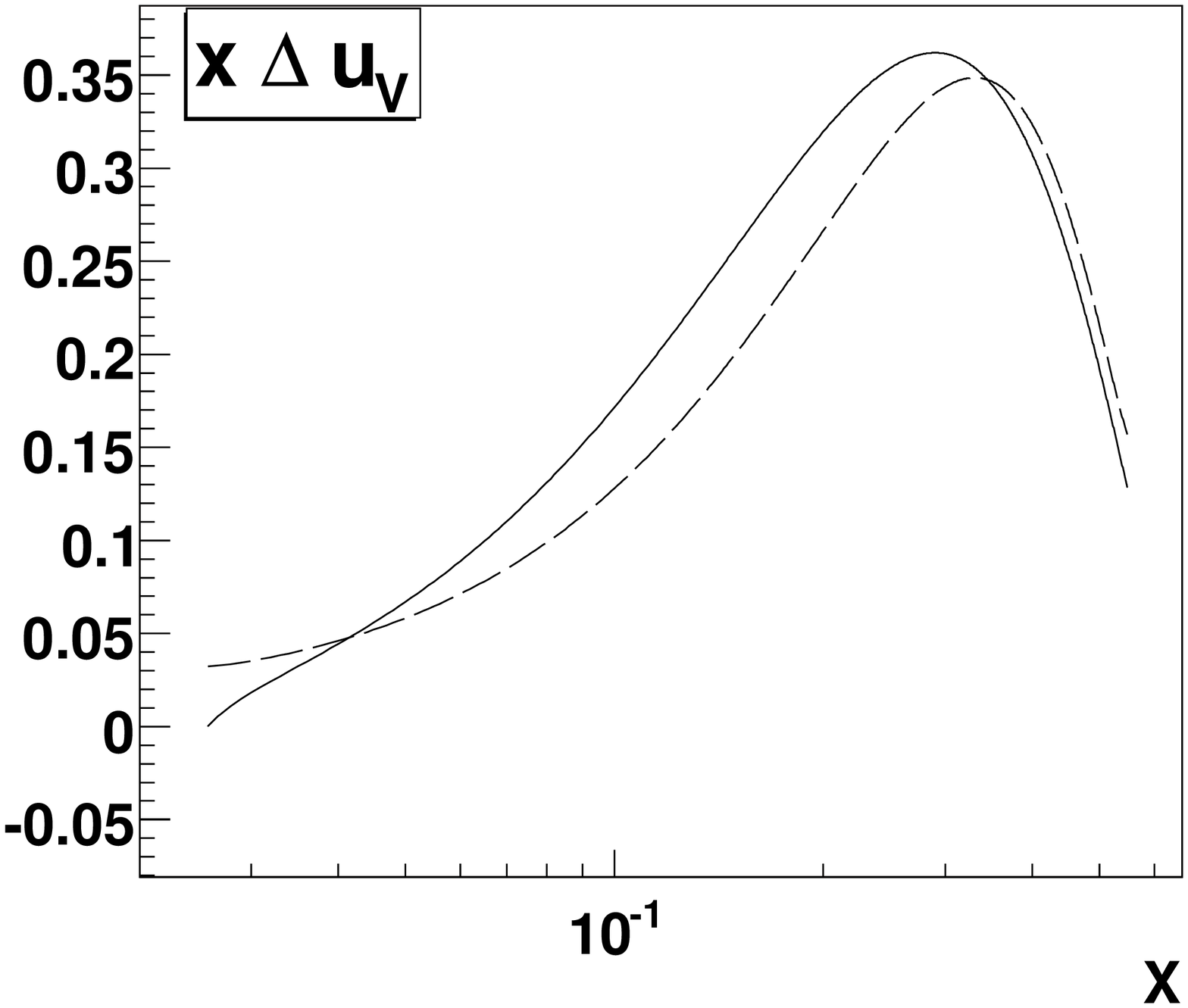}
\includegraphics[height=4.7cm,width=8.7cm]{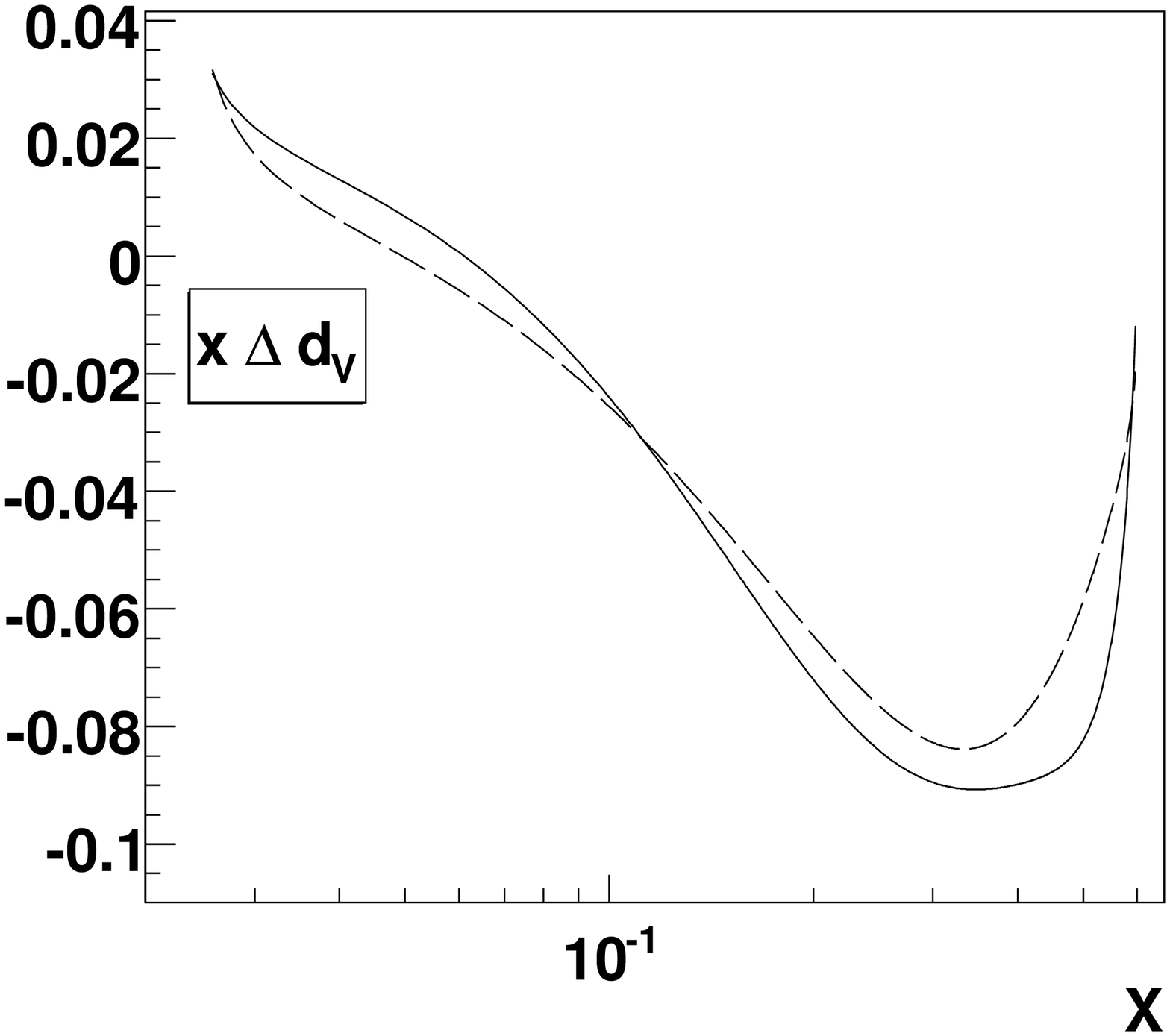}
\includegraphics[height=4.7cm,width=8.7cm]{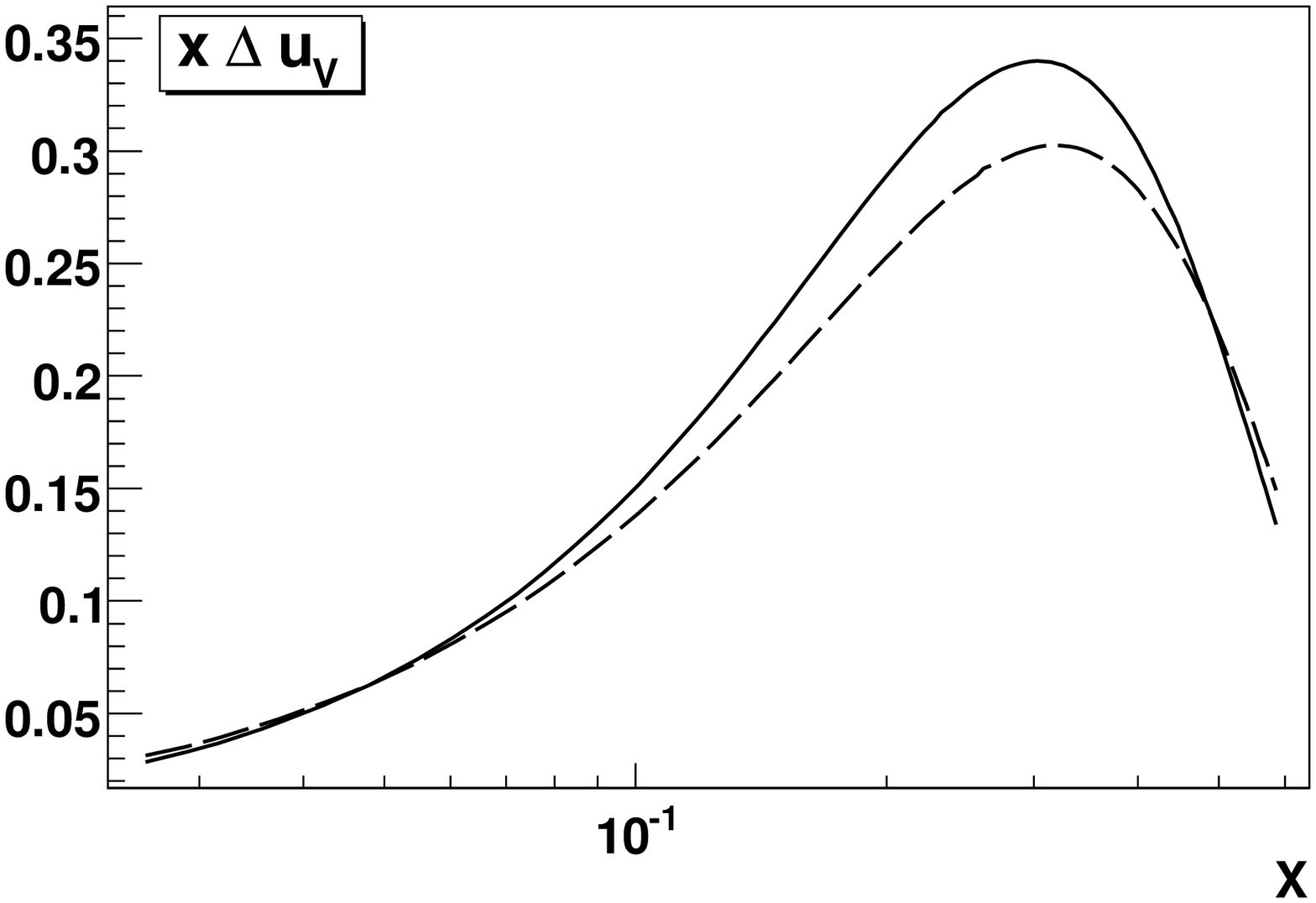}
\includegraphics[height=4.7cm,width=8.7cm]{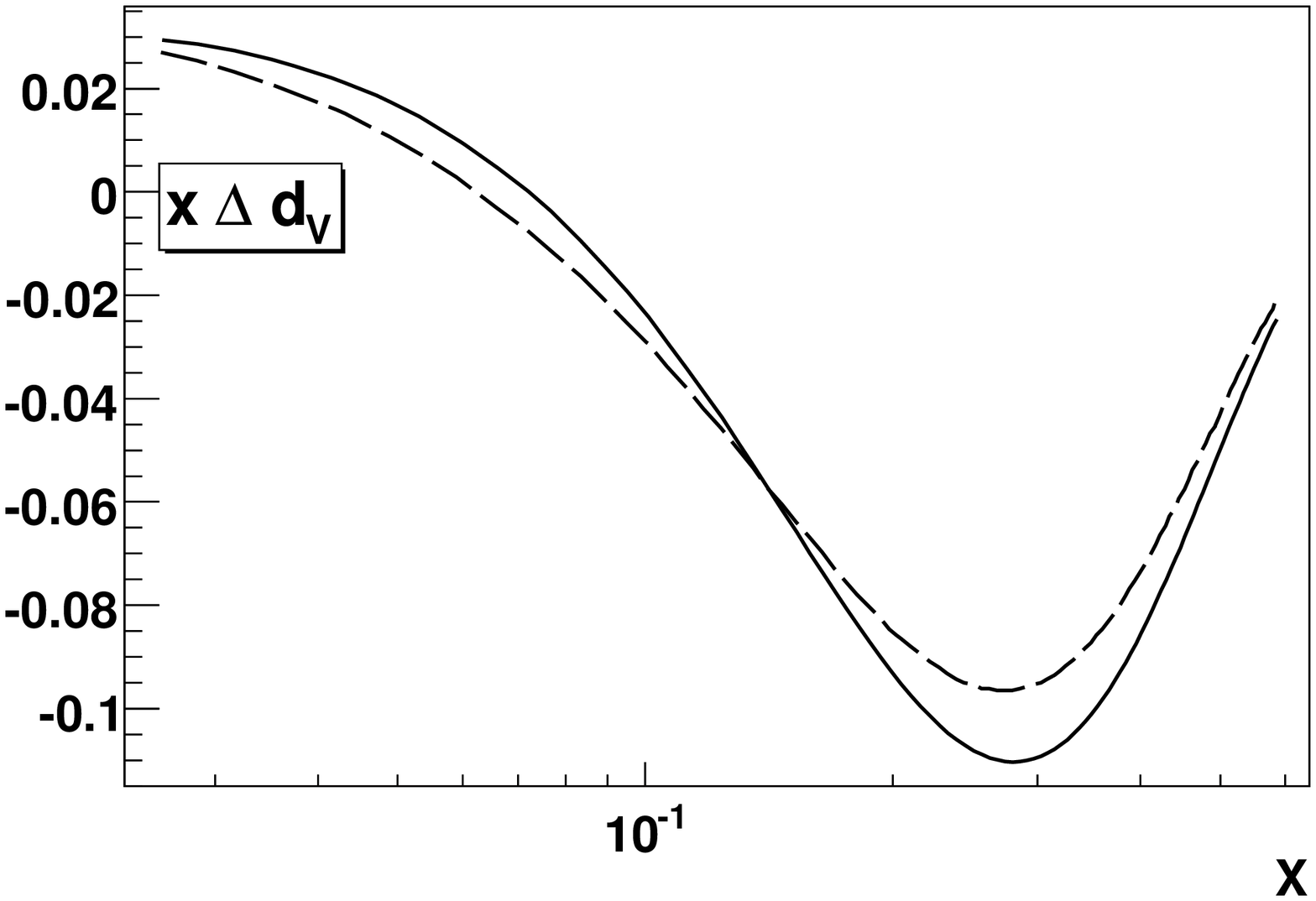}

\label{comb-br}
\end{figure}

\begin{figure}
        \caption{\footnotesize Difference asymmetries constructed with application of Eq. (\ref{expansatz}).}
\includegraphics[height=8.7cm,width=8.7cm]{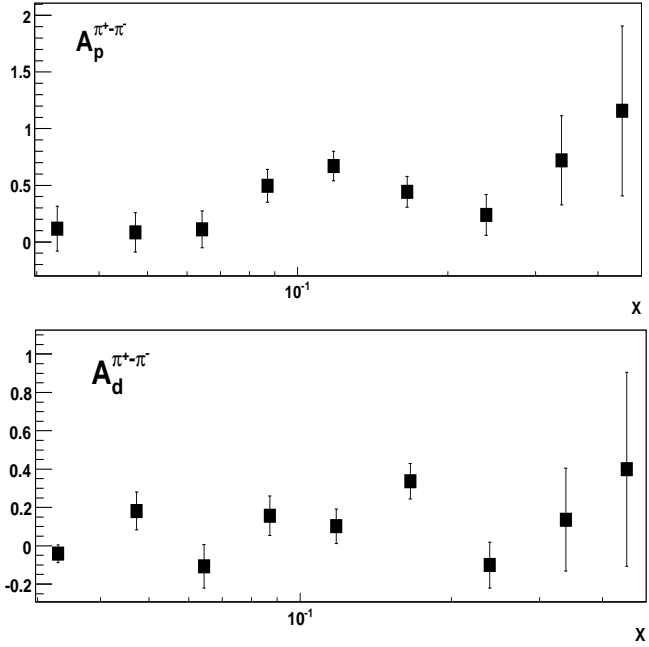}
\label{figasym}
\end{figure}
\begin{figure}
        \caption{\footnotesize (color online) LO extraction of the valence PDFs from the difference asymmetries constructed 
        with Eq. (\ref{expansatz}) (up-oriented triangles) in comparison with the respective published
        HERMES results 
        (down-oriented triangles). The HERMES results are shifted to the right for better visibility. }
\includegraphics[height=4.7cm,width=8.7cm]{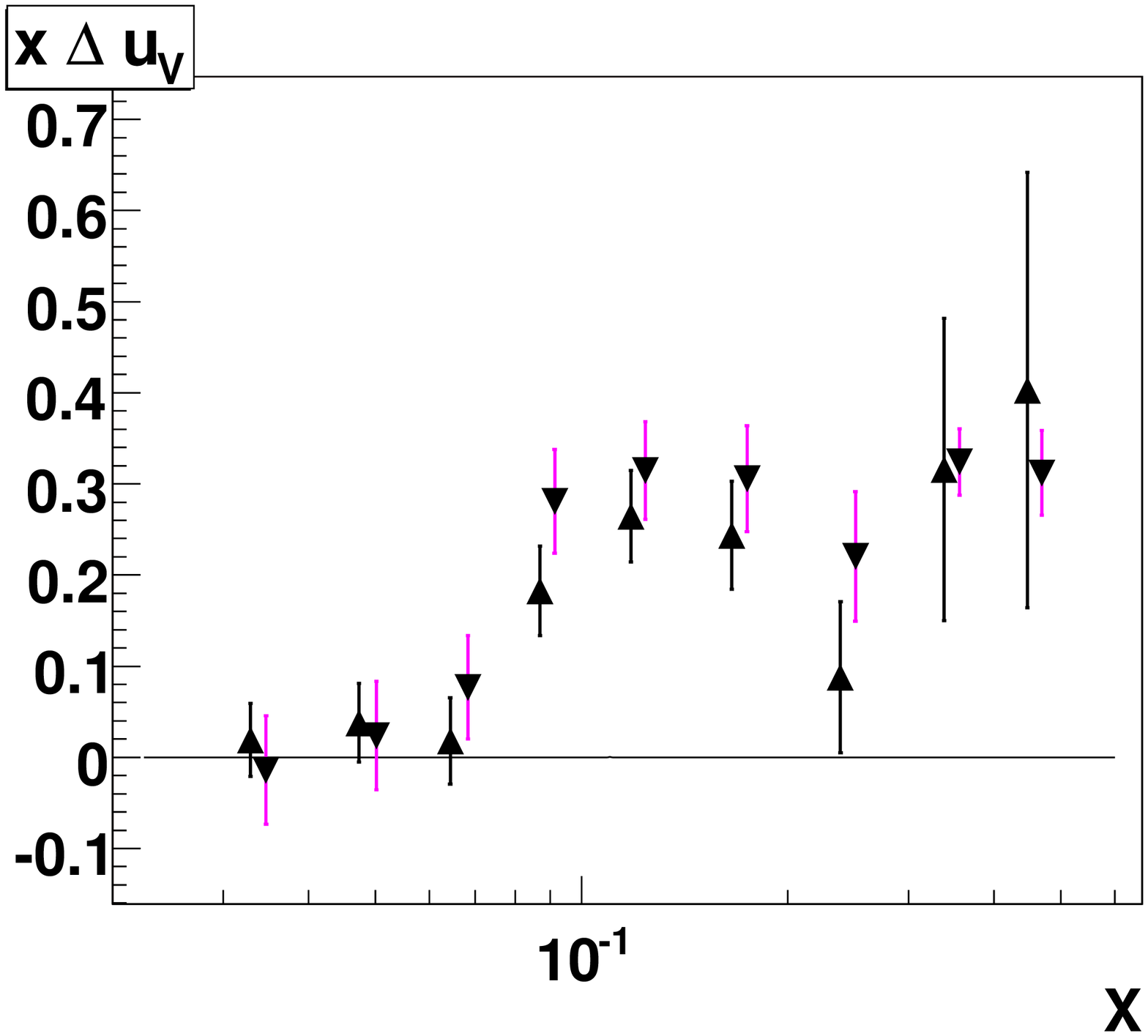}
\includegraphics[height=4.7cm,width=8.7cm]{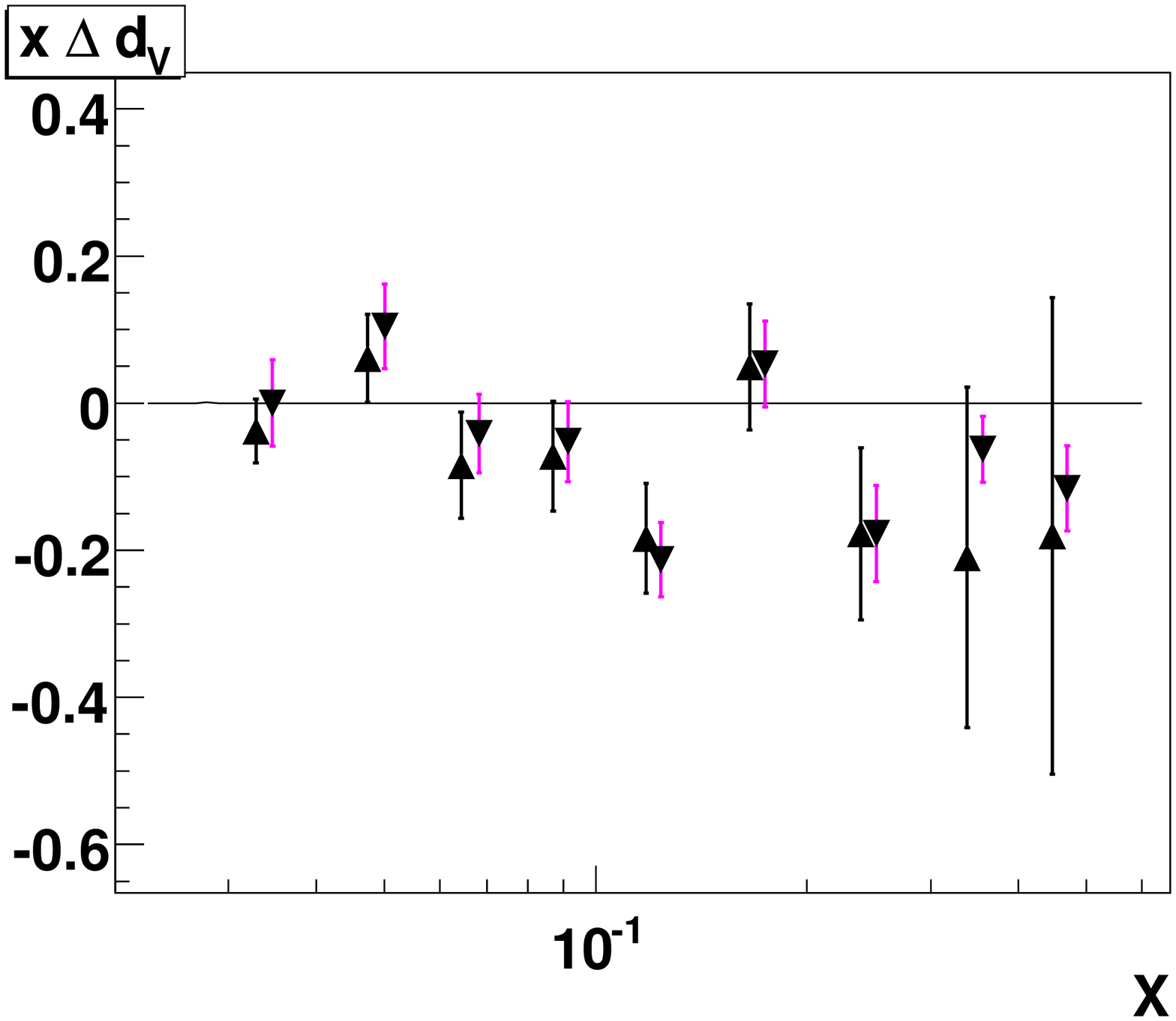}
\label{her-nomjem}
\end{figure}
\begin{figure}
        \caption{\footnotesize (color online) Different LO procedures of the valence PDFs extraction  in comparison.
        Solid line and up-oriented triangles correspond to LO
        extraction with the proposed method and 
         direct extraction with Eq. (\ref{difaslo}), respectively.
        The difference asymmetries constructed with application of Eq. (\ref{expansatz}) are used.
        Down-oriented triangles correspond to LO results of HERMES obtained with  application
        of the purity method to the measured by HERMES usual virtual photon spin asymmetries.
}
\includegraphics[height=4.7cm,width=8.7cm]{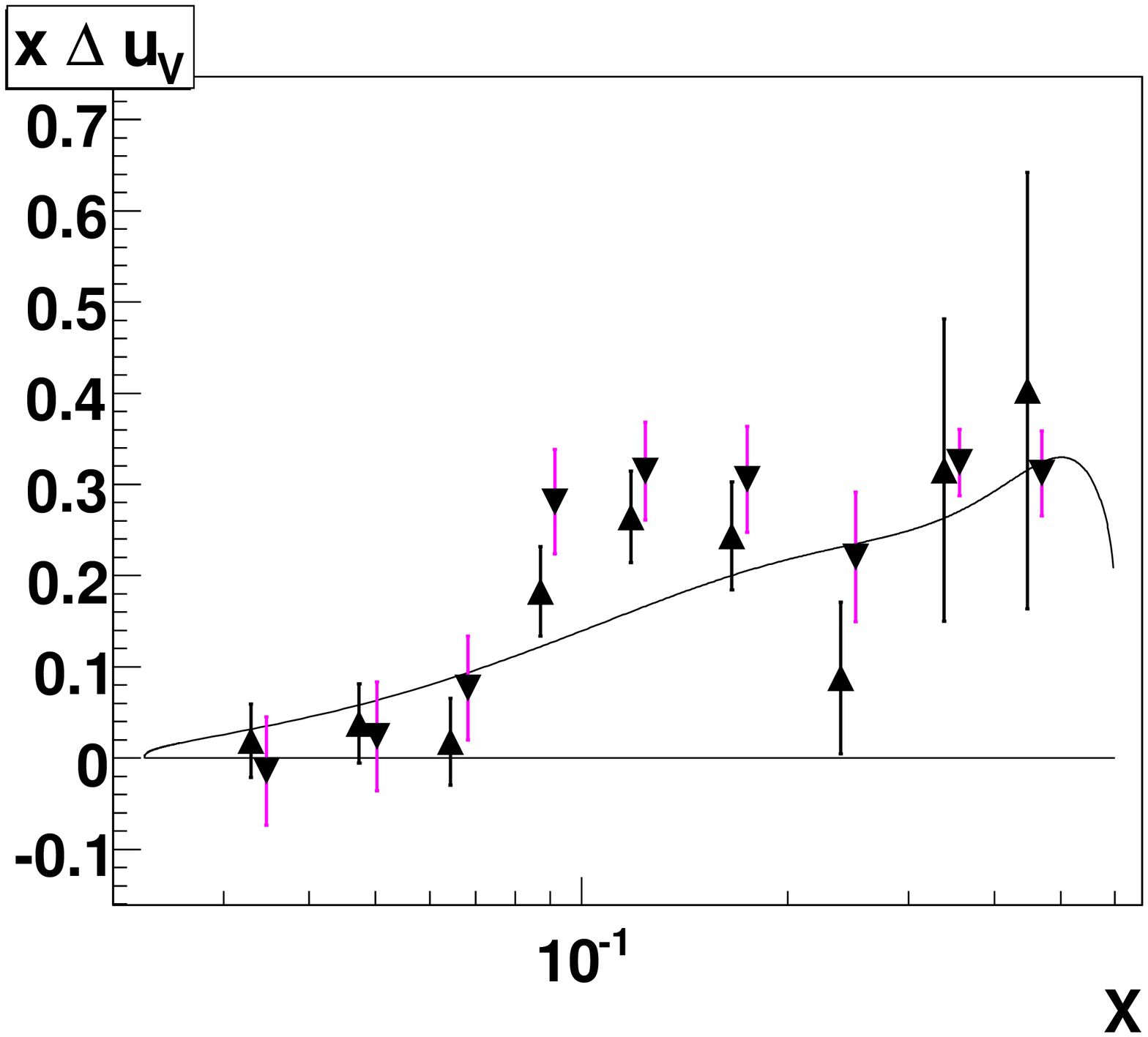}
\includegraphics[height=4.7cm,width=8.7cm]{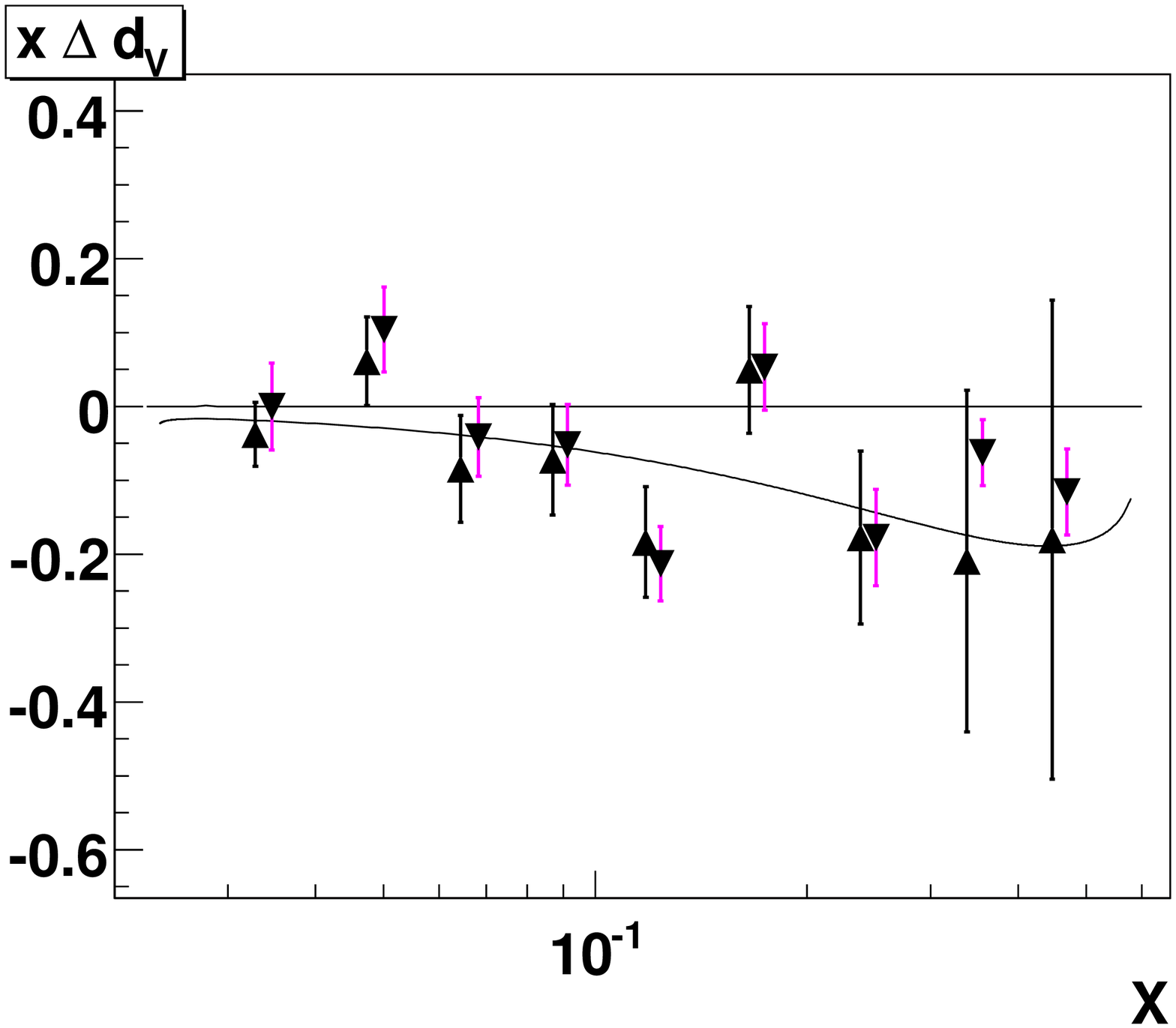}
\label{her-lo-mjem}
\end{figure}

\begin{figure}
        \caption{\footnotesize Results of both LO and NLO analysis of the difference asymmetries constructed 
        with Eq. (\ref{expansatz}). Solid and dashed lines correspond to NLO and LO results, respectively.
       }
\includegraphics[height=4.7cm,width=8.7cm]{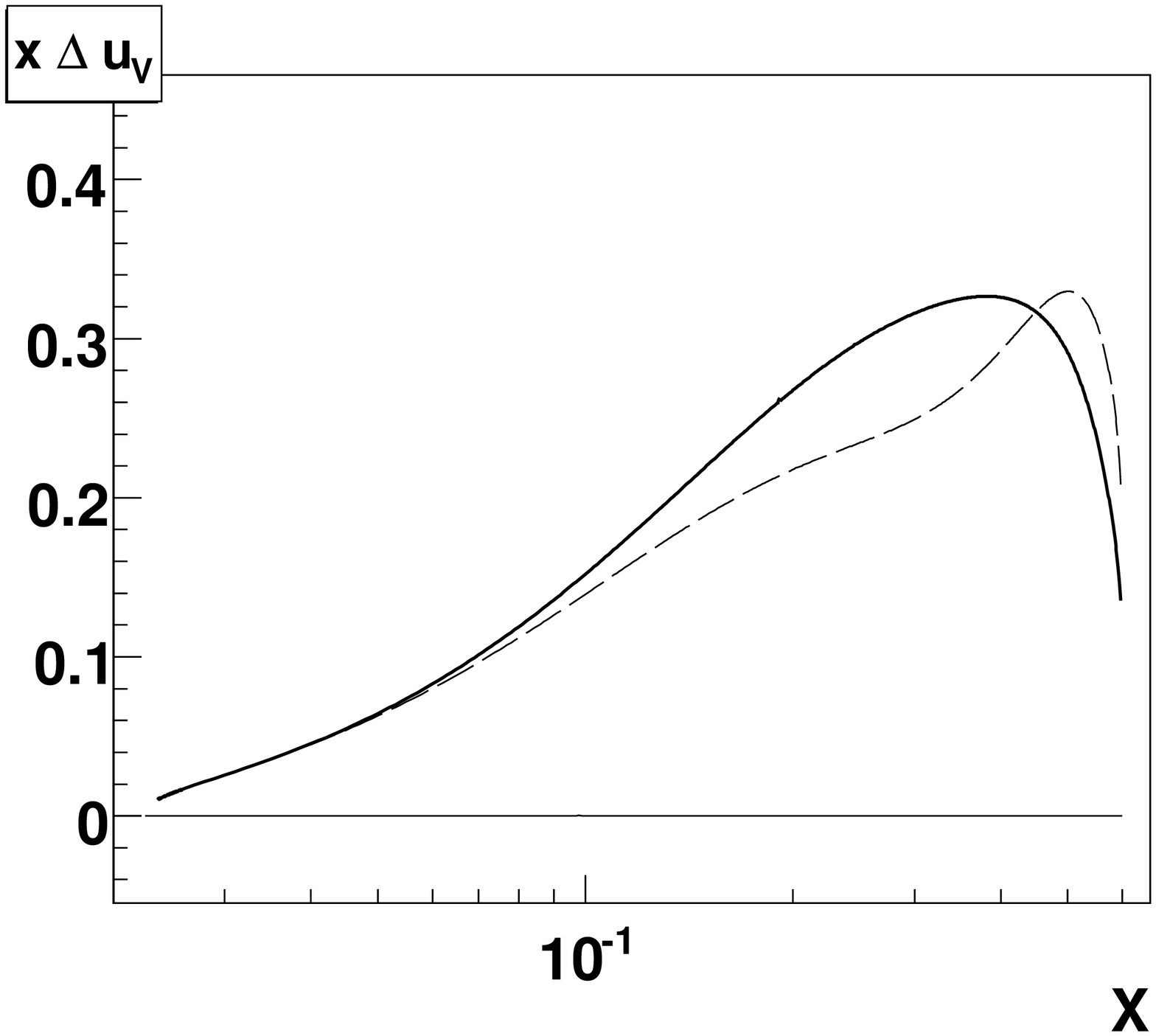}
\includegraphics[height=4.7cm,width=8.7cm]{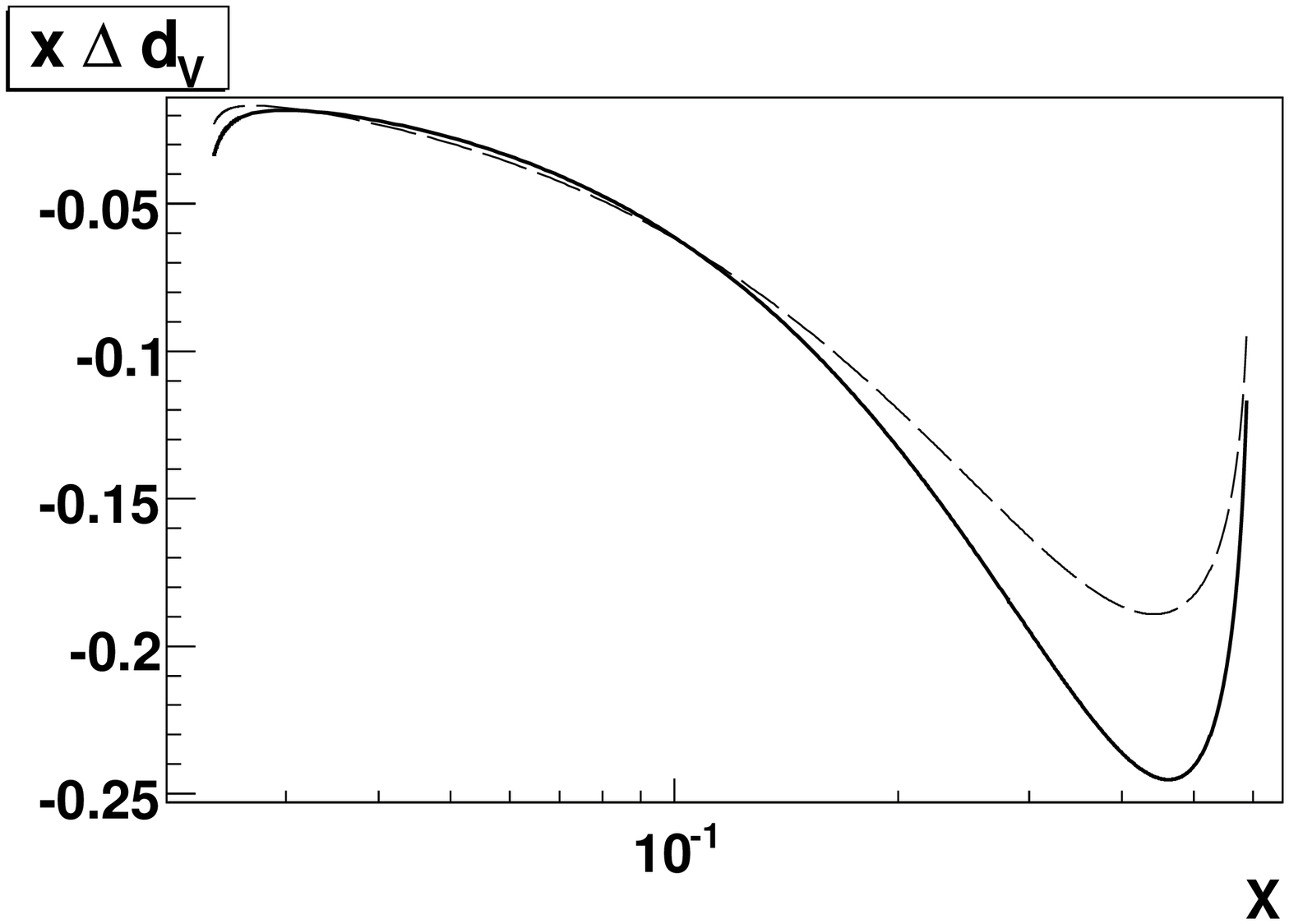}
\label{her-lonlo}
\end{figure}

\begin{figure}
        \caption{\footnotesize (color online) NLO results  obtained with application of MJEM to the moments
        extracted from the difference asymmetries corrected due to evolution. 
Dotted lines correspond to corrections estimated with the different parametrizations.
       Dashed line corresponds to the curve averaged over corrections.  Solid line corresponds to reconstruction without corrections.}
\includegraphics[height=4.7cm,width=8.7cm]{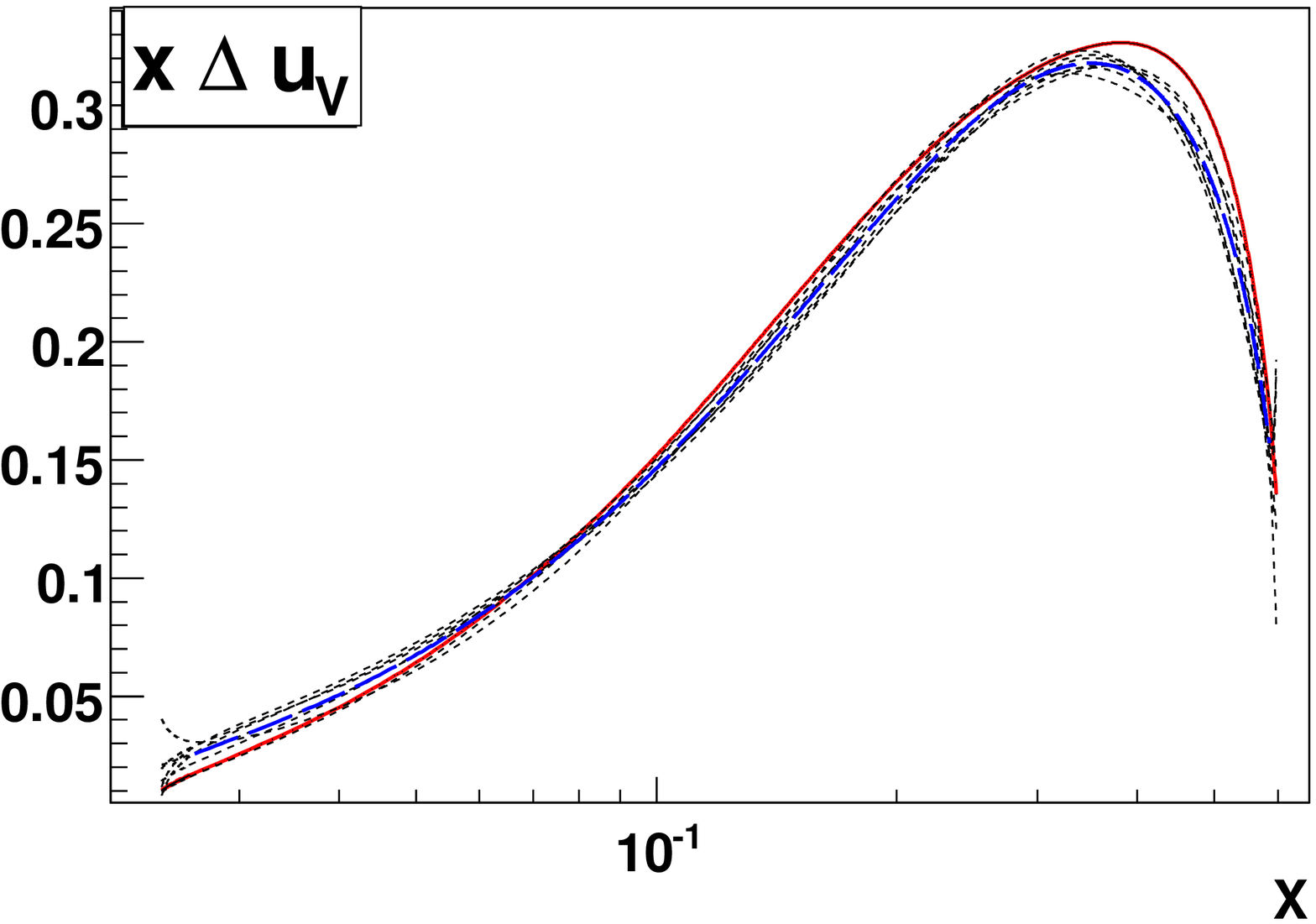}
\includegraphics[height=4.7cm,width=8.7cm]{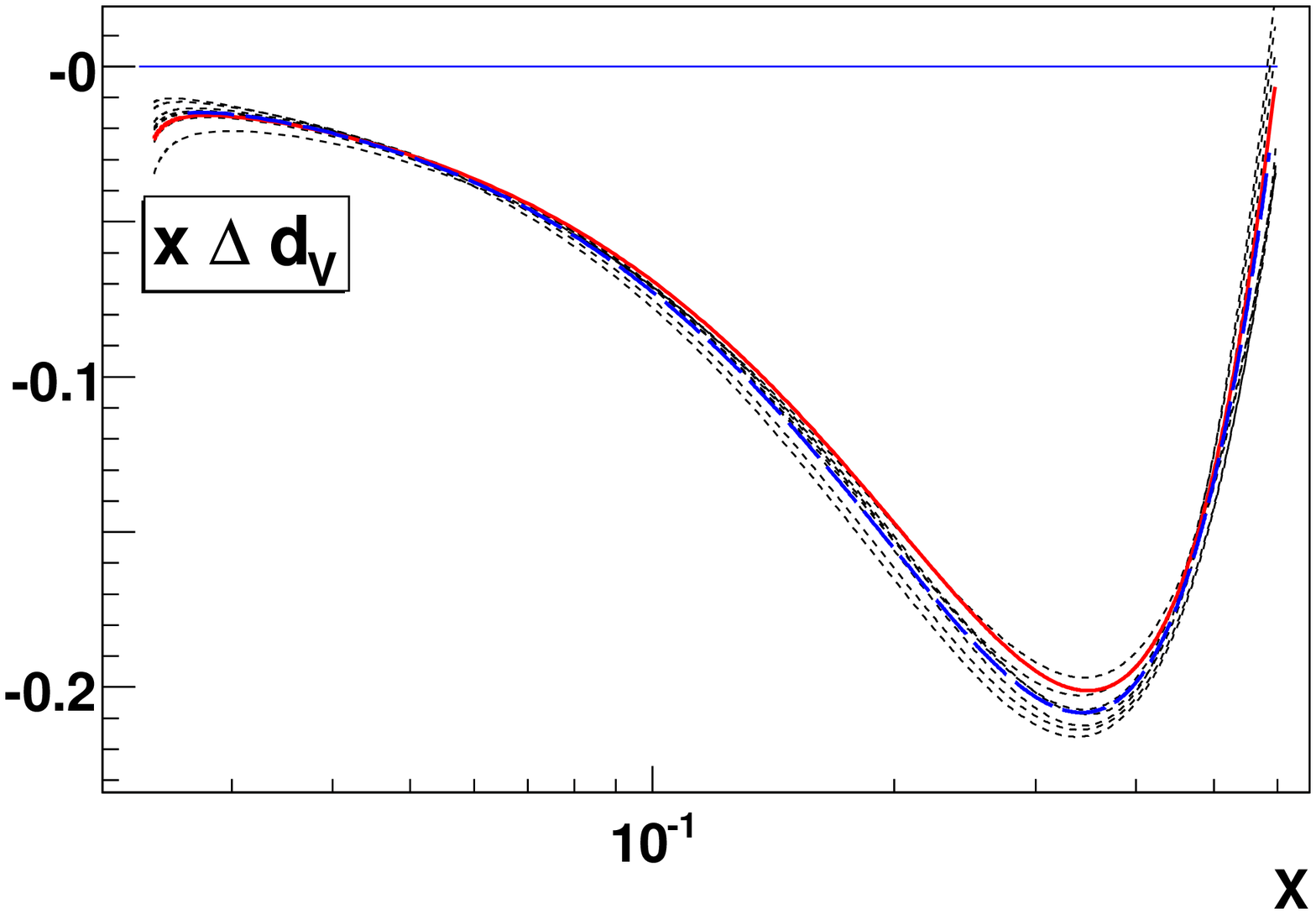}
\label{her-nlo-corrections}
\end{figure}

\clearpage


\begin{thebibliography}{99}
        \bibitem{smc} SMC collaboration (B. Adeva et al.), Phys. Lett. B {\bf 369} (1996) 93 
        \bibitem{hermes} HERMES collaboration (A. Airapetyan et al.), Phys. Rev. {\bf D71} (2005) 012003
        \bibitem{compass} COMPASS collaboration (G. Baum et al.),
                "COMPASS: A proposal for a common muon and proton apparatus
                for structure and spectroscopy", CERN-SPSLC-96-14 (1996).
        \bibitem{our1} A.N. Sissakian, O.Yu. Shevchenko and O.N. Ivanov, Phys. Rev. {\bf D68} (2003) 031502 
        \bibitem{our2} A.N. Sissakian, O.Yu. Shevchenko and O.N. Ivanov, Phys. Rev. {\bf D70} (2004) 074032 
        \bibitem{our3} A.N. Sissakian, O.Yu. Shevchenko and O.N. Ivanov, JETP Lett. {\bf 82} (2005) 53 
        \bibitem{florian-new} D. de Florian, G.A. Navarro, R. Sassot, Phys. Rev. D{\bf 71} (2005) 094018

        \bibitem{frankfurt} L. Frankfurt et al, Phys. Lett. B {\bf 230} (1989) 141  

        \bibitem{christova} E. Christova and E. Leader, hep-ph/0007303.

        \bibitem{sidorov} V.G. Krivokhizhin et al, Z. Phys. {\bf C36} (1987) 51;  JINR-E2-86-564
        \bibitem{parisi} G. Parisi, N. Sourlas, Nucl. Phys. {\bf B151} (1979) 421 
        \bibitem{barker} I.S. Barker, C.S. Langensiepen, G. Shaw, Nucl. Phys. {\bf B186} (1981) 61; CERN-TH-2988
        \bibitem{sidorov2} E. Leader, A.V. Sidorov, D.B. Stamenov, Int. J. Mod. Phys. {\bf A13} (1998) 5573  
        
        \bibitem{grsv2000} M. Gluck, E. Reya, M. Stratmann, W. Vogelsang, Phys. Rev. {\bf D63} (2001) 094005; hep-ph/0011215
        \bibitem{minuit} F. James, M. Roos, Comput. Phys. Commun. {\bf 10} (1975) 343 
        \bibitem{pepsi}  L. Mankiewicz, A. Schafer, M. Veltri, Comput. Phys. Commun. 71 (1992) 305
        \bibitem{LEPTO} G. Ingelman, A. Edin, J. Rathsman, Comput. Phys. Commun. {\bf 101} (1997) 108.
        \bibitem{parametrizations} Asymmetry Analysis collaboration (Y. Goto et al.),  Phys. Rev. D {\bf 62} (2000) 034017;\\
                E. Leader, A. Sidorov, D. Stamenov,  Eur. Phys. J. C{\bf 23} (2002) 479\\
                Asymmetry Analysis Collaboration (M. Hirai et al.), Phys. Rev. {\bf D69} (2004) 054021

        \bibitem{compass-dis} COMPASS Collaboration (E.S. Ageev et al.), Phys. Lett. B{\bf 612} (2005) 154 
        \bibitem{smc-evolution} SMC collaboration (B. Adeva et al), Phys. Rev. D{\bf 58} (1998) 112002 

        \bibitem{florian97}  D. de Florian, O.A. Sampayo, R. Sassot, Phys. Rev. D {\bf57} (1998) 5803 
        \bibitem{jlab} X. Jiang et al., hep-ex/0412010 


\end{thebibliography}
\end{document}